\documentclass[12pt]{iopart}
\usepackage{txfonts,color}
\usepackage{graphicx}
\usepackage[square,sort&compress,comma]{natbib}
\usepackage{iopams}
\usepackage{setstack}
\usepackage{stmaryrd}
\begin{document}

\title[Recurrence interval analysis of high-frequency financial returns]{Recurrence interval analysis of high-frequency financial returns and its application to risk estimation}

\author{Fei Ren $^{1, 2}$ and Wei-Xing Zhou $^{1, 2, 3, 4}$}

\address{$^1$ School of Business, East China University of Science and Technology, Shanghai 200237, China}
\address{$^2$ Research Center for Econophysics, East China University of Science and Technology, Shanghai 200237, China}
\address{$^3$ School of Science, East China University of Science and Technology, Shanghai 200237, China}
\address{$^4$ Research Center on Fictitious Economics \& Data Science, Chinese Academy of Sciences, Beijing 100080, China}%

\ead{wxzhou@ecust.edu.cn (Wei-Xing Zhou)}

\begin{abstract}
We investigate the probability distributions of the recurrence
intervals $\tau$ between consecutive 1-min returns above a positive
threshold $q>0$ or below a negative threshold $q<0$ of two indices
and 20 individual stocks in China's stock market. The distributions
of recurrence intervals for positive and negative thresholds are
symmetric, and display power-law tails tested by three
goodness-of-fit measures including the Kolmogorov-Smirnov (KS)
statistic, the weighted KS statistic and the Cram{\'{e}}r-von Mises
criterion. Both long-term and shot-term memory effects are observed
in the recurrence intervals for positive and negative thresholds
$q$. We further apply the recurrence interval analysis to the risk
estimation for the Chinese stock markets based on the probability
$W_q(\Delta{t},t)$, Value-at-Risk (VaR) analysis and VaR analysis
conditioned on preceding recurrence intervals.
\end{abstract}

\submitto{\NJP}
\maketitle

{\color{blue}{\tableofcontents}}

\section{Introduction}

The recurrence interval analysis has caused a growing interest in
extreme value statistics, and has been extensively studied in a
variety of experimental time series, such as records of climate
\cite{Bunde-Eichner-Havlin-Kantelhardt-2004-PA,Bunde-Eichner-Kantelhardt-Havlin-2005-PRL},
seismic activities \cite{Saichev-Sornette-2006-PRL} and turbulence
\cite{Liu-Jiang-Ren-Zhou-2009-XXX}. In recent years, a great deal of
financial data have been compiled up and thus makes it possible to
do relatively accurate analysis of extreme events in stock markets.
By investigating the recurrence intervals between extreme events in
stock markets, we can better understand the statistical properties
of these extreme events. This is supposed to be of great importance
for risk assessment. The recurrence intervals between volatilities
exceeding a certain threshold $q$ have been carefully studied, and
numerous phenomena have been unveiled
\cite{Kaizoji-Kaizoji-2004a-PA,Yamasaki-Muchnik-Havlin-Bunde-Stanley-2005-PNAS,Wang-Yamasaki-Havlin-Stanley-2006-PRE,Lee-Lee-Rikvold-2006-JKPS,Wang-Weber-Yamasaki-Havlin-Stanley-2007-EPJB,VodenskaChitkushev-Wang-Weber-Yamasaki-Havlin-Stanley-2008-EPJB,Jung-Wang-Havlin-Kaizoji-Moon-Stanley-2008-EPJB,Wang-Yamasaki-Havlin-Stanley-2008-PRE,Qiu-Guo-Chen-2008-PA,Ren-Zhou-2008-EPL,Ren-Guo-Zhou-2009-PA,Ren-Gu-Zhou-2009-PA}.

On the other hand, only a few papers have been devoted to the study
of recurrence intervals between large price returns
\cite{Yamasaki-Muchnik-Havlin-Bunde-Stanley-2006-inPFE,Bogachev-Eichner-Bunde-2007-PRL,Bogachev-Bunde-2008-PRE,Bogachev-Bunde-2009-PRE}.
It has been verified that the long-term correlation of the time
series has a remarkable influence on the recurrence interval
distribution. Numerical simulation studies have shown that for the
linear long-term correlated time series the recurrence interval
distribution follows a stretched exponential
\cite{Bunde-Eichner-Havlin-Kantelhardt-2003-PA,Bunde-Eichner-Kantelhardt-Havlin-2005-PRL,Altmann-Kantz-2005-PRE,Olla-2007-PRE}.
Santhannam {\em{et al.}} subsequently provided an analytical
expression for recurrence intervals distribution in linear long-term
correlated time series, and claimed that the distribution is a
combination of stretched exponential and power law form
\cite{Santhanam-Kantz-2008-PRE}. Recently, Bogachev and Bunde
studied the recurrence interval distribution for the artificial
multifractal signals which have non-linear correlation, and
power-law behavior whose exponent varies with the threshold value
was observed
\cite{Bogachev-Eichner-Bunde-2007-PRL,Bogachev-Eichner-Bunde-2008-EPJST}.
The multifractal structure has proven to be the main feature of
financial records, and consequently the recurrence interval
distribution should have power-law behavior. Power-law behavior with
exponent depending on the threshold value was indeed observed in a
variety of representative financial records
\cite{Kaizoji-Kaizoji-2004a-PA,Bogachev-Bunde-2008-PRE}.

Since the stock market crash occurring in May 2007, the Chinese
stock markets experienced a strongly fluctuating period. The
increasing number of the extreme events with large price returns
offers an opportunity to better understand the statistic properties
of the recurrence intervals. We investigate the reoccurring
intervals between price returns above a threshold $q>0$ or below a
threshold $q<0$ using the high-frequency data in recent years, and
find tails of the recurrence interval distributions obey power-law
scaling, independent of the threshold value. This scaling behavior
is consistent with some empirical results of volatility recurrence
intervals
\cite{Yamasaki-Muchnik-Havlin-Bunde-Stanley-2005-PNAS,Wang-Yamasaki-Havlin-Stanley-2006-PRE,Wang-Weber-Yamasaki-Havlin-Stanley-2007-EPJB,Jung-Wang-Havlin-Kaizoji-Moon-Stanley-2008-EPJB,Qiu-Guo-Chen-2008-PA}.
The present recurrence interval analysis mainly focuses on the
probability distribution and memory effect of the recurrence
intervals. Another purpose of this paper is to make risk estimation
for the Chinese stock markets based on the recurrence interval
analysis. We note that the empirical analysis on the recurrence of
financial returns was carried out using daily data
\cite{Yamasaki-Muchnik-Havlin-Bunde-Stanley-2006-inPFE,Bogachev-Eichner-Bunde-2007-PRL,Bogachev-Bunde-2008-PRE,Bogachev-Bunde-2009-PRE}.
In this work, we investigate the recurrence interval of 1-min
high-frequency returns to gain better statistics.

The paper is organized as follows. In Section 2, we explain the
database analyzed. Section 3 studies the probability distribution of
recurrence intervals using the Kolmogorov-Smirnov (KS) and
Cram\'{e}r-von Mises (CvM) tests. Section 4 further studies the
memory effects of recurrence intervals. In Section 5, we attempt to
perform the risk estimation based on the recurrence interval
analysis. Section 6 concludes.

\section{Data sets}

Our analysis is based on a database of China's stock market
retrieved from GTA Information Technology Co., Ltd
(http://www.gtadata.com/). We study the 1-min intraday data of $20$
liquid stocks actively traded on the Shanghai Stock Exchange and the
Shenzhen Stock Exchange from January 2000 to May 2009, the Shanghai
Stock Exchange Composite Index (SSEC) and the Shenzhen Stock
Exchange Composite Index (SZCI) from January 2003 to April 2009.
Since the sampling time is 1 minute, the number of data points is
about 340000 for each of the two Chinese indices and about 500000
for the individual stocks. These $20$ stocks are actively traded
stocks representative in a variety of industry sectors. Each stock
is uniquely identified with a stock code which is a unique 6-digit
number. A stock with the code initiated with $600$ is traded on the
Shanghai Stock Exchange, while a stock with the code initiated with
$000$ is listed on the Shenzhen Stock Exchange.

\section{Probability distribution of recurrence interval of price returns}

\begin{figure}[htb]
\centering
\includegraphics[width=8cm]{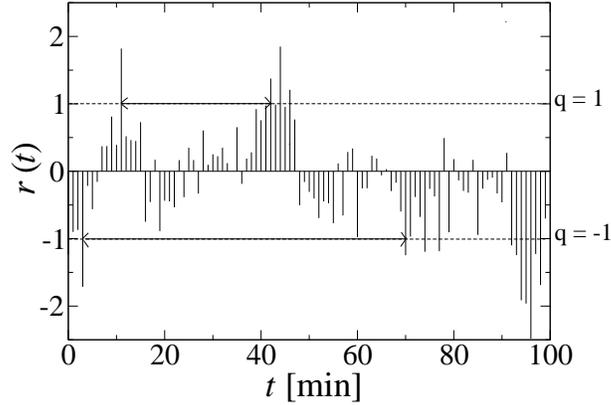}
\caption{\label{Fig:RI:Illus} Illustration of recurrence intervals
of the normalized returns of the SSEC index.}
\end{figure}

We study the recurrence intervals between extreme events with large
positive price returns or negative price returns. The price return
is defined as the logarithmic difference between two consecutive
minutes, that is,
\begin{equation}
R(t)=\ln Y(t)-\ln Y(t-1)
\end{equation}
where $Y(t)$ is minute price at time $t$. Before we calculate the
recurrence interval, we normalize the price return of each stock by
dividing its standard deviation
\begin{equation}
r(t)= \frac{R(t)}{[\langle R(t)^2 \rangle-\langle R(t)
\rangle^2]^{1/2}} . \label{e10}
\end{equation}
We investigate the interval $\tau \equiv \tau_q$ between two
consecutive price returns above a positive threshold $q>0$ or below
a negative threshold $q<0$ as illustrated in figure
\ref{Fig:RI:Illus}, and then study the statistics of these
recurrence intervals between consecutive large price increases or
decreases.

\subsection{PDF of recurrence interval of price returns}

\begin{figure}[htb]
\centering
\includegraphics[width=5cm]{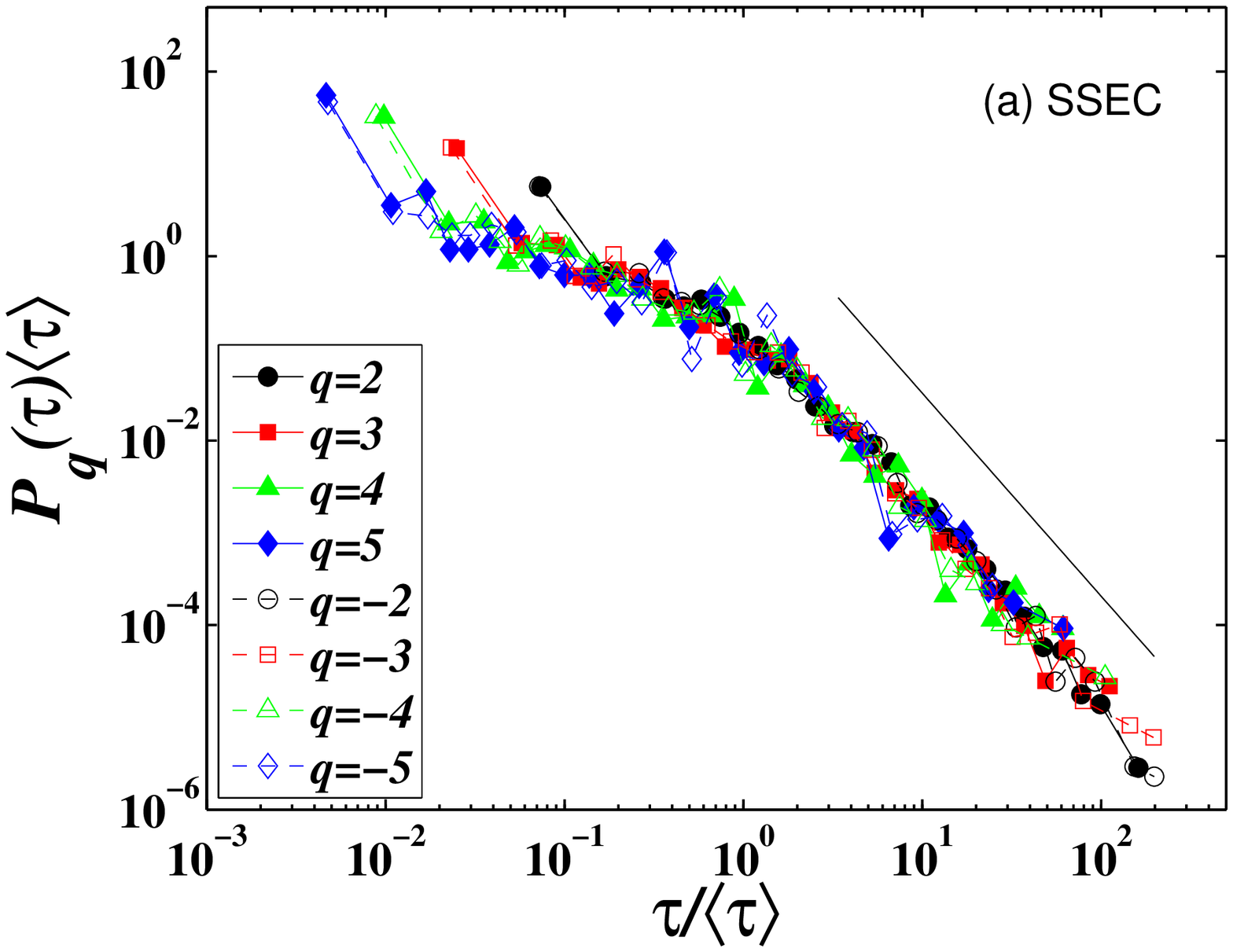}
\includegraphics[width=5cm]{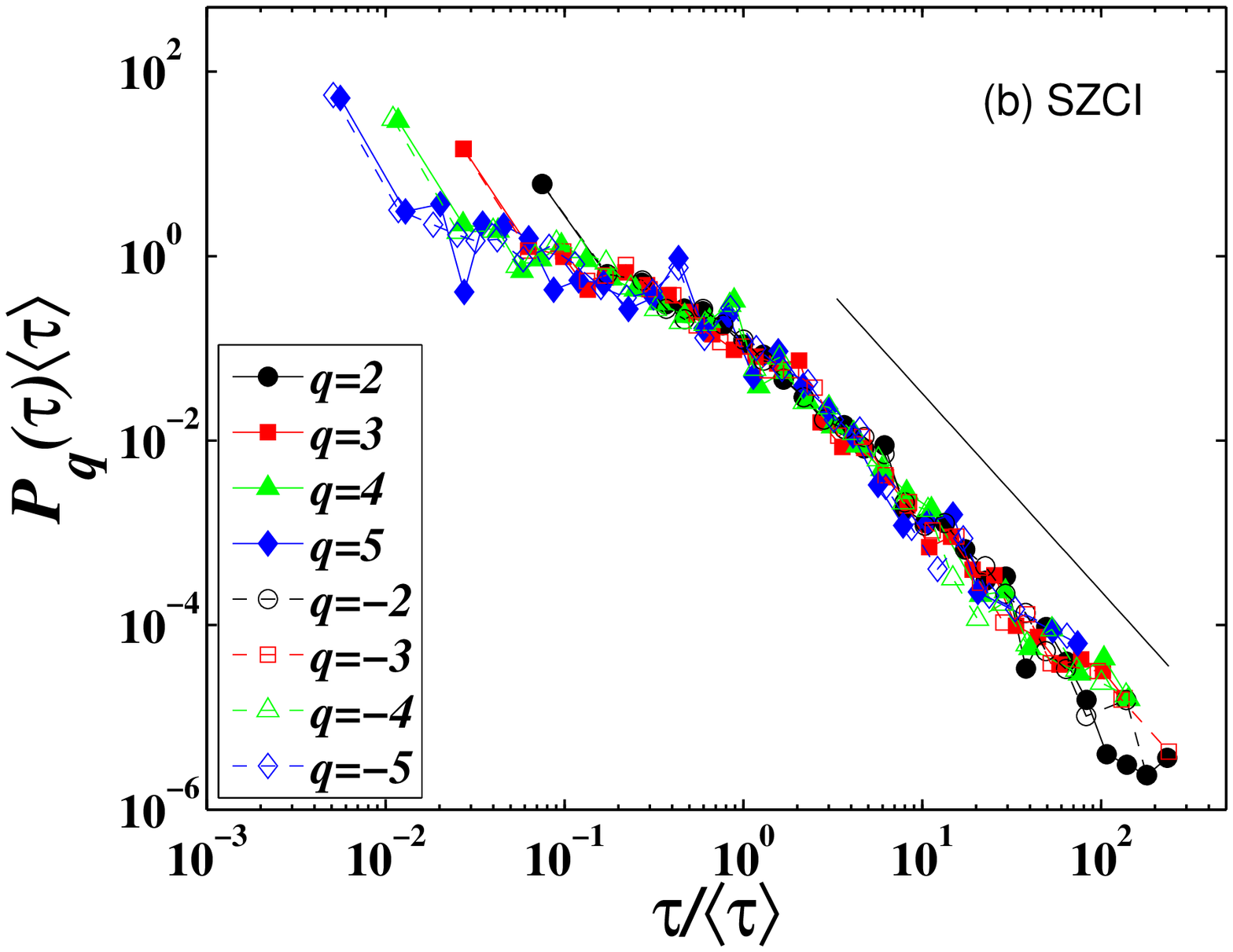}
\includegraphics[width=5cm]{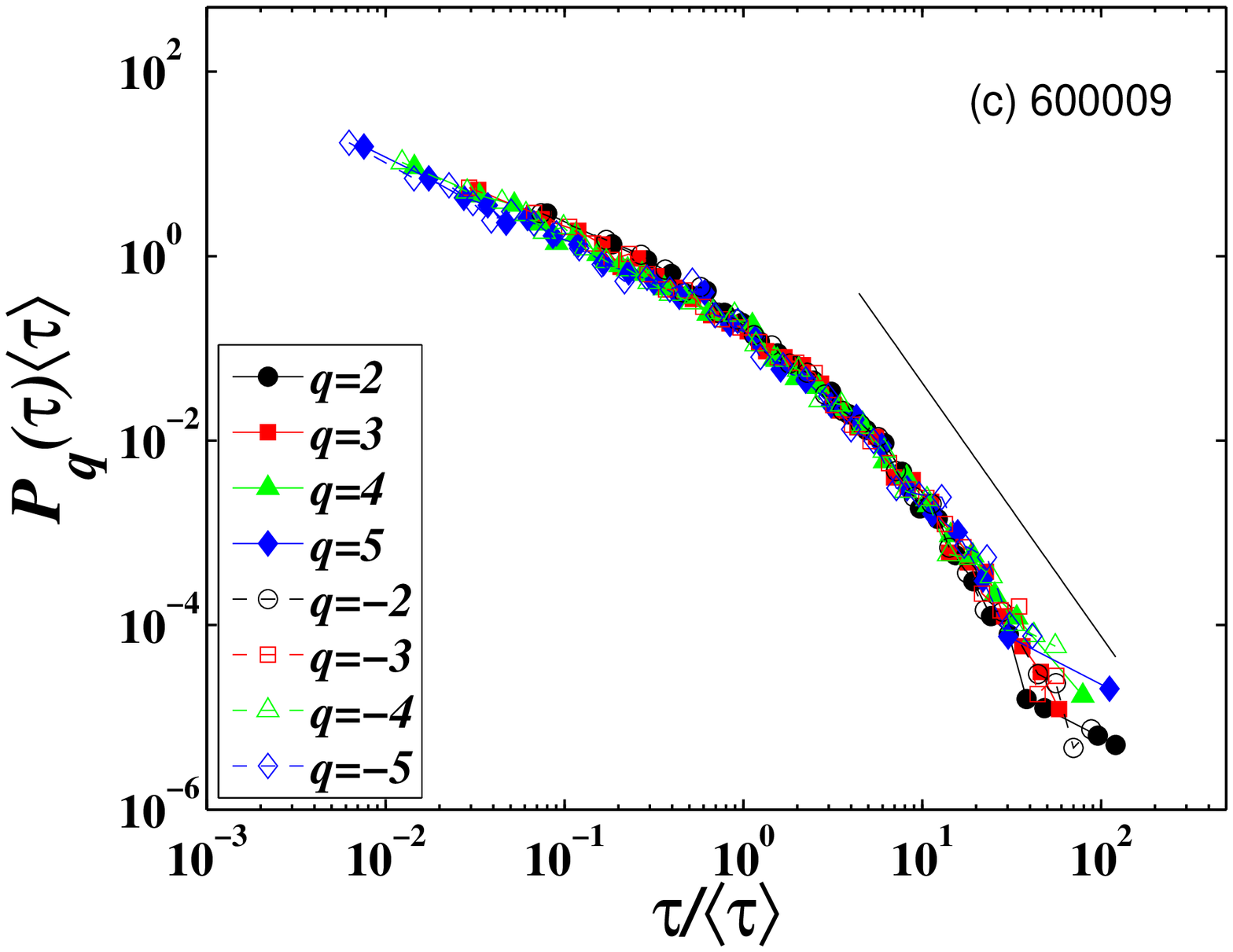}
\includegraphics[width=5cm]{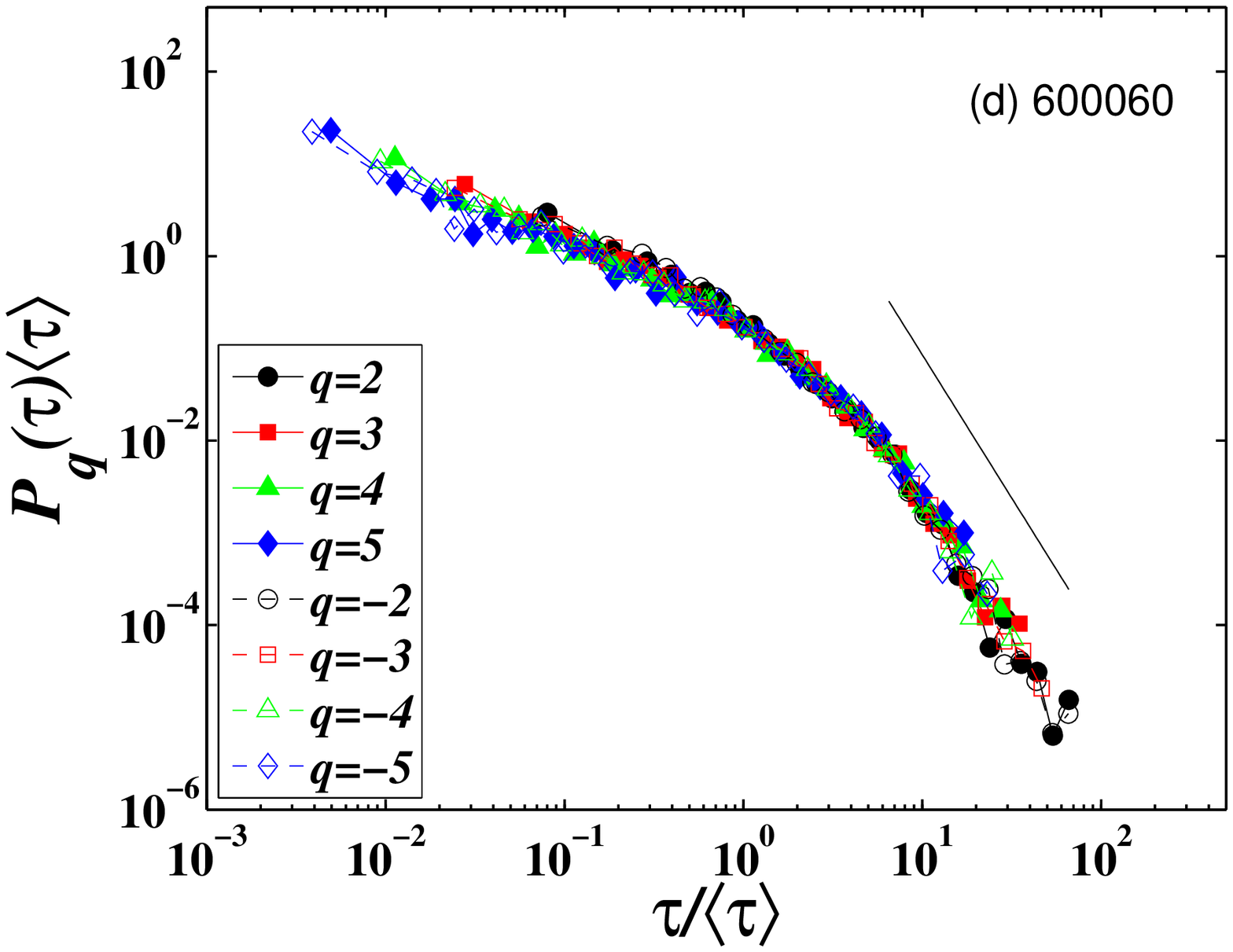}
\includegraphics[width=5cm]{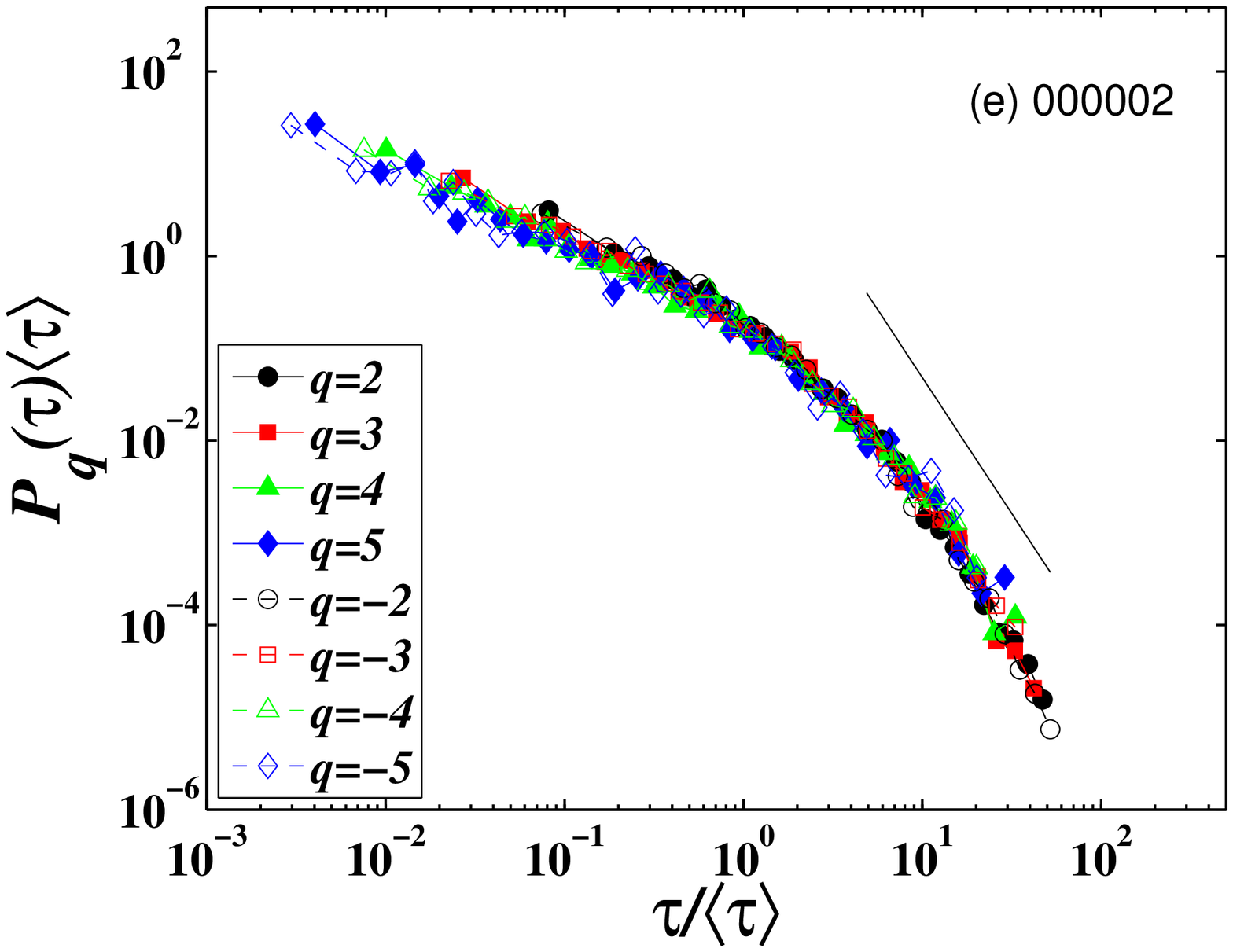}
\includegraphics[width=5cm]{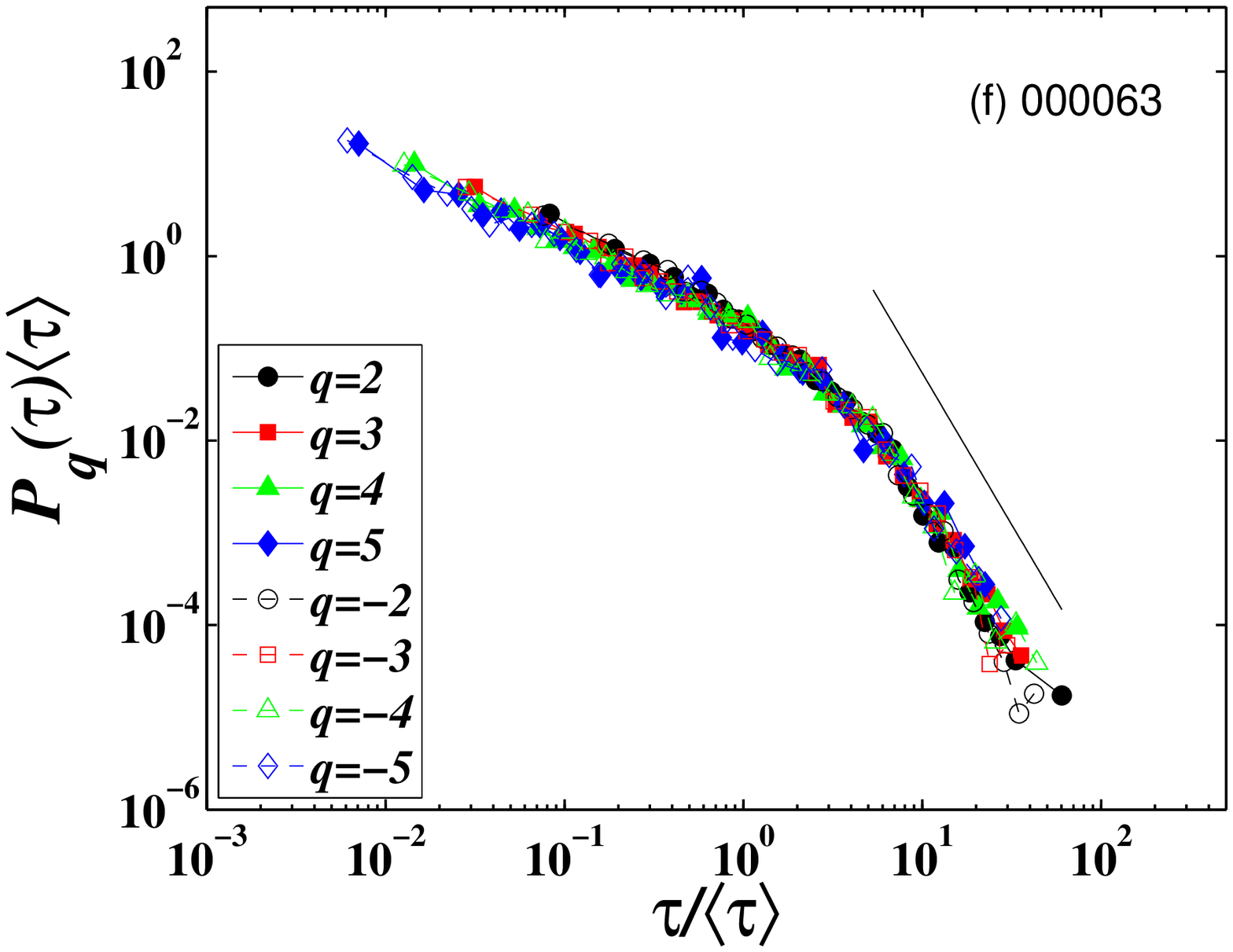}
\caption{\label{Fig:RI:PDF} Empirical probability distributions of
scaled recurrence intervals for different thresholds
$q=-5,-4,-3,-2,2,3,4,5$ for (a) SSEC, (b) SZCI, and (c)-(e) four
representative stocks 600009, 600060, 000002 and 000063. The solid
curves are the fitted functions $c x^{-\delta}$ with parameters
listed in Table~\ref{TB:goodness-of-fit-test}.}
\end{figure}

We first calculate the empirical probability distribution function
(PDF) $P_q(\tau)$ of recurrence intervals of price returns. In
figure \ref{Fig:RI:PDF}, the scaled PDFs $P_q(\tau)\langle \tau
\rangle$  are plotted as a function of the scaled recurrence
intervals $\tau / \langle \tau \rangle$ for both positive and
negative thresholds for the two Chinese indices and four
representative stocks, where $\langle \tau \rangle$ is the mean
recurrence interval. For both the Chinese indices and individual
stocks, one observes $P_q(\tau)\langle \tau \rangle$ for $q>0$ shows
a profile very similar to that for $q<0$ with the same magnitude of
thresholds, e.g. $q=2$ and $q=-2$. This indicates that the
recurrence interval distributions for positive and negative
thresholds are symmetric.

We find that, for small $\tau / \langle \tau \rangle$, the scaled
PDFs for different $q$ values differ from each other, especially for
the two Chinese indices, which is partly due to the discreteness
effect. In contrast, the tails of the scaled PDFs nicely collapse
onto a single curve displaying a scaling behavior. Speaking
differently, $P_q(\tau)\langle \tau \rangle$ only depends on $\tau /
\langle \tau \rangle$ as
\begin{equation}
P_q(\tau)= \frac{1} {\langle \tau \rangle} f ( \tau/\langle \tau
\rangle ), \label{Eq:Pq:f}
\end{equation}
and do not depend on the threshold $q$ when $\tau / \langle \tau
\rangle$ is large enough.

\subsection{Fitting the scaling function of recurrence interval PDFs}

We then focus on the tail of recurrence interval PDF and study the
particular form of the scaling function. The curves in
figure~\ref{Fig:RI:PDF} suggest that the tails of the PDFs may
follow power-law form. We assume that the empirical PDF above
$x_{\min}$ obeys a scaling form as in Eq.~(\ref{Eq:Pq:f}) and the
scaling function follows
\begin{equation}
  f(\tau/\langle \tau \rangle)=f(x)=c x^{-\delta},\ x\geq x_{\min}
  \label{Eq:PL}
\end{equation}
where $x_{\min}$ is the lower bound of the power-law distribution.
Since our hypothesis is that the empirical PDFs for different $q$
values above $x_{\min}$ are coincident with their common best
power-law fit, we aggregate the interval samples for different $q$
values above $x_{\min}$, and fit them using a common power-law
function.

To make an accurate estimation of the parameters for this power law
distribution, we use a method proposed by Clauset, Shalizi and
Newman using maximum likelihood method based on the
Kolmogorov-Smirnov (KS) statistic
\cite{Clauset-Shalizi-Newman-2009-SIAMR}. Supposing $F_q$ is the
cumulative distribution function (CDF) for empirical data and
$F_{\rm{PL}}$ the CDF of power-law fit. The $KS$ statistic is
defined as
\begin{equation}
   KS = \max_{x\geq \hat{x}_{\min}} \left(|F-F_{\rm{PL}}|\right).
   \label{Eq:KS}
\end{equation}
With a simple fundamental idea that making the empirical PDF and the
best power-law fit as similar as possible, the estimate
$\hat{x}_{\min}$ is determined by minimizing the $KS$ statistic. The
parameters $c$ and $\delta$ are estimated using maximum likelihood
method. The fitted power law lines are illustrated in
figure~\ref{Fig:RI:PDF} with the estimated parameters
$\hat{x}_{\min}$, $\delta$, $c$ and resultant $KS$ statistic
depicted in Table~\ref{TB:goodness-of-fit-test}. Among the $20$
individual stocks, there are $16$ stocks which have $x_{\min}
\lesssim 10 $ showing a scaling region larger than one order of
magnitude, and their power-law exponents are estimated to be
$3.0\pm0.3$. The two Chinese indices have scaling regions more than
two orders of magnitude with power-law exponents $2.2\pm0.1$.

\begin{table}[htp]
 \centering
 \caption{$KS$, $KSW$ and $CvM$ tests of the recurrence interval
distributions for different $q$ values by comparing empirical data
with the best fitted distribution and synthetic data with the best
fitted distribution. } \label{TB:goodness-of-fit-test}
\begin{tabular}{cccccccc}
  \hline
   Code &  $\hat{x}_{\min}$ & $\delta$ & $c$ & $KS$ & $p_{KS}$ & $p_{KSW}$ & $W^2$ \\
  \hline
   SSEC  & $3.39$ & $2.20(0.03)$ & $0.29$ & $0.03$ & $0.033$ & $\color{red}{\bf 0.000}$ & $0.31$\\%
   SZCI  & $3.33$ & $2.15(0.03)$ & $0.24$ & $0.04$ & $0.010$ & $0.103$ & $0.53$\\%
 600000  & $4.95$ & $2.78(0.04)$ & $1.30$ & $0.05$ & $\color{red}{\bf 0.000}$ & $0.058$ & $\color{red}{\bf 1.37}$\\%
 600009  & $4.44$ & $2.75(0.04)$ & $1.12$ & $0.02$ & $0.035$ & $0.356$ & $0.39$\\%
 600016  & $4.86$ & $3.01(0.05)$ & $1.97$ & $0.03$ & $0.016$ & $0.216$ & $0.49$\\%
 600058  & $5.61$ & $2.95(0.05)$ & $1.87$ & $0.04$ & $\color{red}{\bf0.005}$ & $0.098$ & $0.71$\\%
 600060  & $6.51$ & $3.11(0.07)$ & $2.57$ & $0.03$ & $0.143$ & $0.282$ & $0.20$\\%
 600073  &$18.36$ & $3.72(0.21)$ &$28.76$ & $0.03$ & $0.968$ & $0.719$ & $0.03$\\%
 600088  & $4.97$ & $2.94(0.05)$ & $1.73$ & $0.03$ & $0.023$ & $0.163$ & $0.48$\\%
 600098  &$10.11$ & $3.65(0.12)$ &$14.33$ & $0.03$ & $0.374$ & $0.519$ & $0.16$\\%
 600100  & $4.90$ & $2.84(0.04)$  &$1.56$ & $0.05$ & $\color{red}{\bf 0.000}$ & $0.054$ & $\color{red}{\bf 1.30}$\\%
 600104  & $5.15$ & $3.04(0.05)$ & $2.25$ & $0.03$ & $\color{red}{\bf0.006}$ & $0.013$ & $0.74$\\%
 000001  & $9.31$ & $3.23(0.09)$ & $4.94$ & $0.04$ & $0.097$ & $0.650$ & $0.22$\\%
 000002  & $4.89$ & $2.95(0.05)$ & $1.79$ & $0.03$ & $0.013$ & $0.134$ & $0.45$\\%
 000021  & $5.41$ & $3.26(0.06)$ & $3.49$ & $0.03$ & $0.027$ & $0.170$ & $0.40$\\%
 000027  & $5.25$ & $3.18(0.05)$ & $2.98$ & $0.03$ & $0.011$ & $0.283$ & $0.30$\\%
 000029  & $4.72$ & $2.76(0.04)$ & $1.14$ & $0.03$ & $0.010$ & $0.256$ & $0.38$\\%
 000031  &$11.67$ & $3.90(0.15)$ &$30.43$ & $0.04$ & $0.492$ & $0.713$ & $0.08$\\%
 000039  & $5.16$ & $3.22(0.06)$ & $2.99$ & $0.03$ & $0.010$ & $0.121$ & $0.39$\\%
 000060  &$11.28$ & $3.38(0.12)$ & $7.11$ & $0.02$ & $0.963$ & $0.597$ & $0.04$\\%
 000063  & $5.32$ & $3.29(0.06)$ & $3.71$ & $0.02$ & $0.102$ & $0.460$ & $0.18$\\%
 000066  &$10.65$ & $3.44(0.10)$ & $9.94$ & $0.03$ & $0.475$ & $0.444$ & $0.12$\\%
  \hline
\end{tabular}
\end{table}

We have shown how to fit the empirical recurrence interval PDFs and
provide good estimation of the parameters. It is necessary to test
how good the power law fits the empirical PDF. We further perform
the goodness-of-fit test using $KS$ statistic. In doing so, the
bootstrapping approach is adopted
\cite{Clauset-Shalizi-Newman-2009-SIAMR,Gonzalez-Hidalgo-Barabasi-2008-Nature}.
We first generate 1000 synthetic samples from the power-law
distribution that best fits the empirical distribution, and then
reconstruct the cumulative distribution $F_{\rm{sim}}$ of each
simulated sample and its CDF $F_{\rm{sim,PL}}$ of the best power-law
fit. We calculate the $KS$ statistic for the synthetic data from
\begin{equation}
   KS_{\rm{sim}} = \max\left(|F_{\rm{sim}}-F_{\rm{sim,PL}}|\right).
   \label{Eq:KS2:sim}
\end{equation}
This $KS$ statistic is relatively insensitive on the edges of two
cumulative distributions. To avoid this problem we use a weighted
$KS$ statistic defined as
\cite{Gonzalez-Hidalgo-Barabasi-2008-Nature}
\begin{equation}
KSW_{\rm{sim}} = \max
 \left( \frac{|F_{\rm{sim}}-F_{\rm{sim,PL}}|} {\sqrt{F_{\rm{sim,PL}}(1-F_{\rm{sim,PL}})}} \right). \label{Eq:KSw}
\end{equation}
Thus it could be uniformly sensitive across the whole range. The
$p$-value is determined by the frequency that $KS_{\rm{sim}}>KS$ or
$KSW_{\rm{sim}}>KSW$, where $KSW$ is the weighted statistic for the
empirical data. The tests are carried out for the two Chinese
indices and the 20 individual stocks, and the resultant $p$-values
are depicted in Table \ref{TB:goodness-of-fit-test}.

A $p$-value close to $1$ indicates that the empirical PDFs for
different $q$ values are coincident with their common power-law fit
as good as the synthetic data generated from the power-law fit,
whereas a relative small $p$ suggests that the empirical PDFs could
not be well described by their common power-law fit. We consider the
significance level of 1\%. If the $p$-value of an individual stock
is less than 1\%, then the null hypothesis that the empirical PDFs
of this stock can be well fitted by their common power-law fit is
rejected. As shown in Table~\ref{TB:goodness-of-fit-test}, the null
hypothesis is rejected for four stocks (600000, 600058, 600100,
600104) using the $KS$ statistic and for SSEC using the $KSW$
statistic. Based upon the fact that $16$ stocks (out of 20 stocks
analyzed) pass the test using both $KS$ and $KSW$ statistics, we can
conclude that for most of the stocks the tails of recurrence
interval PDFs obey scaling behavior and the scaling function could
be nicely fitted by a power law.

To further test the goodness of this power-law fit, we use another
goodness-of-fit measure based on the Cram\'{e}r-von Mises (CvM)
statistic
\begin{equation}
   W^2 = N \int_{-\infty}^{\infty} ( F-F_{PL})^2 d F_{PL},
   \label{Eq:CvM}
\end{equation}
where $F$ is the CDF of empirical data, $F_{PL}$ is the CDF of the
power-law fit, and $N$ is the total number of scaled interval
samples $x=\tau / \langle \tau \rangle$
\cite{Pearson-Stephens-1962-Bm,Stephens-1964-Bm,Stephens-1970-JRSSB}.
For a sequences of scaled interval samples  $x_1, x_2,\cdots,x_N$,
arranged in ascending order, $W^2$ could be calculated from
\begin{equation}
   W^2 = \frac{1}{12N} + \sum_{i=1}^{N} \left( x_i - \frac{2i-1}{2N} \right)^2.
   \label{Eq:CvM:computation}
\end{equation}
Consider the significance level of $1\%$ and $N\gg1$, if the $W^2$
statistic is greater than a critical value $0.743$ (see the critical
values for different significance levels in
Refs.~\cite{Pearson-Stephens-1962-Bm,Stephens-1964-Bm}), the
hypothesis that the empirical PDFs coincident with their common
power-law fit is rejected. The CvM tests are carried out for the two
Chinese indices and the $20$ stocks, and the corresponding values of
statistic $W^2$ are depicted in the last column of
Table~\ref{TB:goodness-of-fit-test}. Two stocks ($600000$, $600100$)
show $W^2$ greater than the critical value, thus fail in the test
and the tails of their empirical PDFs consequently could not be
approximated by the power-law distribution. $18$ stocks (out of 20
stocks analyzed) and two Chinese indices pass the CvM test, which
further confirms our results that for most of the stocks the tails
of recurrence interval PDFs could be nicely fitted by the power-law
distribution. The CvM test provides similar results to the KS test,
and principally the $W^2$ statistic is smaller when the $p$-value of
a stock is larger.

\section{Memory effect in recurrence interval of price returns}

To fully understand the statistical properties of the recurrence
intervals of price returns, we further investigate the temporal
correlation between them. Empirical studies have revealed that there
exists a memory effect in the volatility recurrence intervals of
various stock markets
\cite{Wang-Yamasaki-Havlin-Stanley-2006-PRE,Wang-Weber-Yamasaki-Havlin-Stanley-2007-EPJB,Jung-Wang-Havlin-Kaizoji-Moon-Stanley-2008-EPJB,Qiu-Guo-Chen-2008-PA,Ren-Guo-Zhou-2009-PA}.
In contrast, to the best of our knowledge, the memory effects of the
recurrence intervals of financial returns have not been investigated
\cite{Yamasaki-Muchnik-Havlin-Bunde-Stanley-2006-inPFE,Bogachev-Eichner-Bunde-2007-PRL,Bogachev-Bunde-2008-PRE}.
Indeed, we do observe memory effects in the recurrence intervals
between large positive and negative returns in the Chinese stock
markets. Since the results of the recurrence intervals between
positive returns show very similar behavior to that of the
recurrence intervals between negative returns, we mainly show the
results of the recurrence intervals between consecutive price
returns below negative thresholds $q<0$.

\subsection{Conditional PDF}
\label{S2:CondPDF}

\begin{figure}[t]
\centering
\includegraphics[width=5cm]{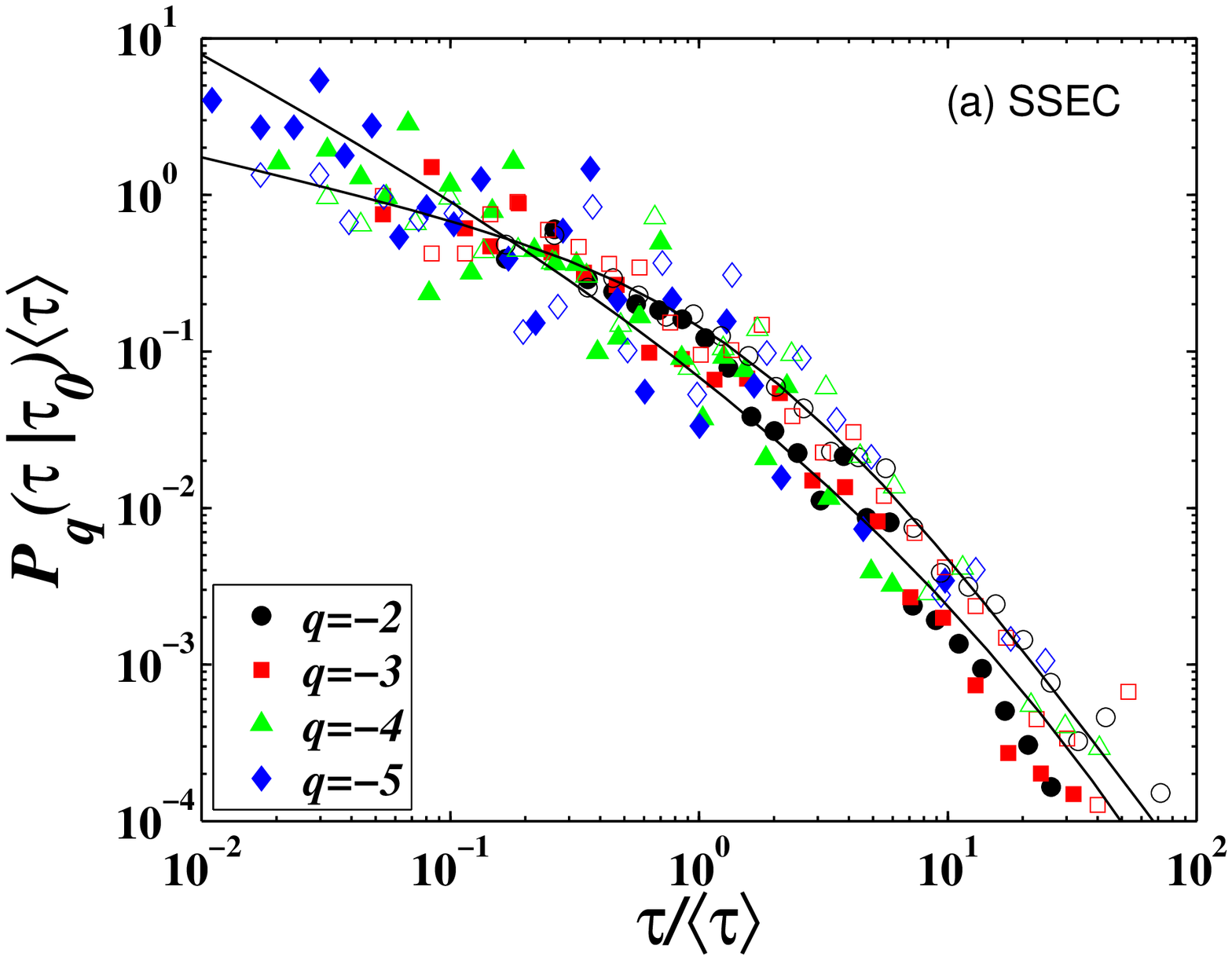}
\includegraphics[width=5cm]{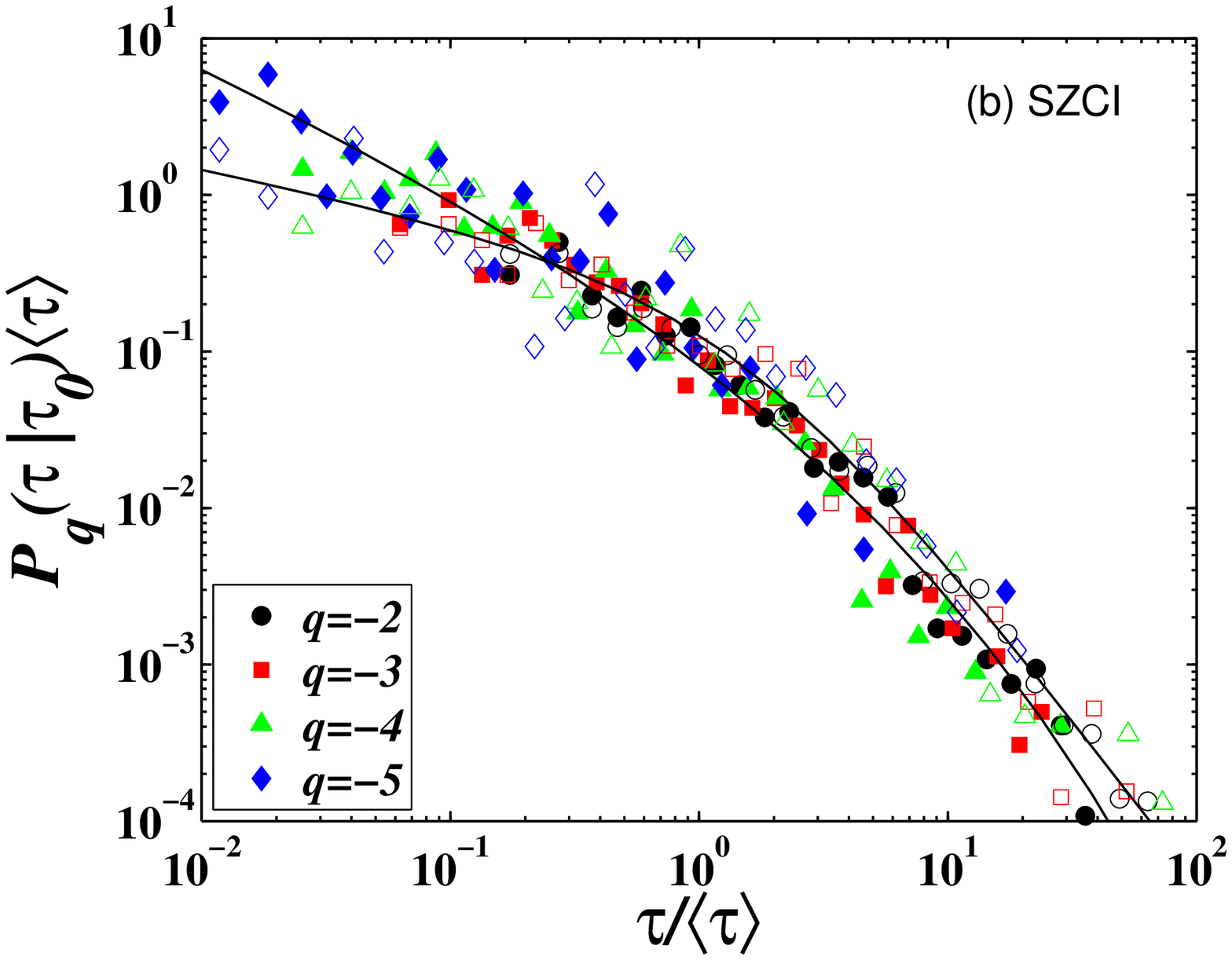}
\includegraphics[width=5cm]{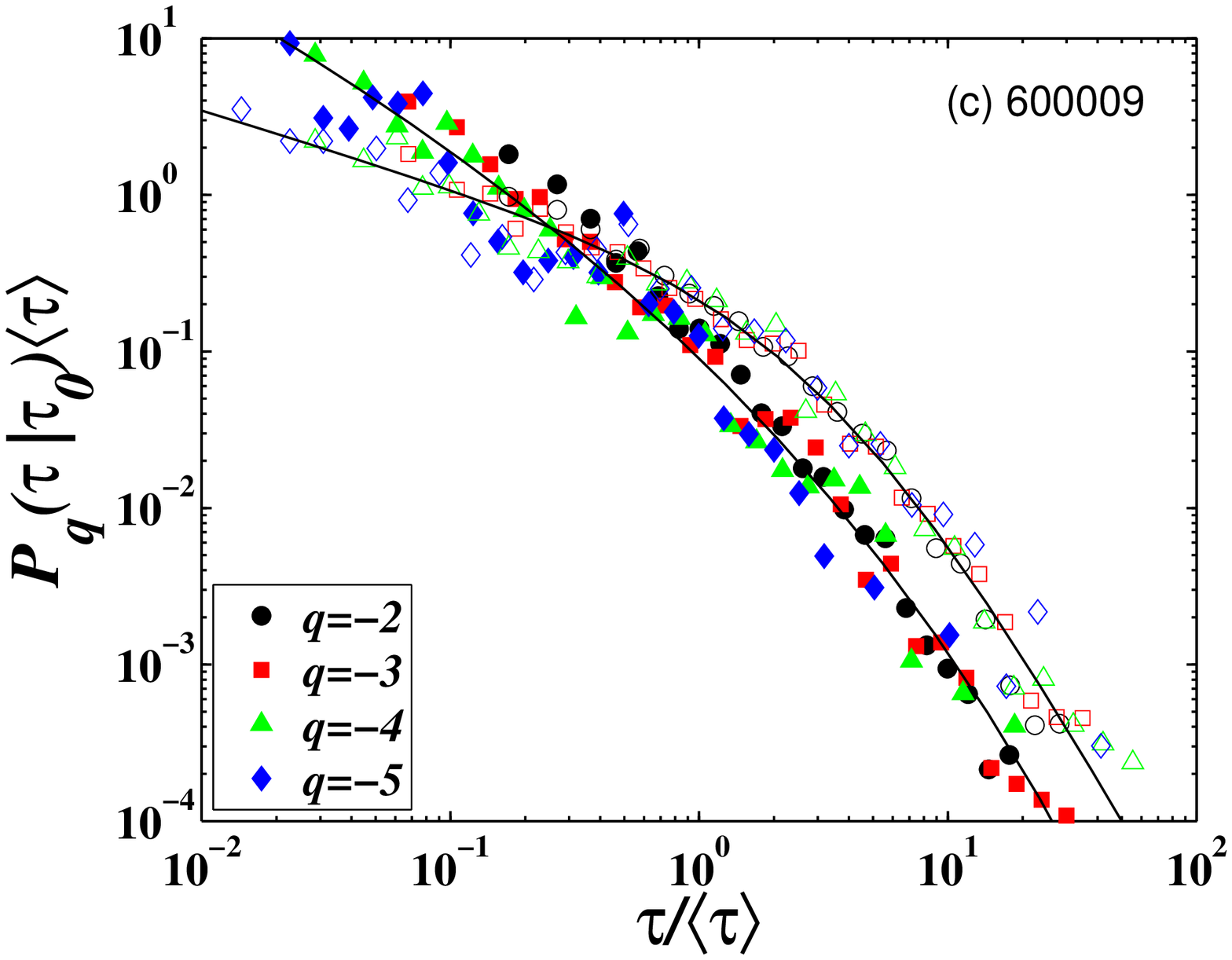}
\includegraphics[width=5cm]{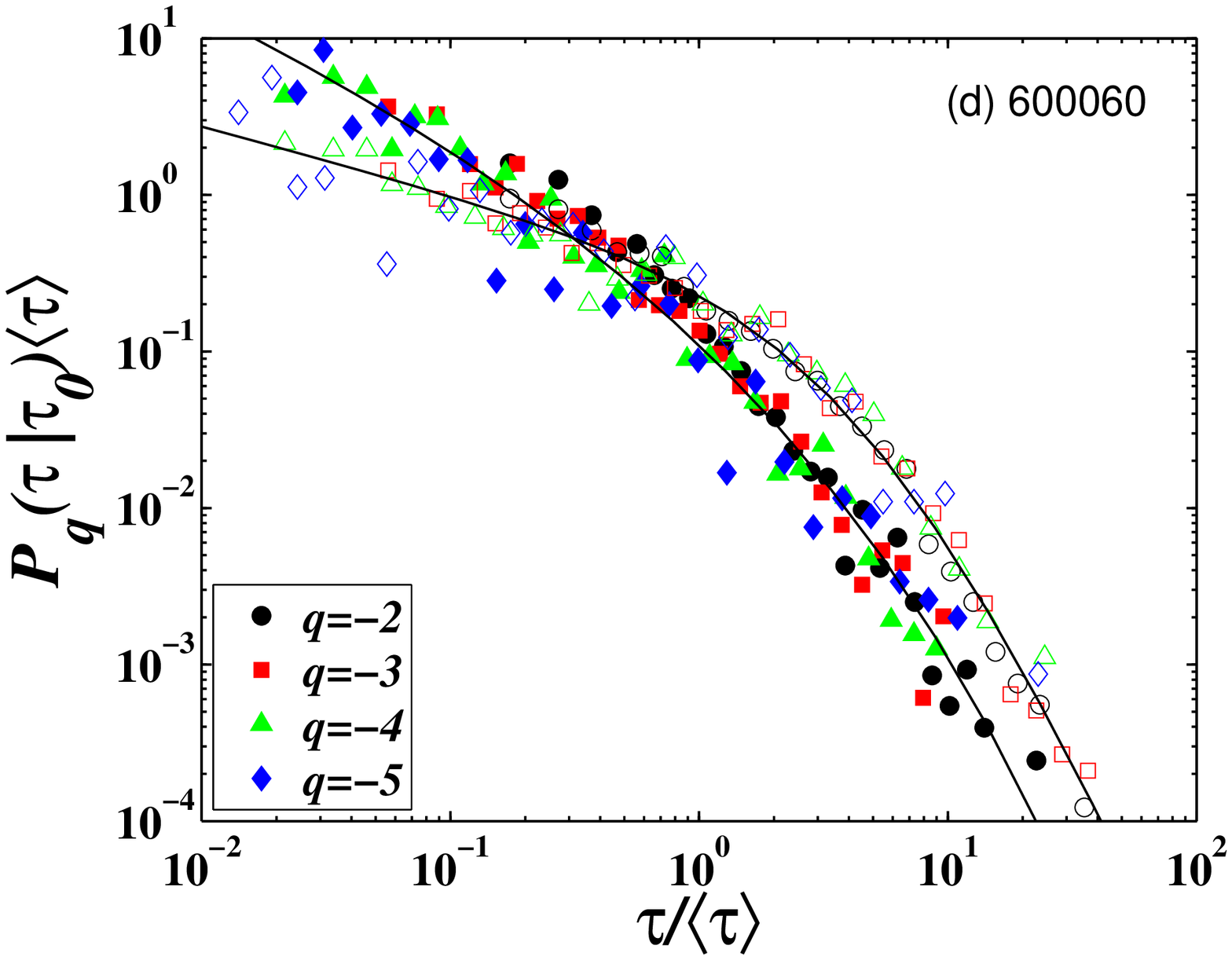}
\includegraphics[width=5cm]{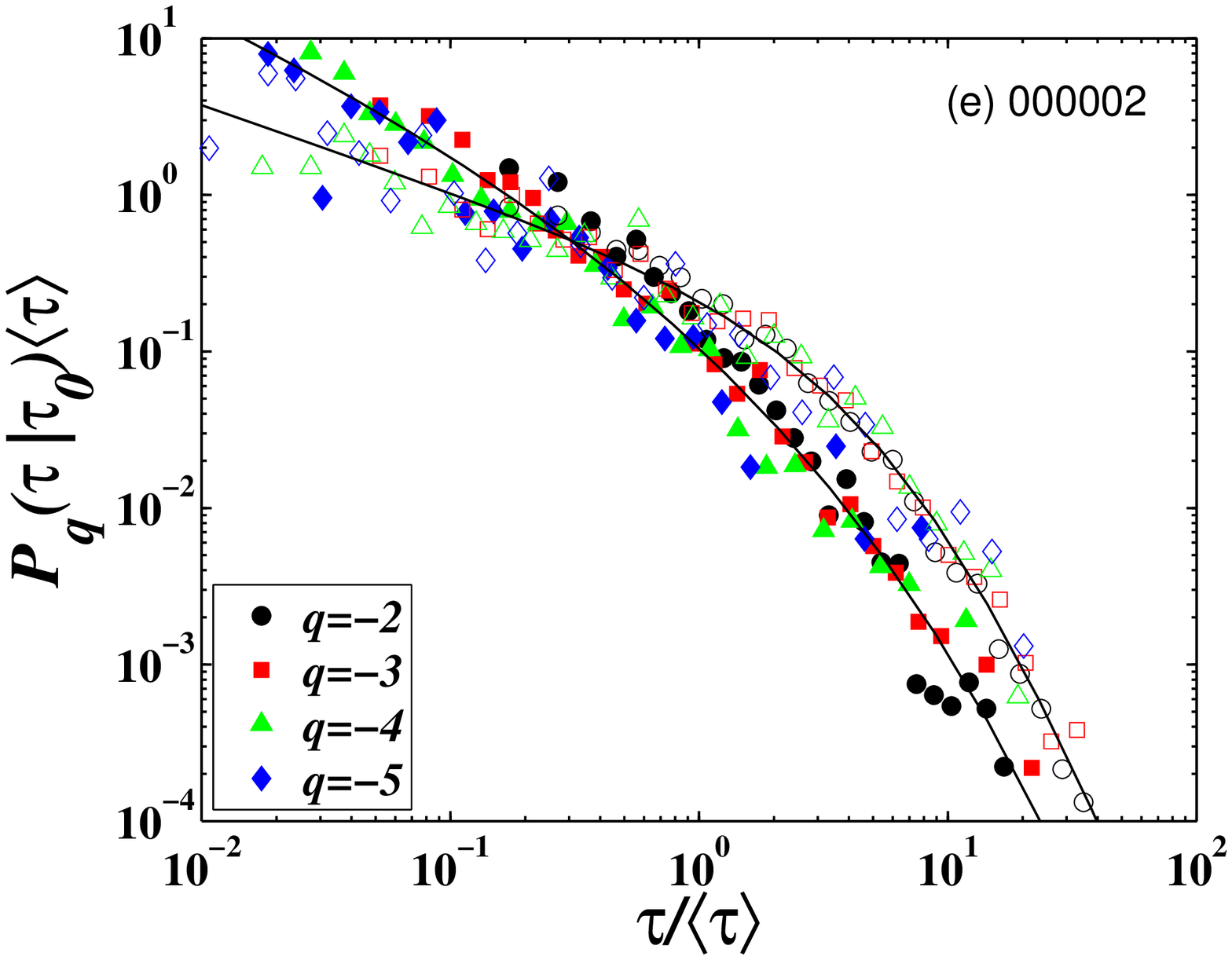}
\includegraphics[width=5cm]{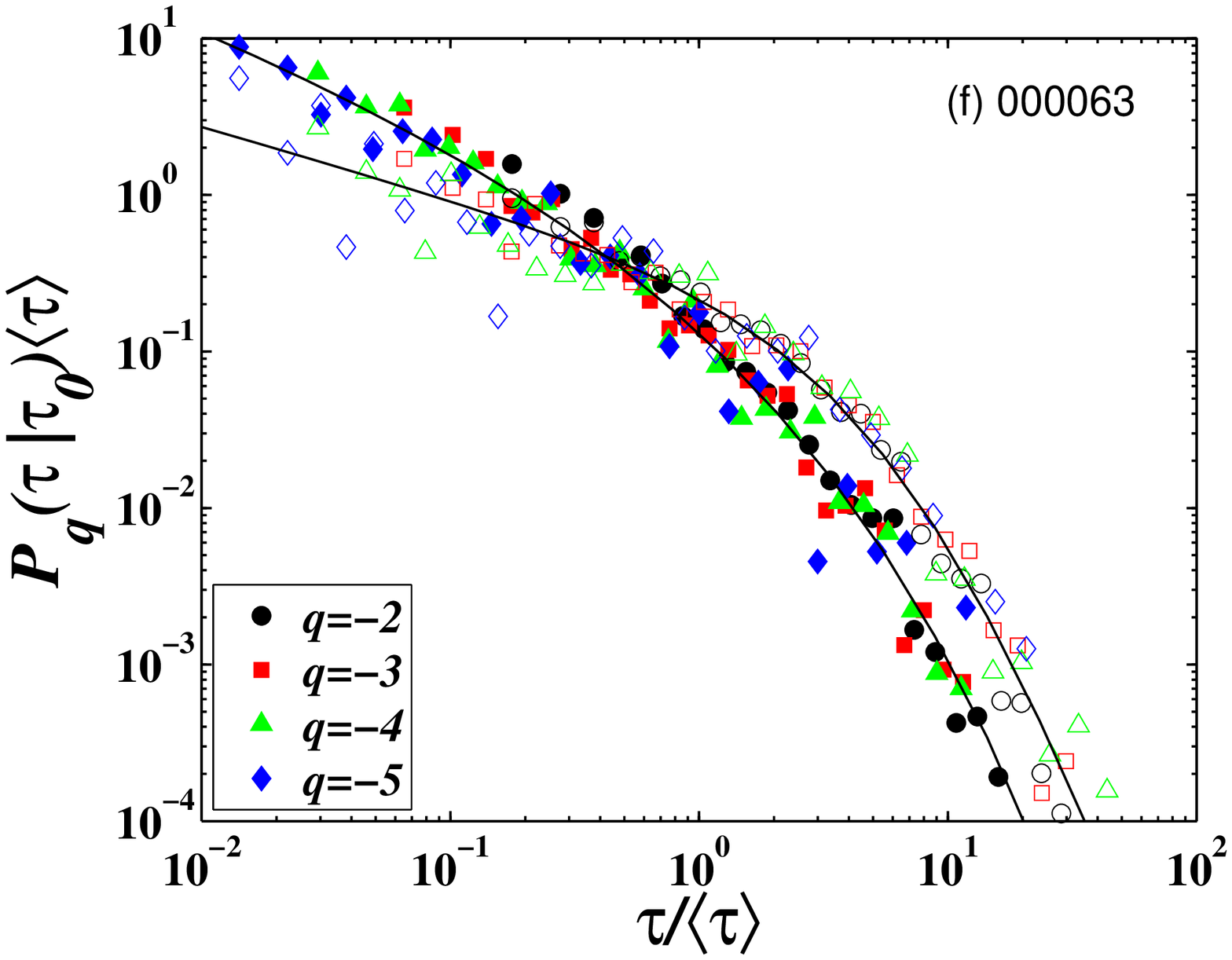}
\caption{\label{Fig:RI:ConPDF} Scaled conditional PDF $P_q(\tau |
\tau_0) \langle \tau \rangle$ of scaled recurrence intervals
$\tau/\langle \tau \rangle$ with $\tau_0$ in the largest $1/4$
subset (open symbols) and smallest $1/4$ subset (filled symbols) for
(a) SSEC, (b) SZCI, and (c)-(e) four representative stocks 600009,
600060, 000002 and 000063. The solid lines are guidelines for eyes.}
\end{figure}

We first investigate the short-term memory by calculating the
conditional PDF $P_q(\tau|\tau_0)$ of recurrence intervals.
$P_q(\tau|\tau_0)$ is defined as the probability of finding interval
$\tau$ conditioned on the preceding interval $\tau_0$. We study the
conditional PDF for a bin of $\tau_0$ in order to get better
statistics. The entire interval sequences are arranged in ascending
order and partitioned to four bins with equal size.
Figure~\ref{Fig:RI:ConPDF} plots the scaled conditional PDF
$P_q(\tau|\tau_0) \langle \tau \rangle$ of the two Chinese indices
and four representative stocks as a function of the scaled
recurrence intervals $\tau/\langle \tau \rangle$ for $\tau_0$ in the
smallest and biggest bins, marked with filled and open symbols,
respectively. The symbols for all negative thresholds for $\tau_0$
in the smallest and biggest bins approximately collapse onto two
separate solid curves as shown in figure~\ref{Fig:RI:ConPDF}. This
may further confirm the scaling behavior of the recurrence interval
PDFs. Moreover, for large $\tau/\langle \tau \rangle$,
$P_q(\tau|\tau_0)$ with large $\tau_0$ is larger than that with
small $\tau_0$, and for small $\tau/\langle \tau \rangle$,
$P_q(\tau|\tau_0)$ with small $\tau_0$ is larger than that with
large $\tau_0$. We find large (small) preceding intervals $\tau_0$
tend to be followed by large (small) intervals $\tau$, and this
observation indicates that there exists a short-term memory in the
recurrence intervals of the Chinese stock markets.

\subsection{Detrended fluctuation analysis}
\label{S2:DFA}

To investigate the long-term memory of the recurrence intervals, we
adopt the detrended fluctuation analysis (DFA) method
\cite{Peng-Buldyrev-Havlin-Simons-Stanley-Goldberger-1994-PRE,Kantelhardt-Bunde-Rego-Havlin-Bunde-2001-PA,Chen-Hu-Carpena-Bernaola-Galvan-Stanley-Ivanov-2005-PRE,Chen-Ivanov-Hu-Stanley-2002-PRE,Hu-Ivanov-Chen-Carpena-Stanley-2001-PRE},
known as a general method of examining the long-term correlation in
time series analysis. The DFA method computes the detrended
fluctuation $F(l)$ of the time series within a window of $l$ points
after removing a linear trend. For long-term power-law correlated
time series, $F(l)$ is expected to scales as a power law with
respect to the time scale $l$
\begin{equation}
   F(l)\sim l ^ \alpha,
   \label{Eq:DFA}
\end{equation}
where the DFA scaling exponent $\alpha$ is supposed to be equal to
the Hurst exponent when $\alpha \leq 1$
\cite{Coronado-Carpena-2005-JBP}: if $0.5<\alpha<1$ the time series
are long-term correlated, and if $\alpha=0.5$ the time series are
uncorrelated
\cite{Peng-Buldyrev-Havlin-Simons-Stanley-Goldberger-1994-PRE,Kantelhardt-Bunde-Rego-Havlin-Bunde-2001-PA}.

\begin{figure}[htb]
\centering
\includegraphics[width=4.9cm]{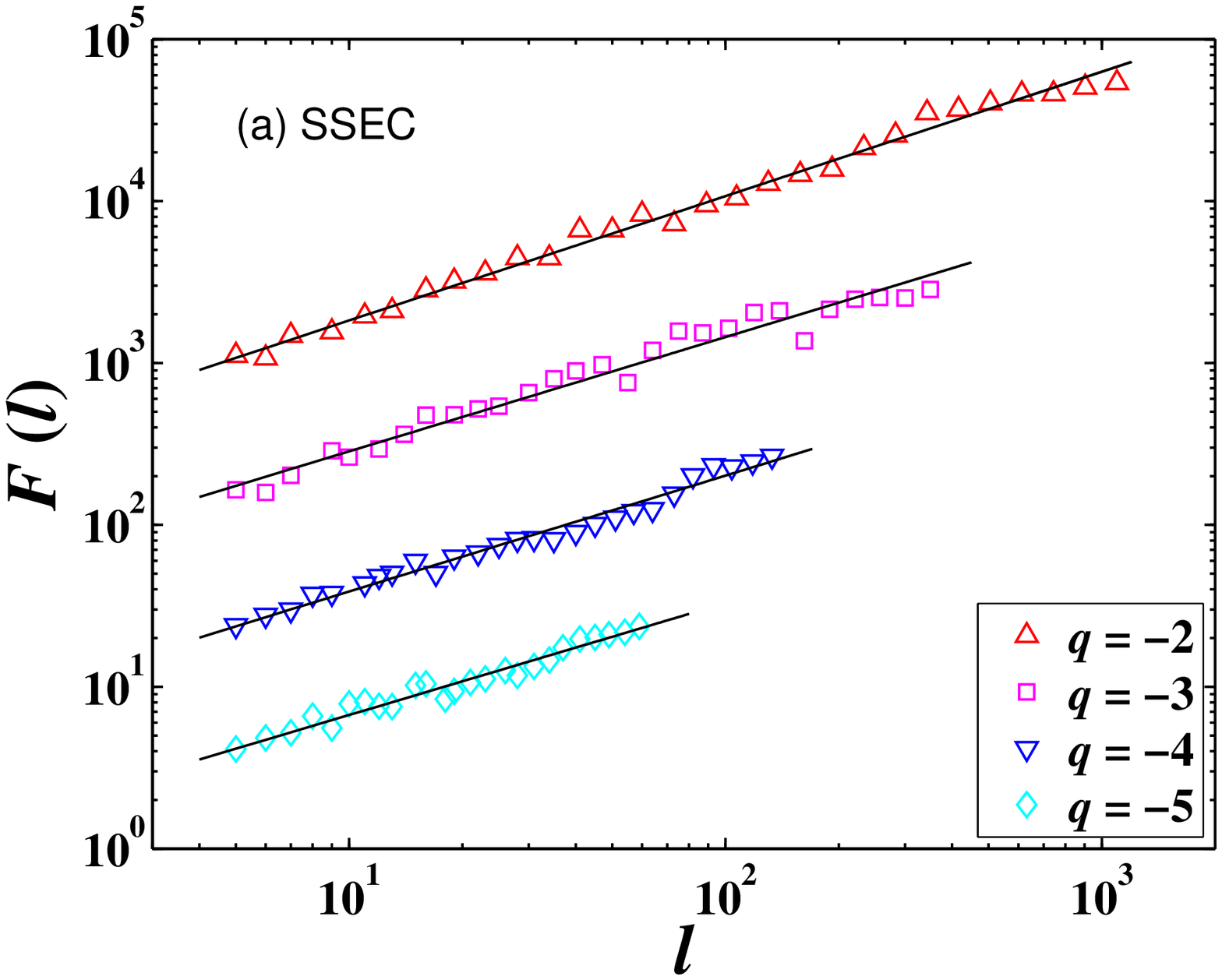}
\includegraphics[width=4.9cm]{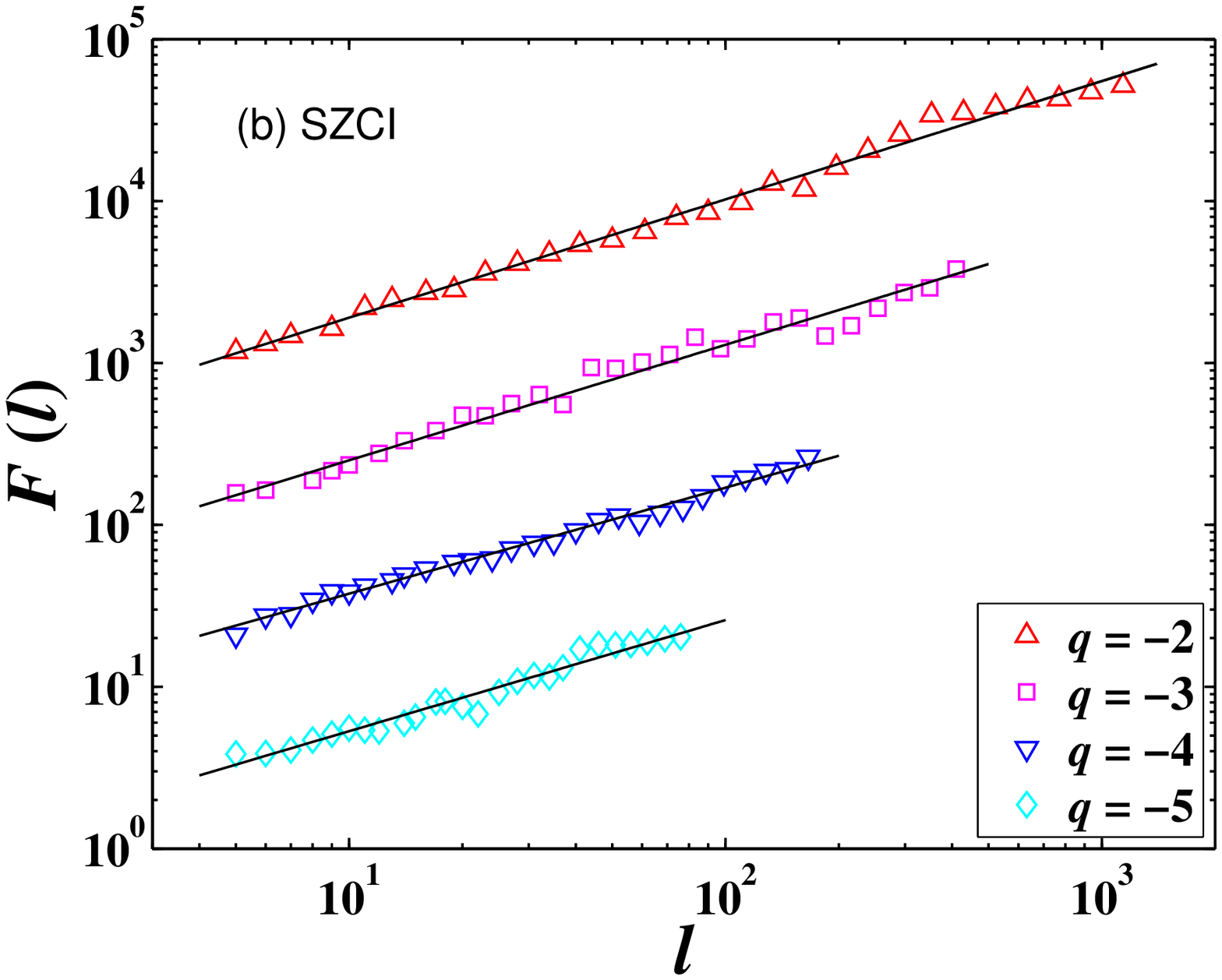}
\includegraphics[width=5.2cm]{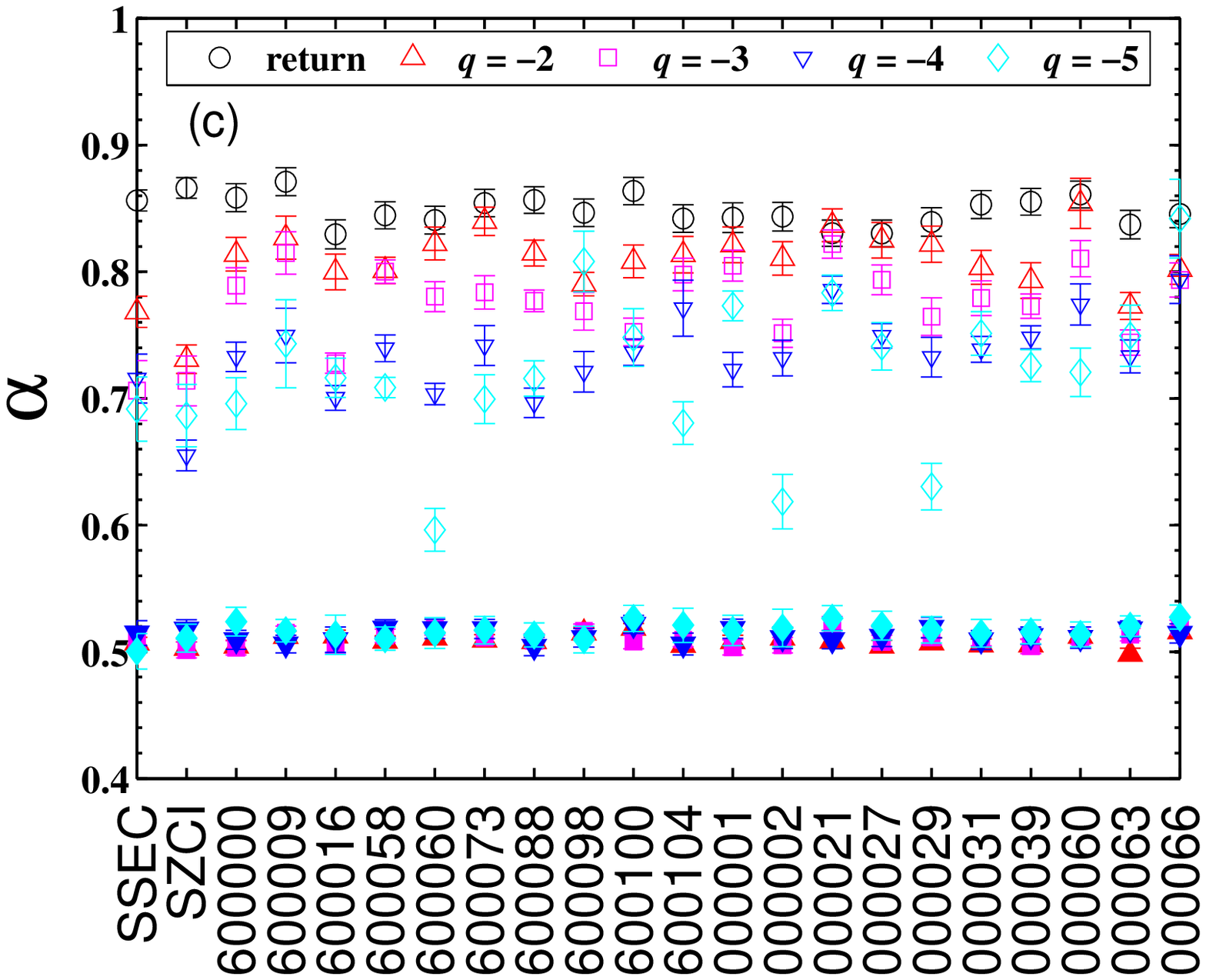}
\caption{\label{Fig:RI:DFA} Detrended fluctuation $F(l)$ of
recurrence intervals for (a) SSEC and (b) SZCI. The curves are
vertically shifted for clarity. (c) Exponents $\alpha$ of negative
returns (black circles), recurrence intervals (open symbols) and
shuffled recurrence intervals (filled symbols) for $2$ Chinese
indices and $20$ stocks.}
\end{figure}

Figure~\ref{Fig:RI:DFA} plots the detrended fluctuation $F(l)$ of
the recurrence intervals for different values of negative threshold
$q$ for the two Chinese indices. The curves show linear behavior in
the double logarithmic plot, which indicates $F(l)$ obeys a scaling
form as in Eq.~(\ref{Eq:DFA}). The DFA exponent $\alpha$ is
estimated by fitting the slope of the curves in
figure~\ref{Fig:RI:DFA}. We do the same calculation to obtain the
DFA scaling exponents $\alpha$ for the $20$ individual stocks, and
plot the exponents in the right panel of figure~\ref{Fig:RI:DFA}.
The exponent $\alpha$ shows a decreasing tendency as the decrease of
negative threshold $q$, but shows values apparently larger than
$0.5$. This suggests that the recurrence intervals are long-term
correlated.

Empirical studies have shown the long-term memory of the recurrence
intervals may arise from the long-term memory of its original time
series
\cite{Pennetta-2006-EPJB,Olla-2007-PRE,Eichner-Kantelhardt-Bunde-Havlin-2006-PRE,Eichner-Kantelhardt-Bunde-Havlin-2007-PRE,Bogachev-Eichner-Bunde-2007-PRL}.
To verify this, we calculate $F(l)$ of the negative return series,
remaining the position of the positive returns but ignoring its
contribution to the calculation of $F(l)$. As it is shown in
Figure~\ref{Fig:RI:DFA}, the exponents $\alpha$ of the negative
return series for all the $20$ stocks and the two Chinese indices
(represented by black circles) are significantly larger than $0.5$,
and therefore the negative return series are long-term correlated.
We then calculate the exponent $\alpha$ of the recurrence interval
of the shuffled price returns which are artificially uncorrelated.
The exponents $\alpha$ for all the individual stocks and the two
Chinese indices tend to be very close to $0.5$, and the recurrence
intervals of the shuffled data are consequently uncorrelated. This
observation provides direct evidence to confirm our assumption that
the long-term memory of the recurrence intervals may due to the
long-term memory of the price returns (either positive returns or
negative returns).

\section{Risk estimation}
\label{S1:Risk}

The study of recurrence interval between extreme events in stock
markets has draw much attention of scientists and economists. Most
of studies focus on investigating the statistical properties of
recurrence interval sequences and understanding its fundamental
dynamics. Not much work has been done applying this recurrence
interval analysis to risk estimation for real stock markets. In the
following paper we attempt to do some risk estimation for Chinese
stock markets based on this recurrence interval analysis following
Ref.~\cite{Yamasaki-Muchnik-Havlin-Bunde-Stanley-2006-inPFE,Bogachev-Eichner-Bunde-2007-PRL,Bogachev-Bunde-2008-PRE,Bogachev-Bunde-2009-PRE}.

\subsection{Probability $W_q(\Delta{t}|t)$}

\begin{figure}[t]
\centering
\includegraphics[width=5cm]{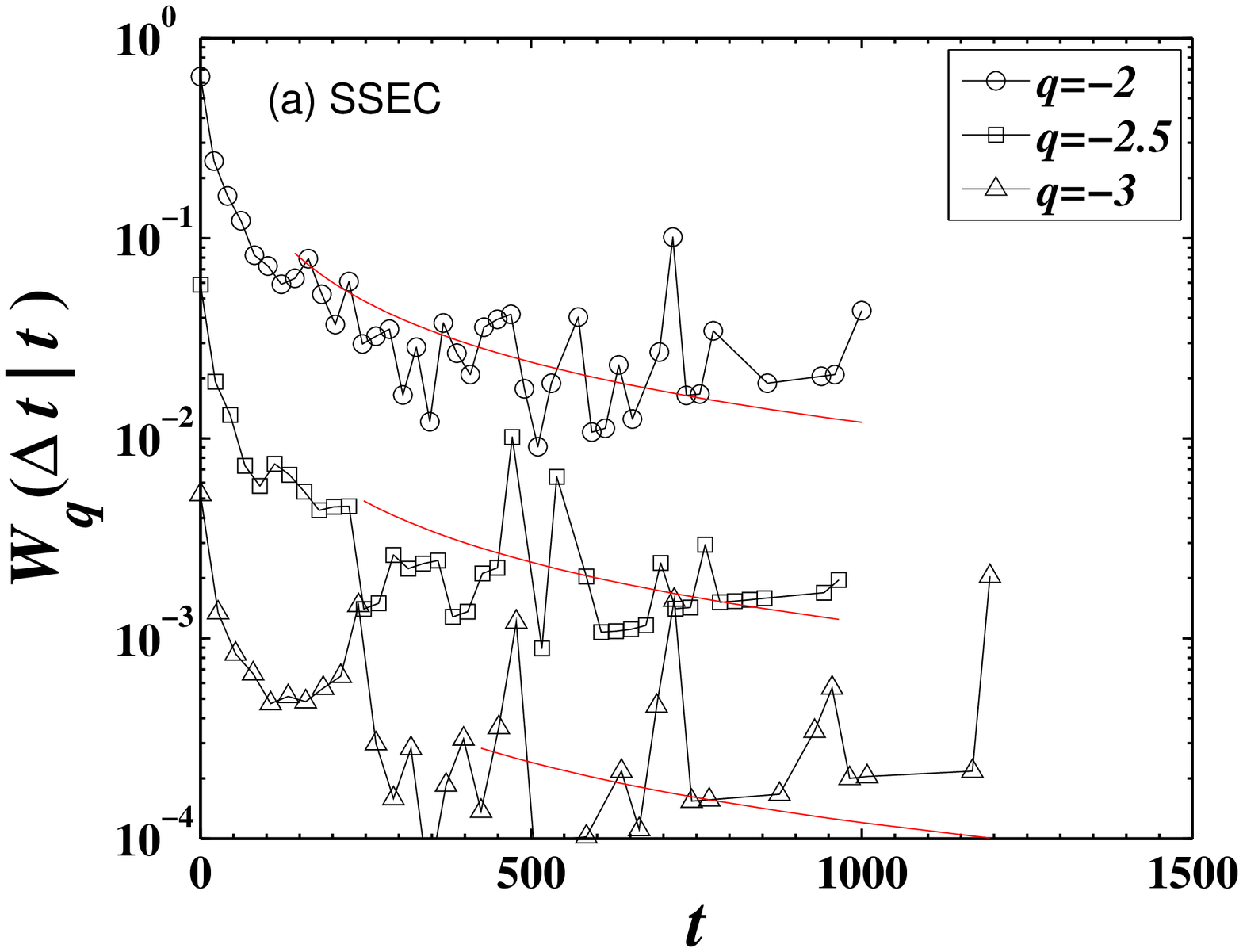}
\includegraphics[width=5cm]{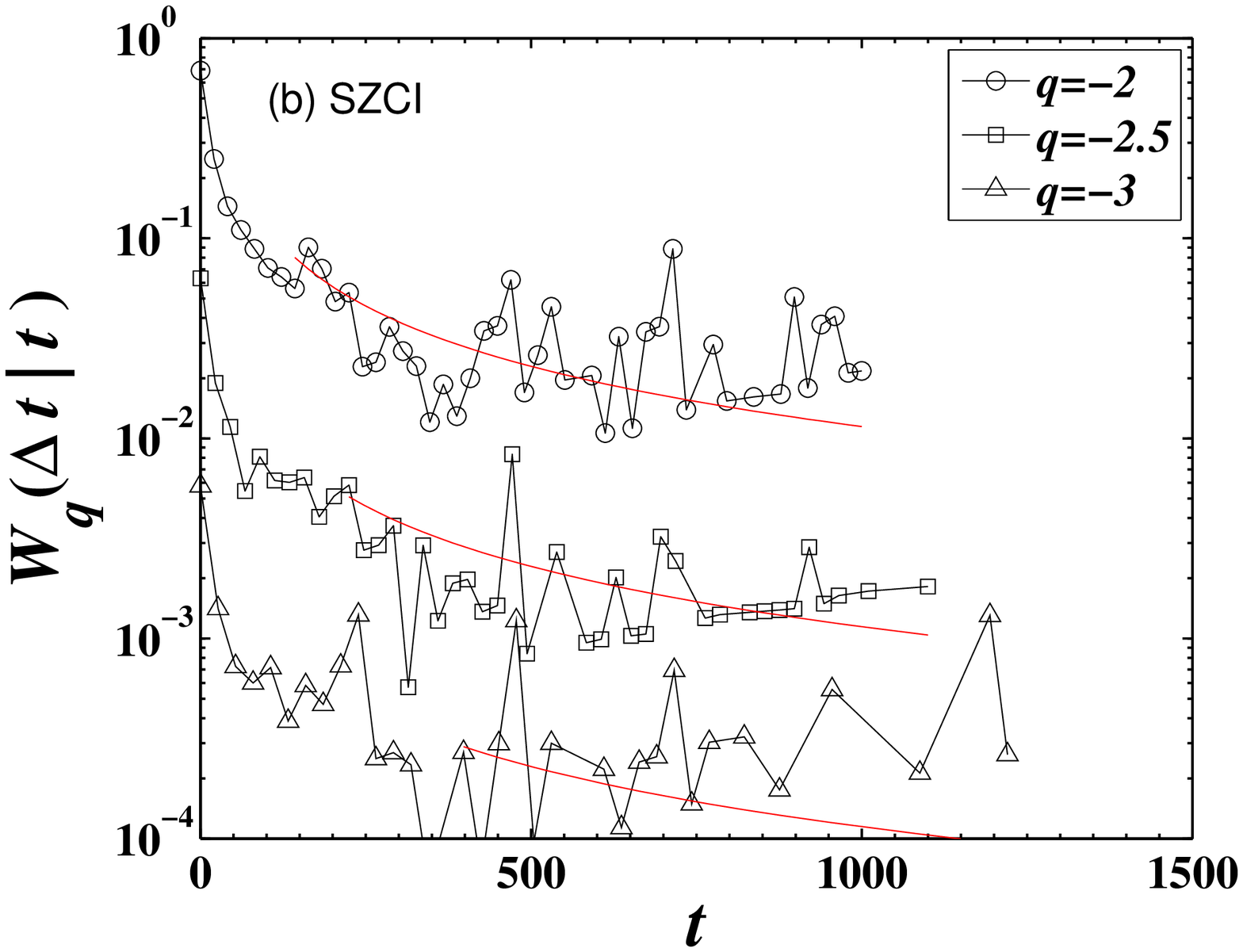}
\includegraphics[width=5cm]{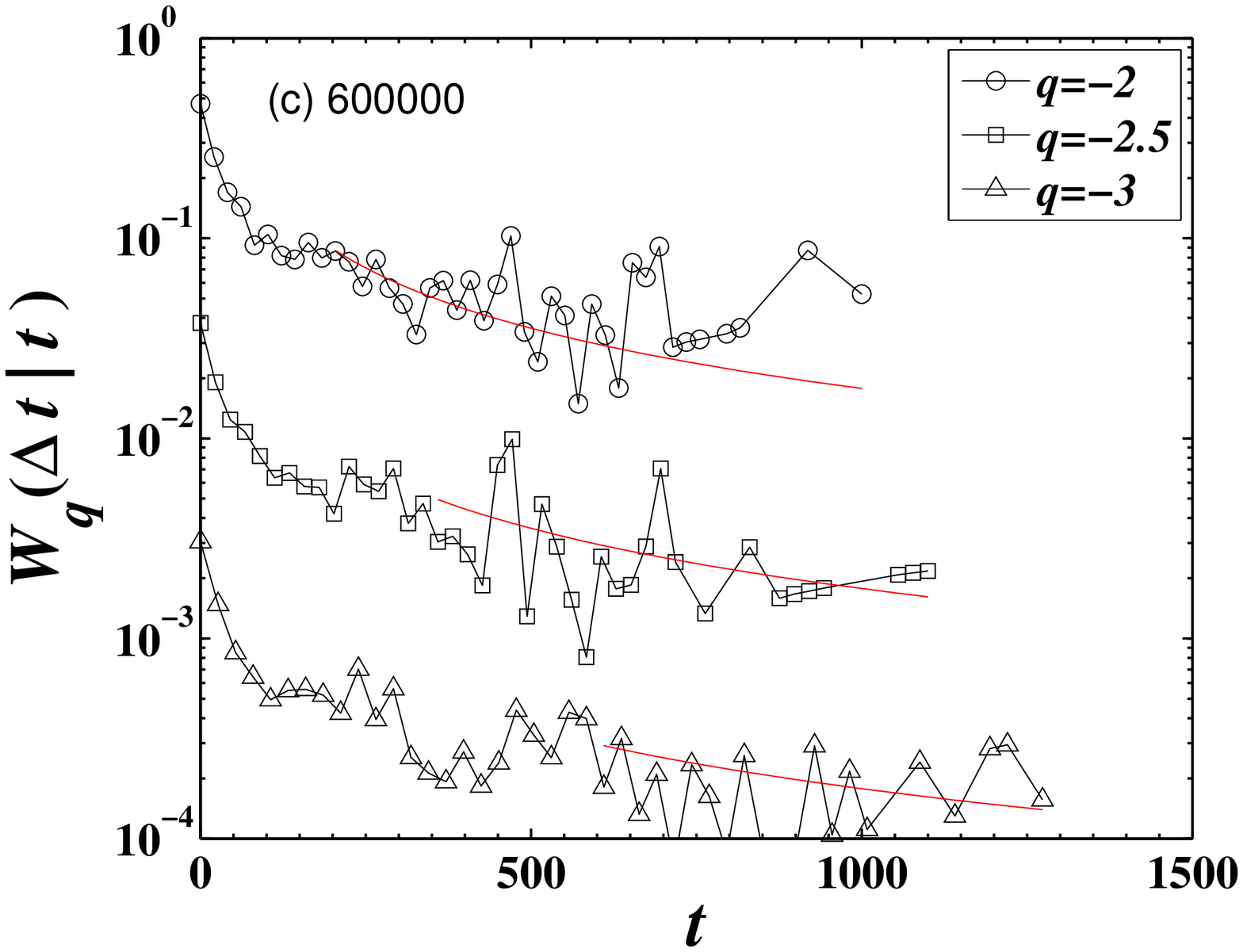}
\includegraphics[width=5cm]{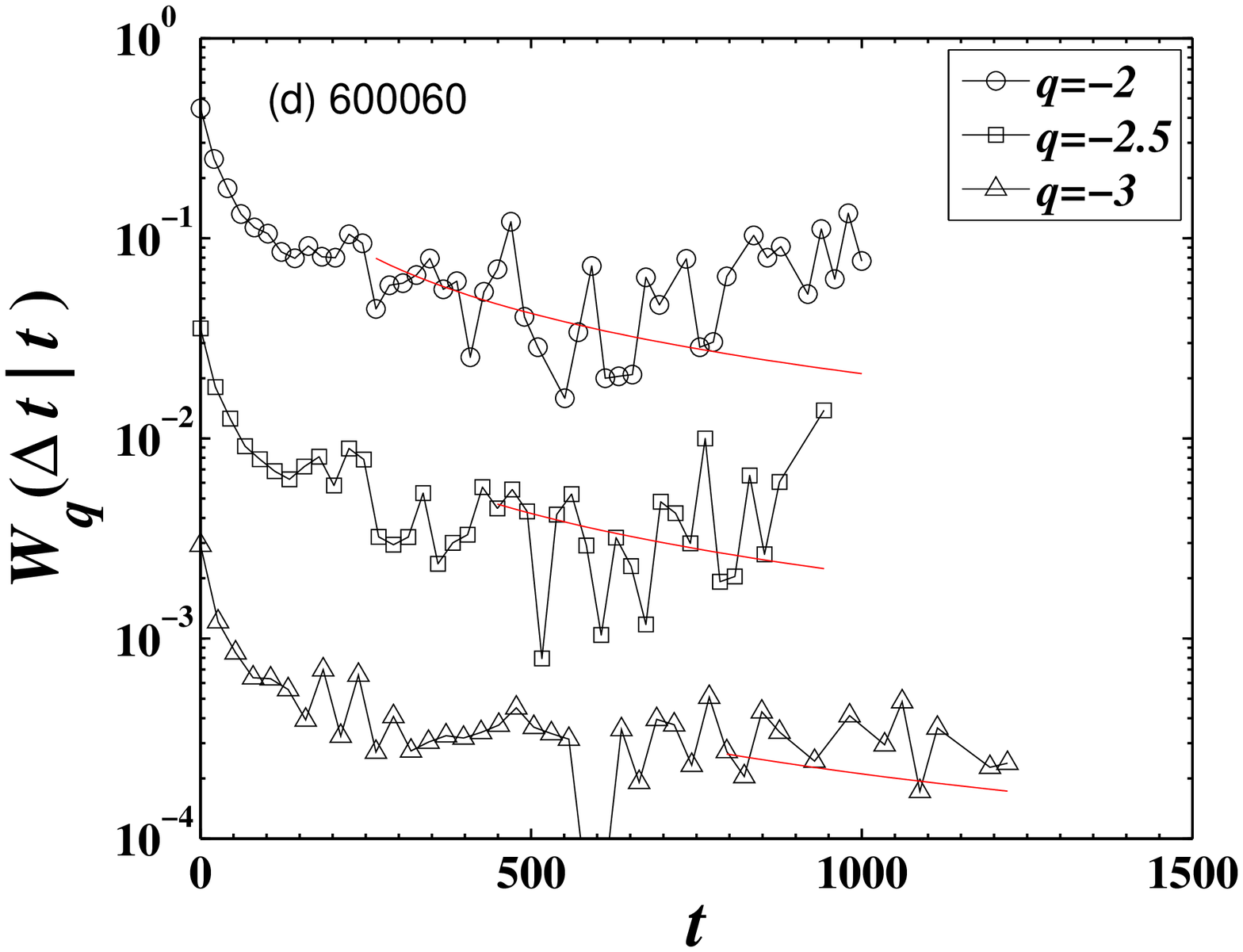}
\includegraphics[width=5cm]{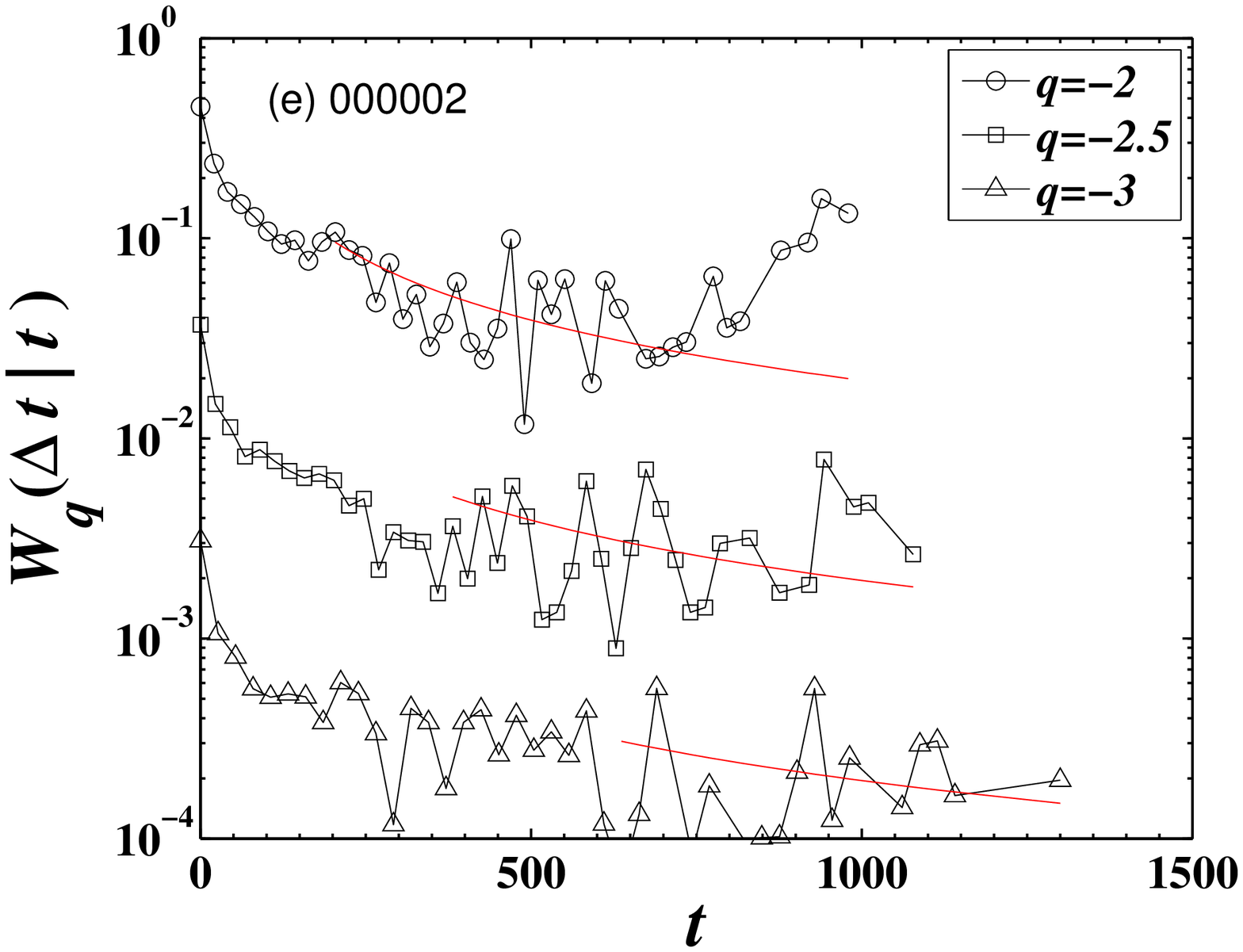}
\includegraphics[width=5cm]{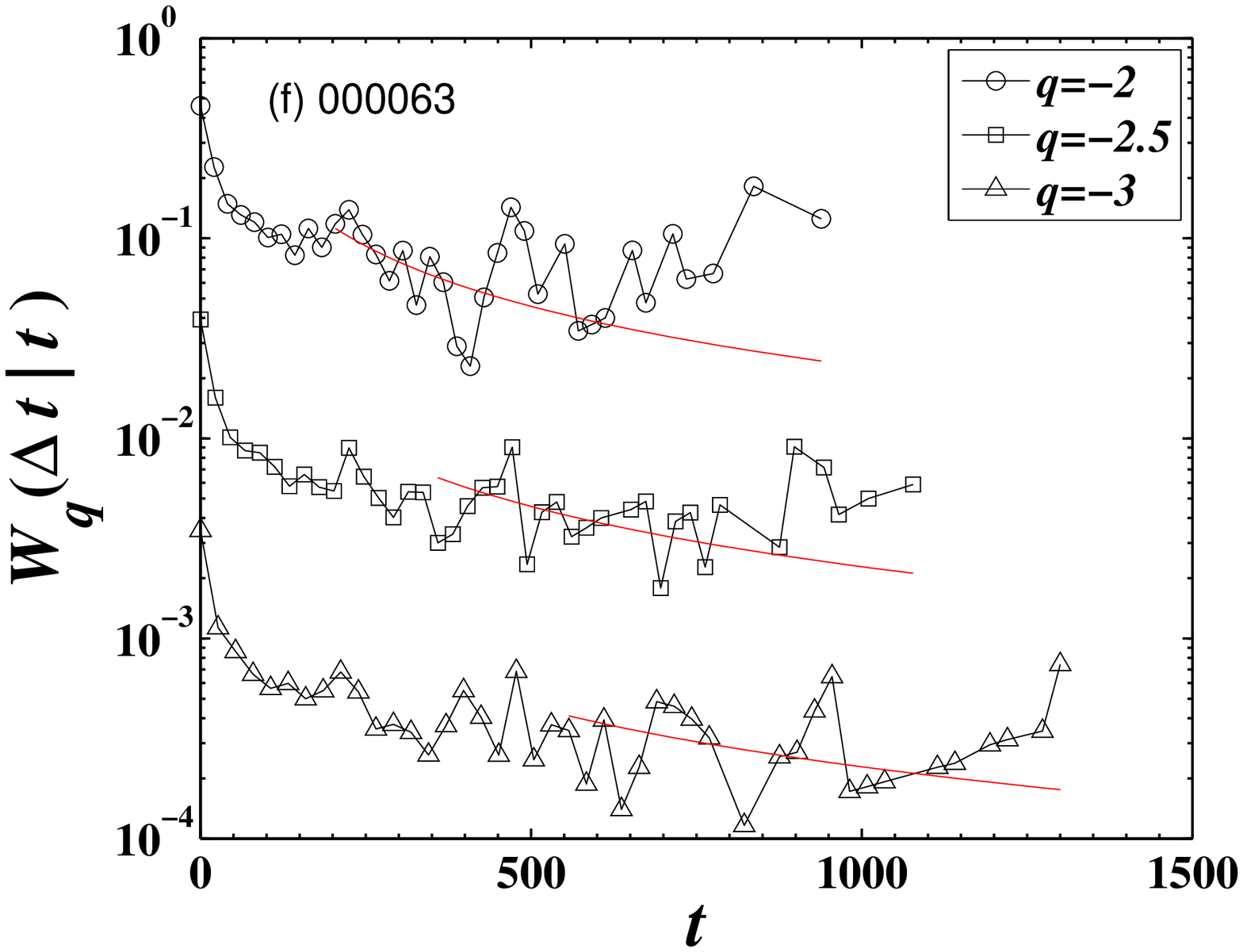}
\caption{\label{Fig:RI:WQ}  Comparison of $W_q(\Delta{t}=10|t)$
estimated empirically with that obtained from Eq.~(\ref{Eq:Wq2}) for
(a) SSEC, (b) SZCI, and (c)-(e) four representative stocks 600009,
600060, 000002 and 000063. The curves for $q=-2.5$ and $q=-3$ have
been shifted vertically by factors of 0.1 and 0.01 for clarity.}
\end{figure}

In risk estimation, a quantity of great importance is the
probability $W_q(\Delta{t}|t)$, that an extreme event with price
return below $q<0$ occurs within a short time interval $\Delta{t}\ll
\langle \tau \rangle$, conditioned on an elapsed time $t$ after the
occurrence of the previous extreme event
\cite{Bogachev-Eichner-Bunde-2007-PRL}:
\begin{equation}
 W_q(\Delta{t}|t)=\frac{\int_t^{t+\Delta{t}}P_q(\tau)d\tau}{\int_t^{\infty}P_q(\tau)d\tau}.
 \label{Eq:Wq}
\end{equation}
Previous study has shown that $P_q(\tau)$ obeys a power law as in
Eq.~(\ref{Eq:PL}) for $\tau/\langle \tau \rangle > \hat{x}_{\min}$.
When $t> \hat{x}_{\min} \langle \tau \rangle$, it can be
algebraically derived that
\begin{equation}
 W_q(\Delta{t}|t)\simeq(\delta-1){\Delta{t}}/{t}.
 \label{Eq:Wq2}
\end{equation}
$W_q(\Delta{t}|t)$ is supposed to be proportional to $\Delta{t}$ and
inversely proportional to $t$. Figure~\ref{Fig:RI:WQ} plots
$W_q(\Delta{t}=10|t)$ for the two Chinese indices and four
representative stocks, and apparently Eq.~(\ref{Eq:Wq2}) fits the
probability $W_q(\Delta{t}|t)$ well for $t> x_{\min} \langle \tau
\rangle$. It is worth pointing out that $W_q(\Delta{t}|t)$ shows an
increasing tendency for extreme large $t$ because of the poor
statistics of rare events with large recurrence intervals.
Measurements for other values of $\Delta{t}$ show qualitatively
similar results. Here an intriguing feature is that
$W_q(\Delta{t}|t)$ is independent of the threshold $q$, which is a
direct consequence of the scaling behavior of $P_q(\tau)$ shown in
Eq.~(\ref{Eq:Pq:f}).

\subsection{Loss probability $p^*$}

\begin{figure}[htb]
\centering
\includegraphics[width=5cm]{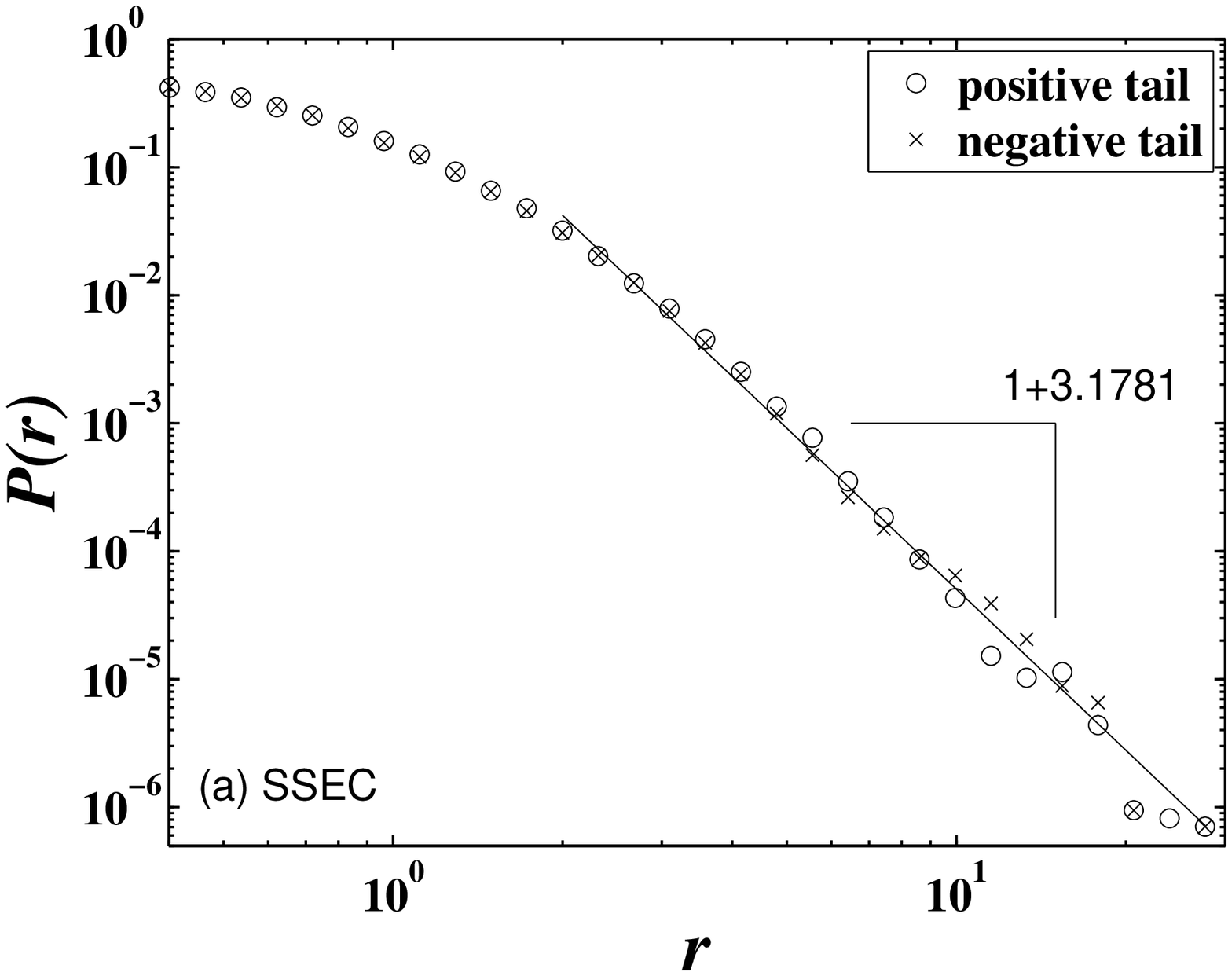}
\includegraphics[width=5cm]{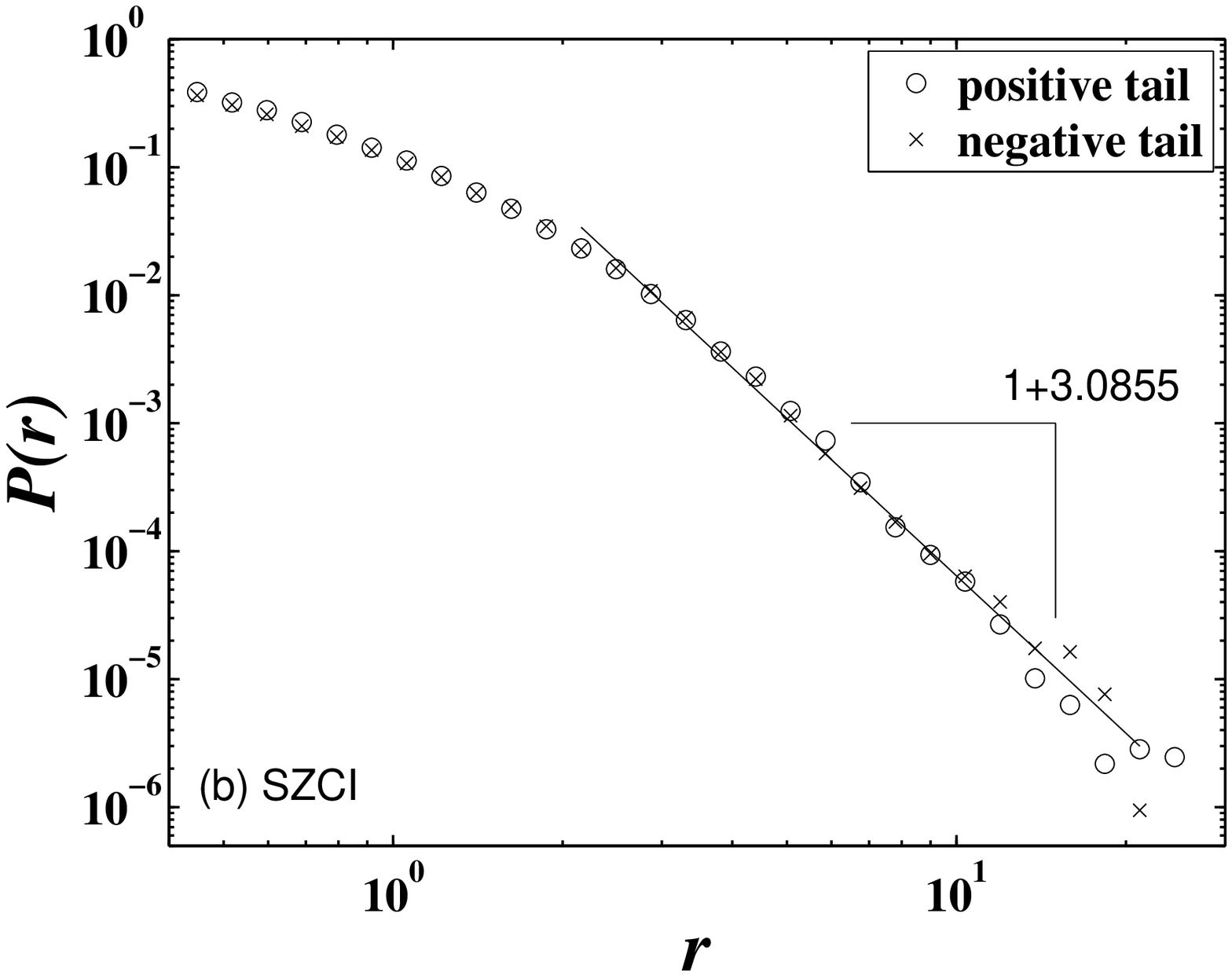}
\includegraphics[width=5cm]{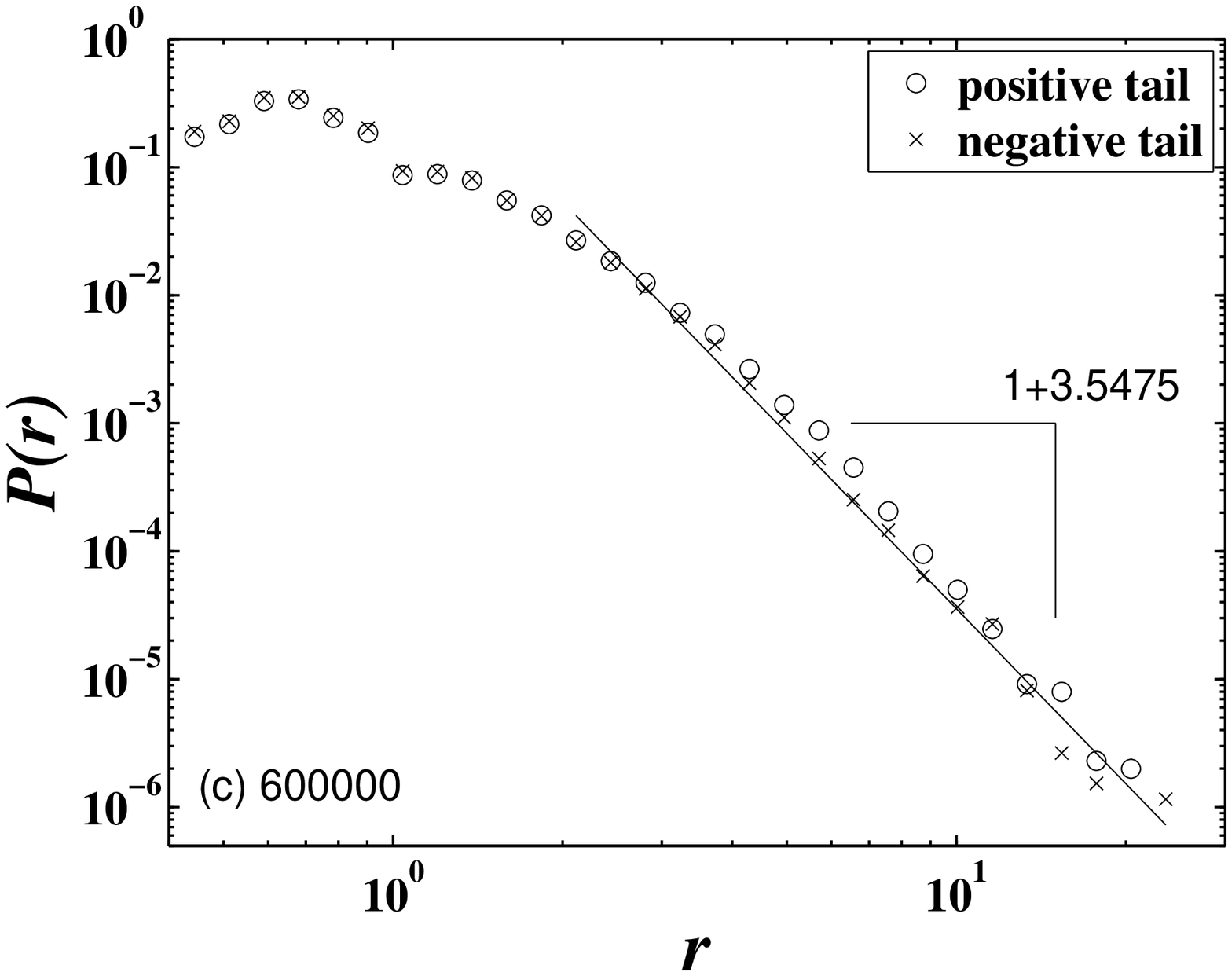}
\includegraphics[width=5cm]{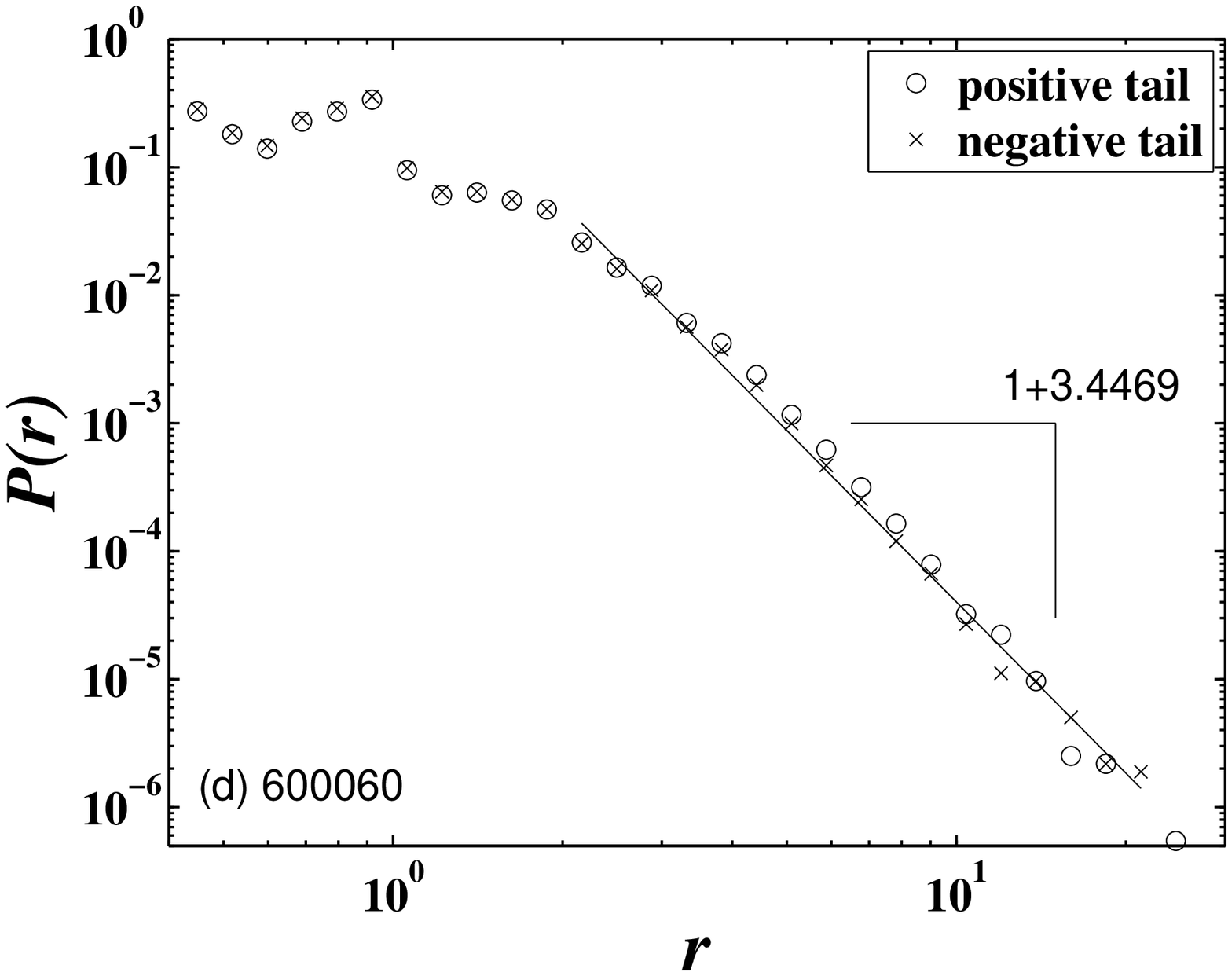}
\includegraphics[width=5cm]{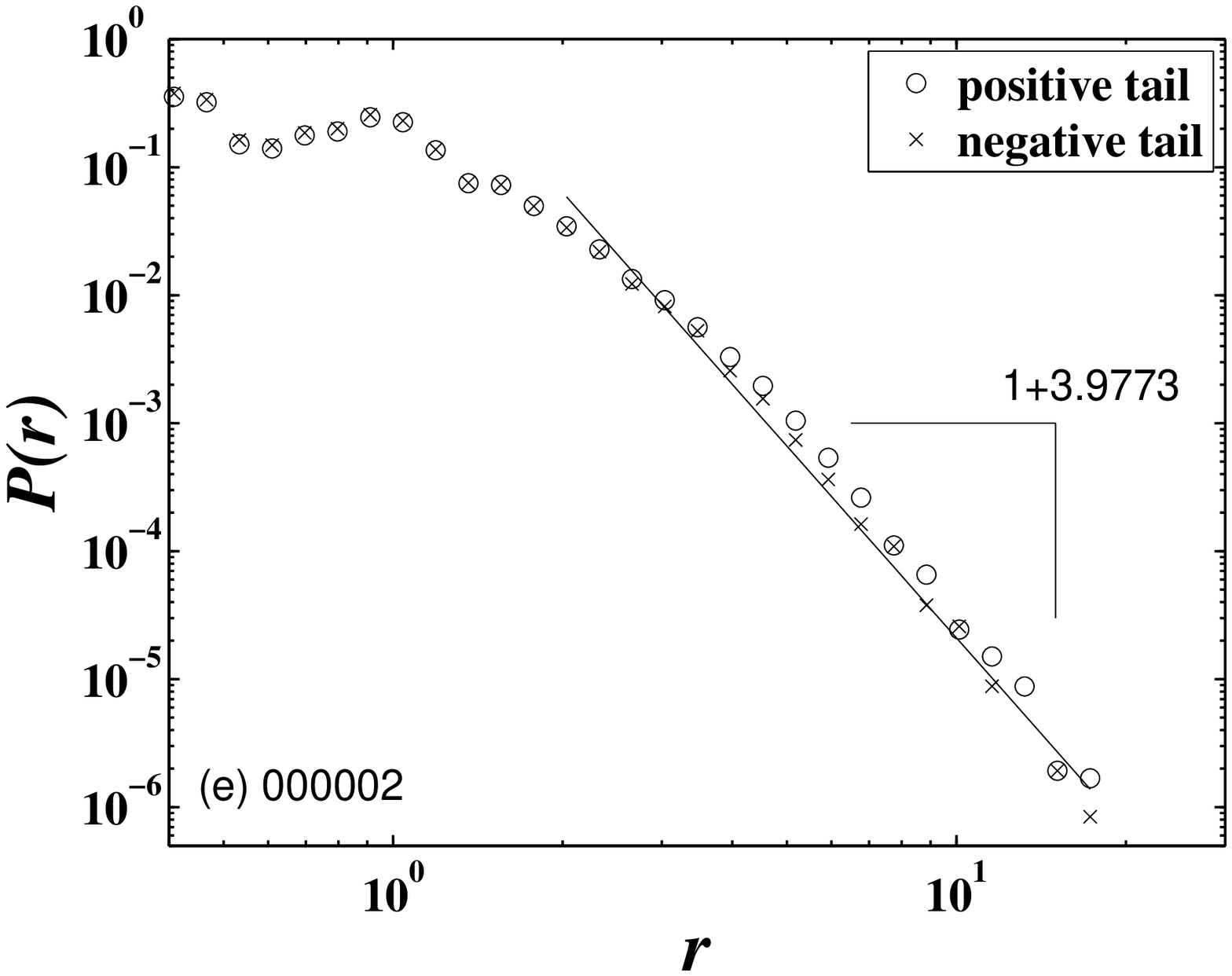}
\includegraphics[width=5cm]{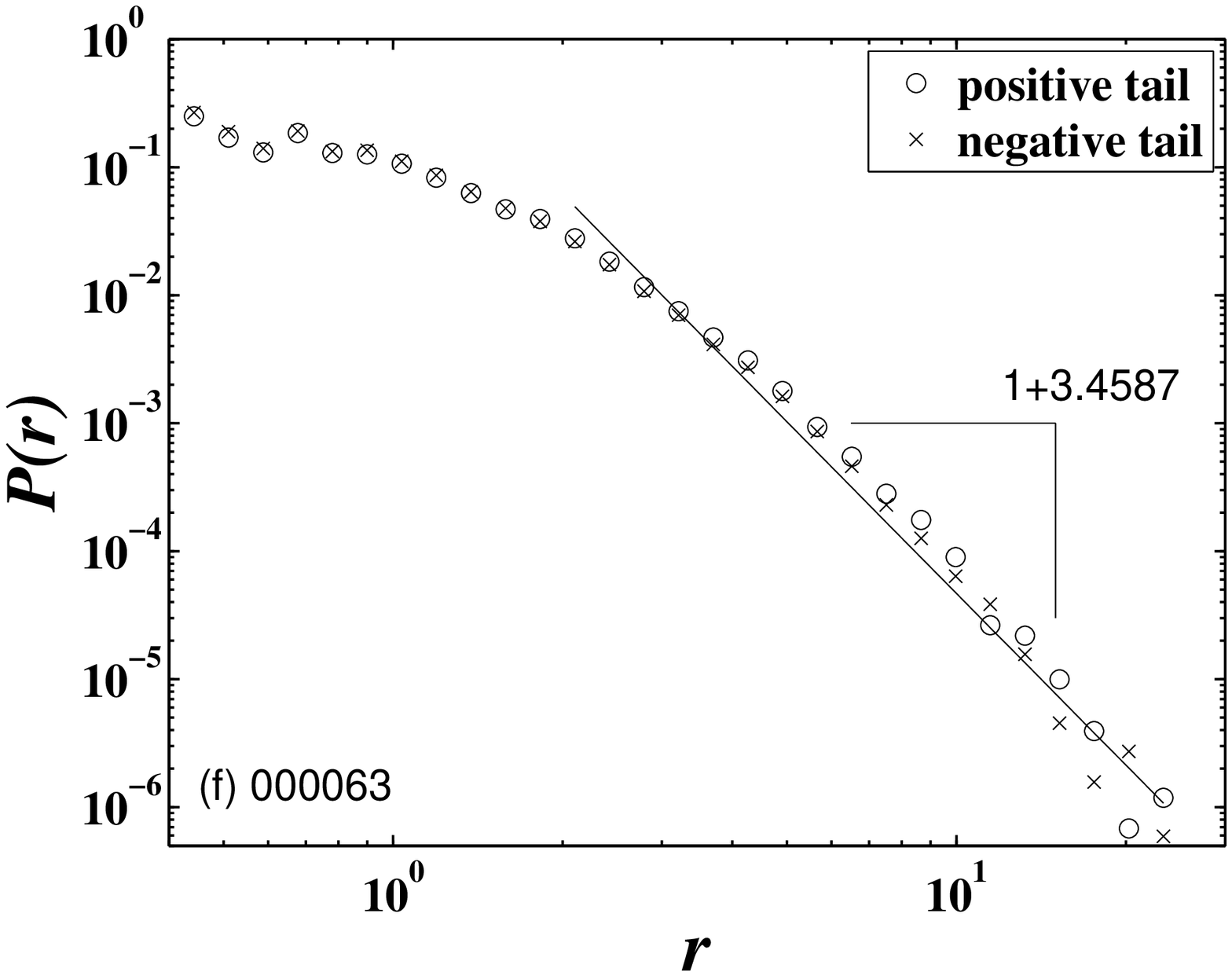}
\caption{\label{Fig:Return:PDF}  Probability distribution $P(r)$ of
normalized return $r$ for (a) SSEC, (b) SZCI, (c)-(e) 4
representative stocks 600009, 600060, 000002 and 000063. The solid
lines are power-law fits for $P(r<0)$.}
\end{figure}

It is well known that the intraday returns $r_{\Delta{t}}$ of stocks
and indexes are distributed according to a Student distribution
\cite{Gu-Chen-Zhou-2008a-PA}, whose tails follow the inverse cubic
law \cite{Gopikrishnan-Meyer-Amaral-Stanley-1998-EPJB}. Hence, its
tail obeys a power law as
\begin{equation}
 p(r) = k |r|^{-(\beta+1)},
 \label{Eq:PDF:r:tail}
\end{equation}
where the exponent $\beta$ increases with the increase of
$\Delta{t}$. When $\Delta{t}$ equals one minute, the inverse cubic
law holds \cite{Gu-Chen-Zhou-2008a-PA}, that is $\beta\approx3$
\cite{Gopikrishnan-Meyer-Amaral-Stanley-1998-EPJB}. Empirical
studies show that the tails of the return distributions for
$|r|\geq2$ follow Eq.~(\ref{Eq:PDF:r:tail}) and the exponent $\beta$
displays values close to $3$ as illustrated in
figure~\ref{Fig:Return:PDF}.

\begin{figure}[htb]
\centering
\includegraphics[width=5cm]{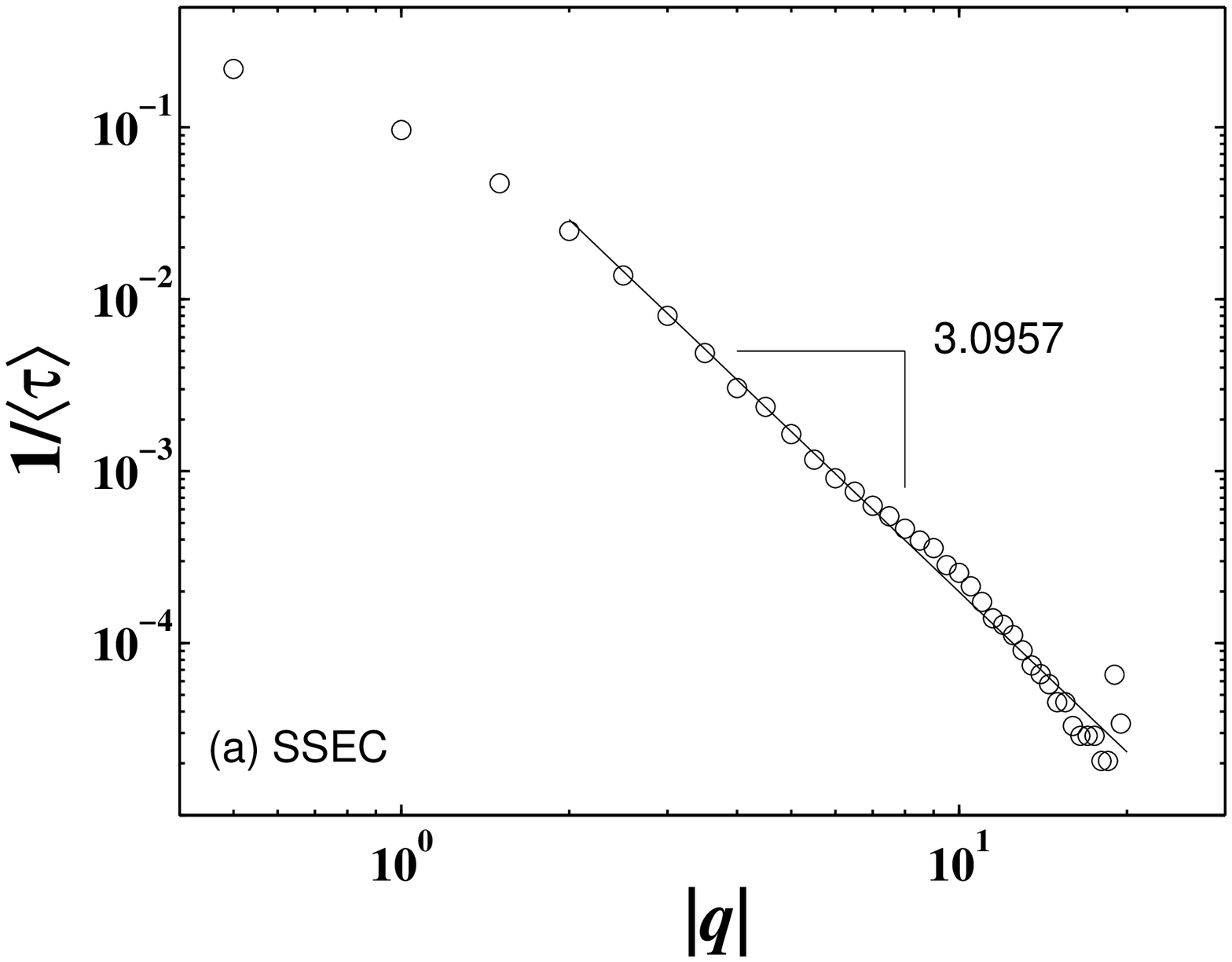}
\includegraphics[width=5cm]{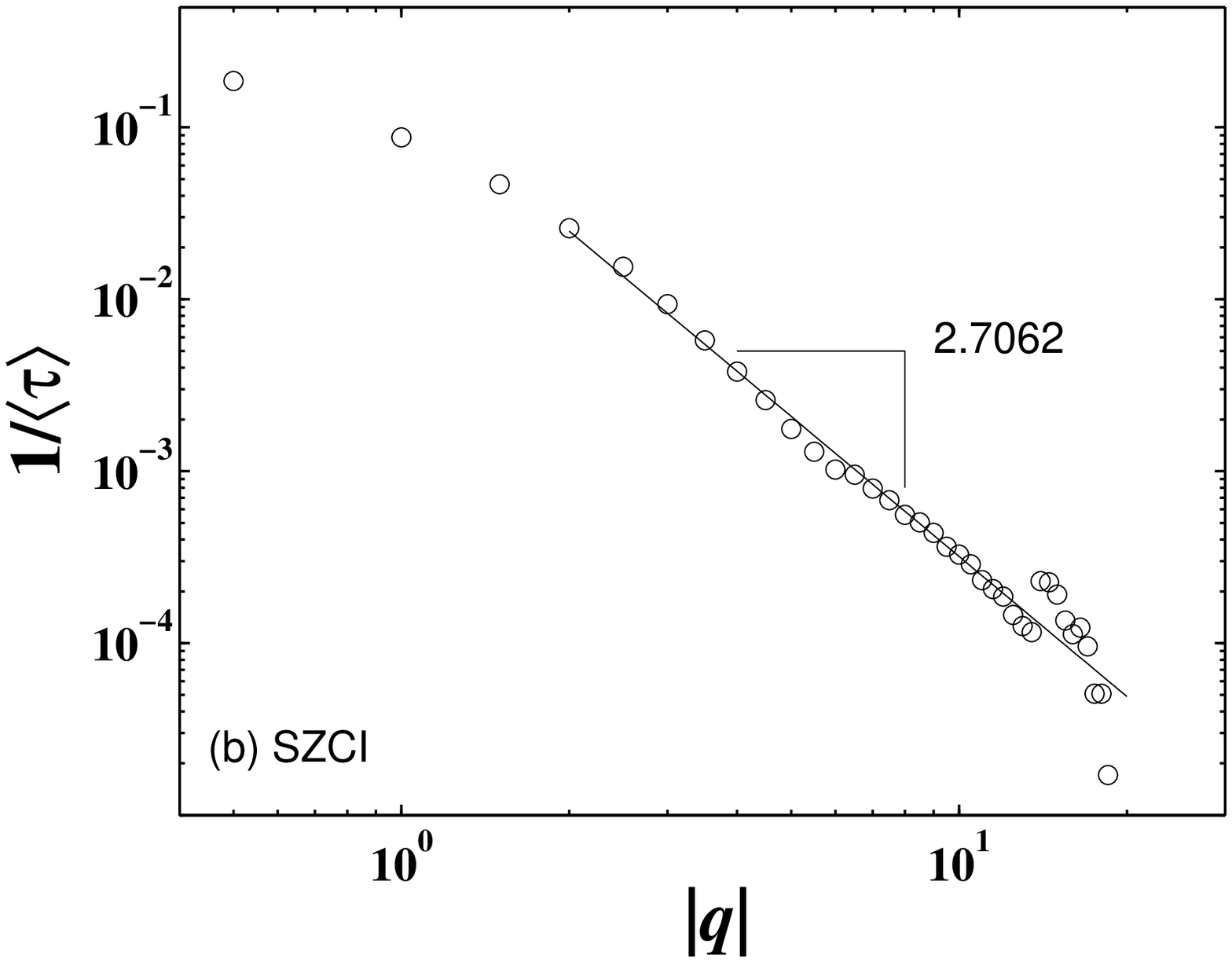}
\includegraphics[width=5cm]{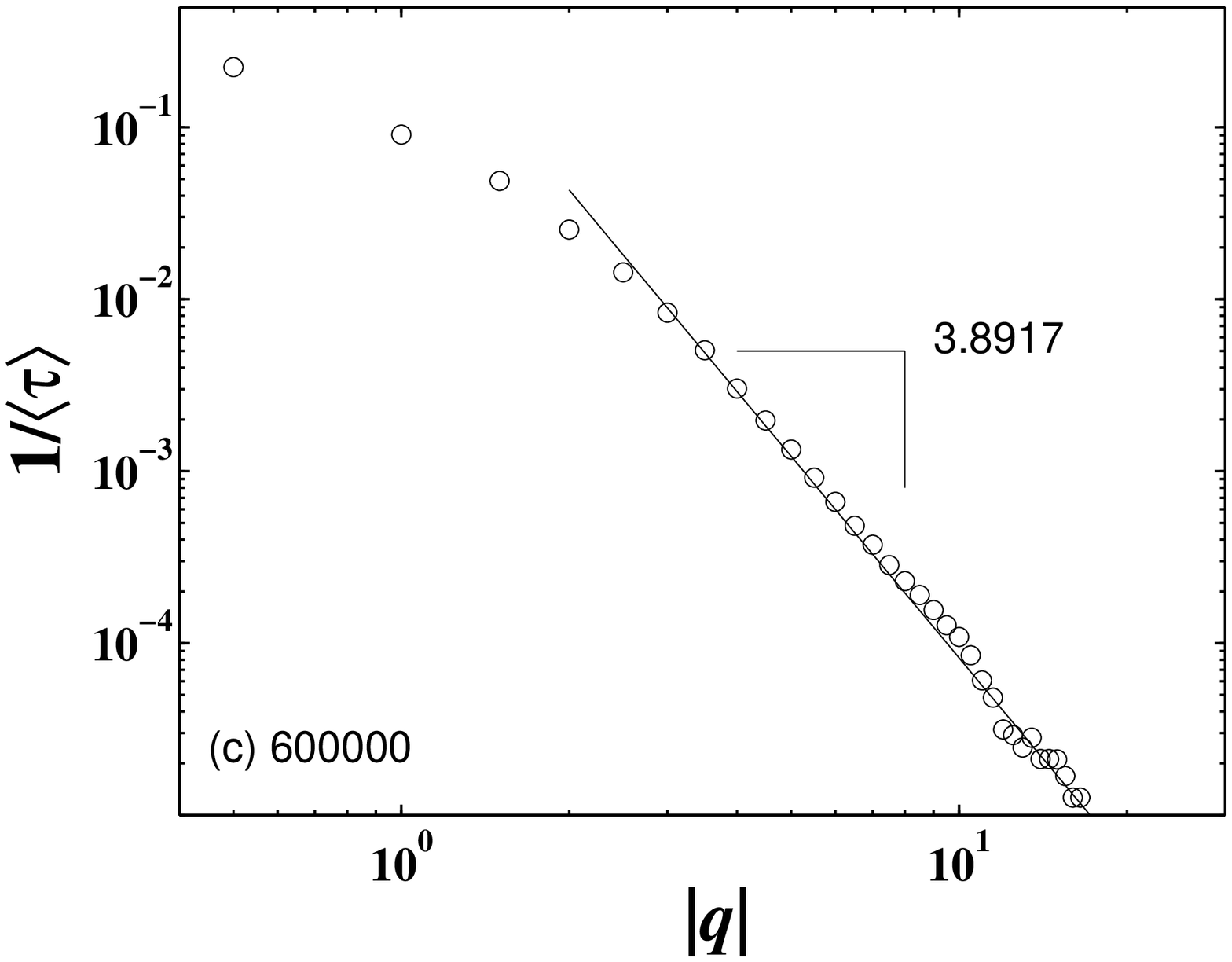}
\includegraphics[width=5cm]{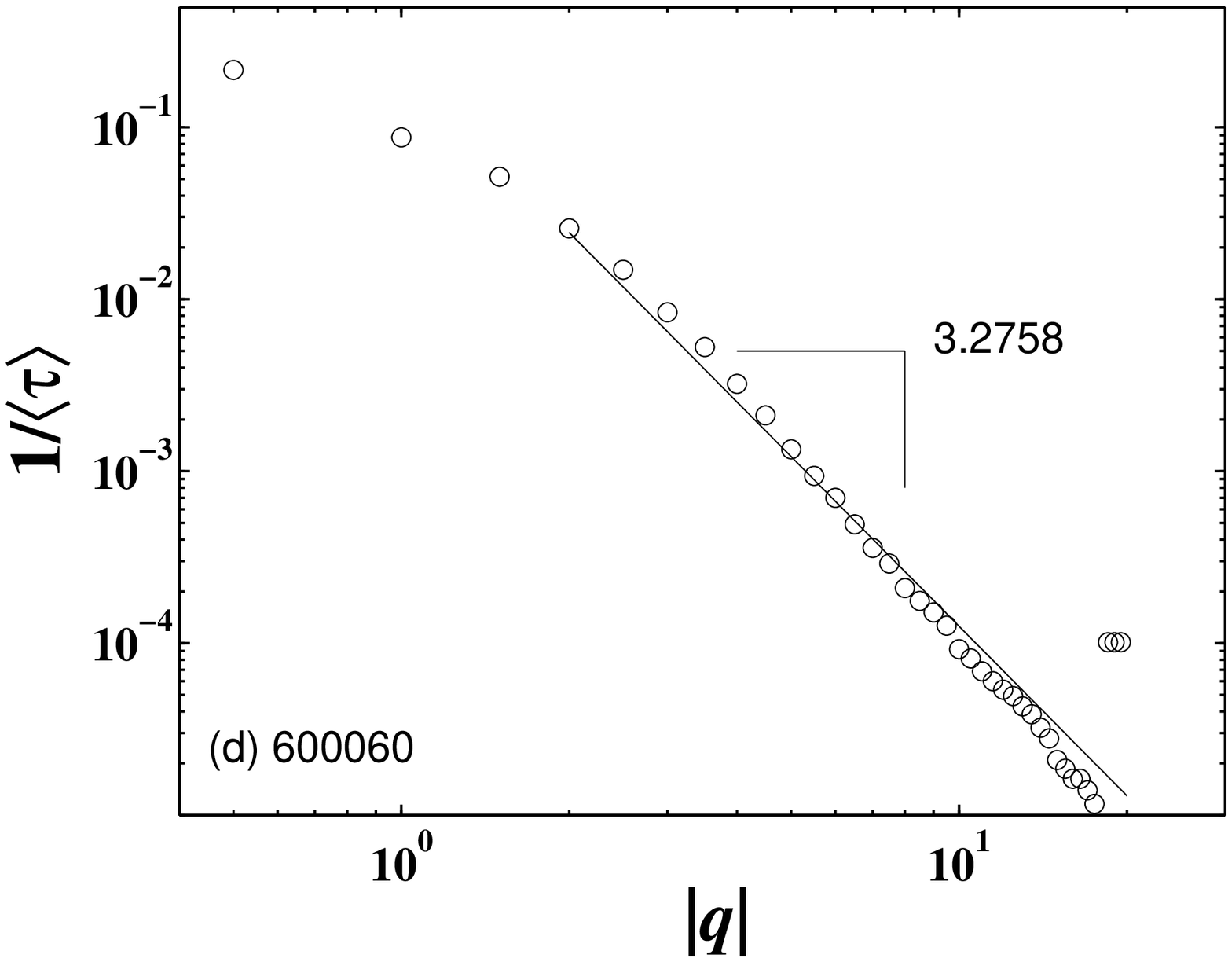}
\includegraphics[width=5cm]{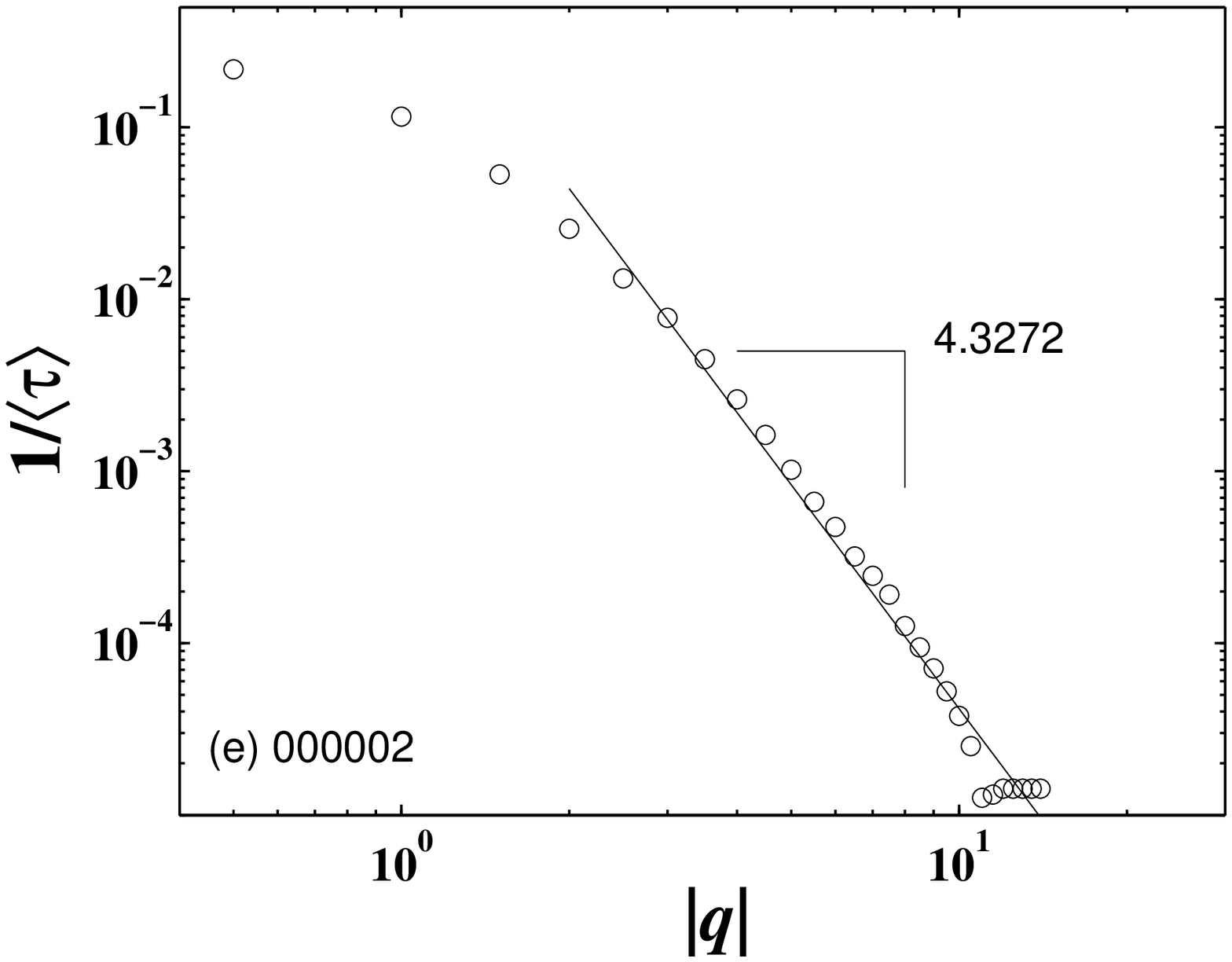}
\includegraphics[width=5cm]{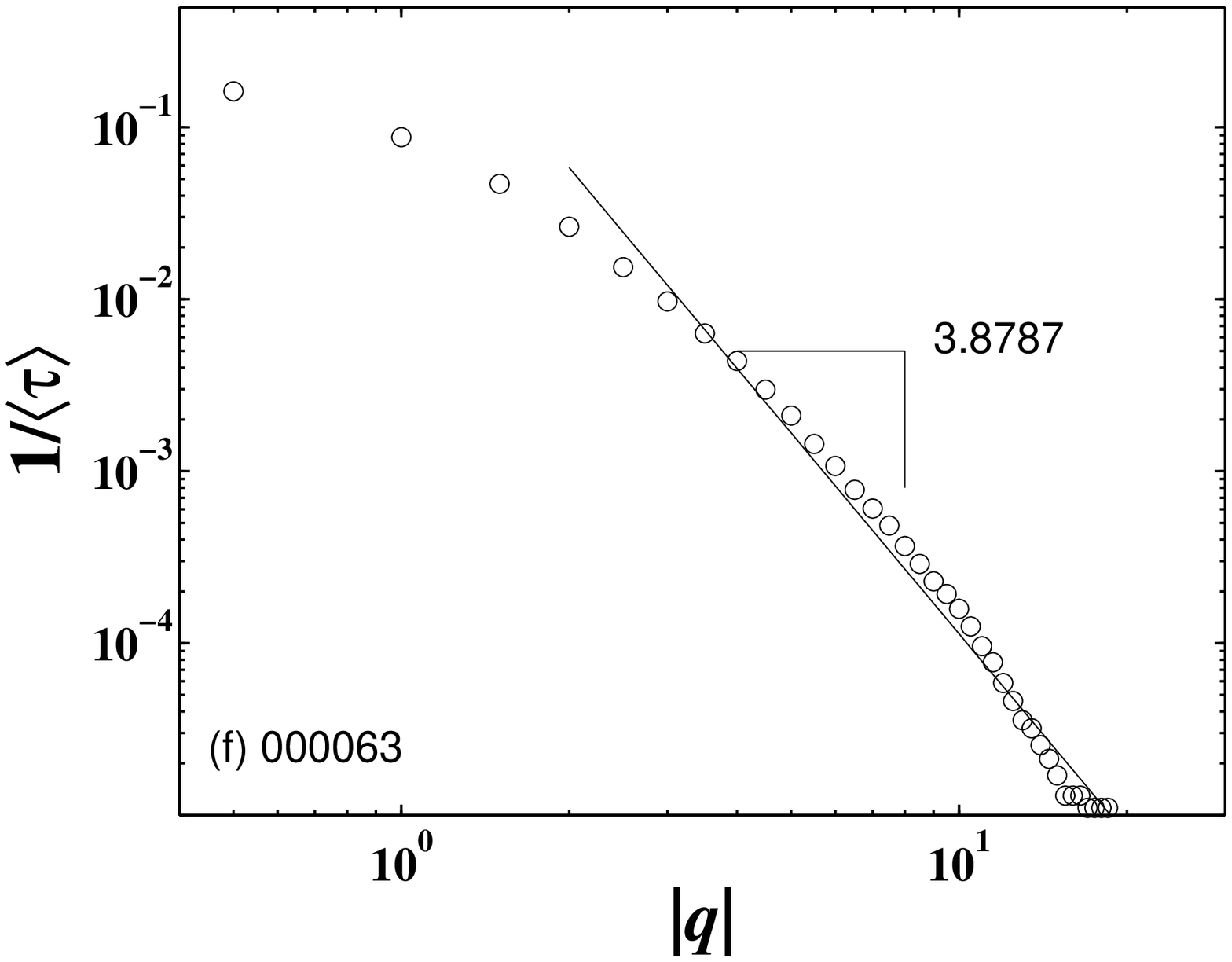}
\caption{\label{Fig:MeanRI:q}  Reciprocal of mean recurrence
interval $1/\langle \tau \rangle$ as a function of absolute
threshold $|q|$ for (a) SSEC, (b) SZCI, and (c)-(e) four
representative stocks 600009, 600060, 000002 and 000063. The solid
lines are fitted power-law curves.}
\end{figure}

For a given risk level of $VaR=q<0$, the probability $p^*$ of loss
is
\begin{equation}
 p^*=\int_{-\infty}^{q}p(r)dr
    = \frac{k}{\beta}|q|^{-\beta},\ q\leq-2.
 \label{Eq:VaR}
\end{equation}
The mean interval $\langle \tau \rangle$ is defined as
\begin{equation}
 \langle \tau \rangle=\frac{total\ number\ of\ return\ samples} {number\ of\ return\ samples\
that\ r<q}.
 \label{Eq:mean:tau}
\end{equation}
By definition, the inverse mean return represents the probability
$p^*$ and should follow
\begin{equation}
 p^*=\frac{1}{\langle \tau \rangle}
    = \frac{k}{\beta}|q|^{-\beta},\ q\leq-2.
 \label{Eq:mean:tau:q}
\end{equation}
Figure~\ref{Fig:MeanRI:q} plots $1/\langle \tau \rangle$ as a
function of $|q|$ for the two Chinese indices and four
representative stocks, and the estimated exponents show values very
close to that of the return distributions which verifies the
relation in Eq.~(\ref{Eq:mean:tau:q}).

\subsection{Conditional loss probability $p^*$}

\begin{figure}[htb]
\centering
\includegraphics[width=5cm]{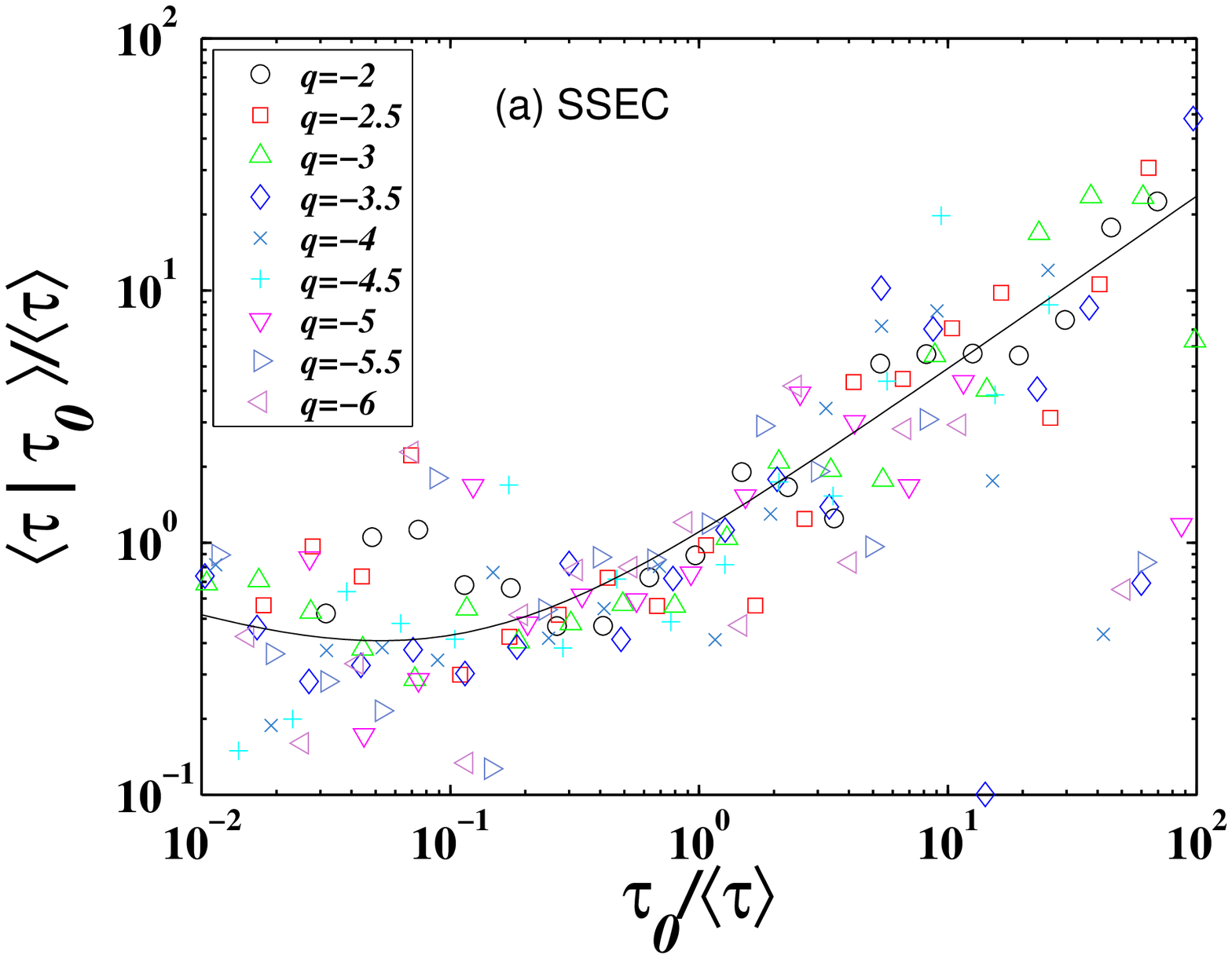}
\includegraphics[width=5cm]{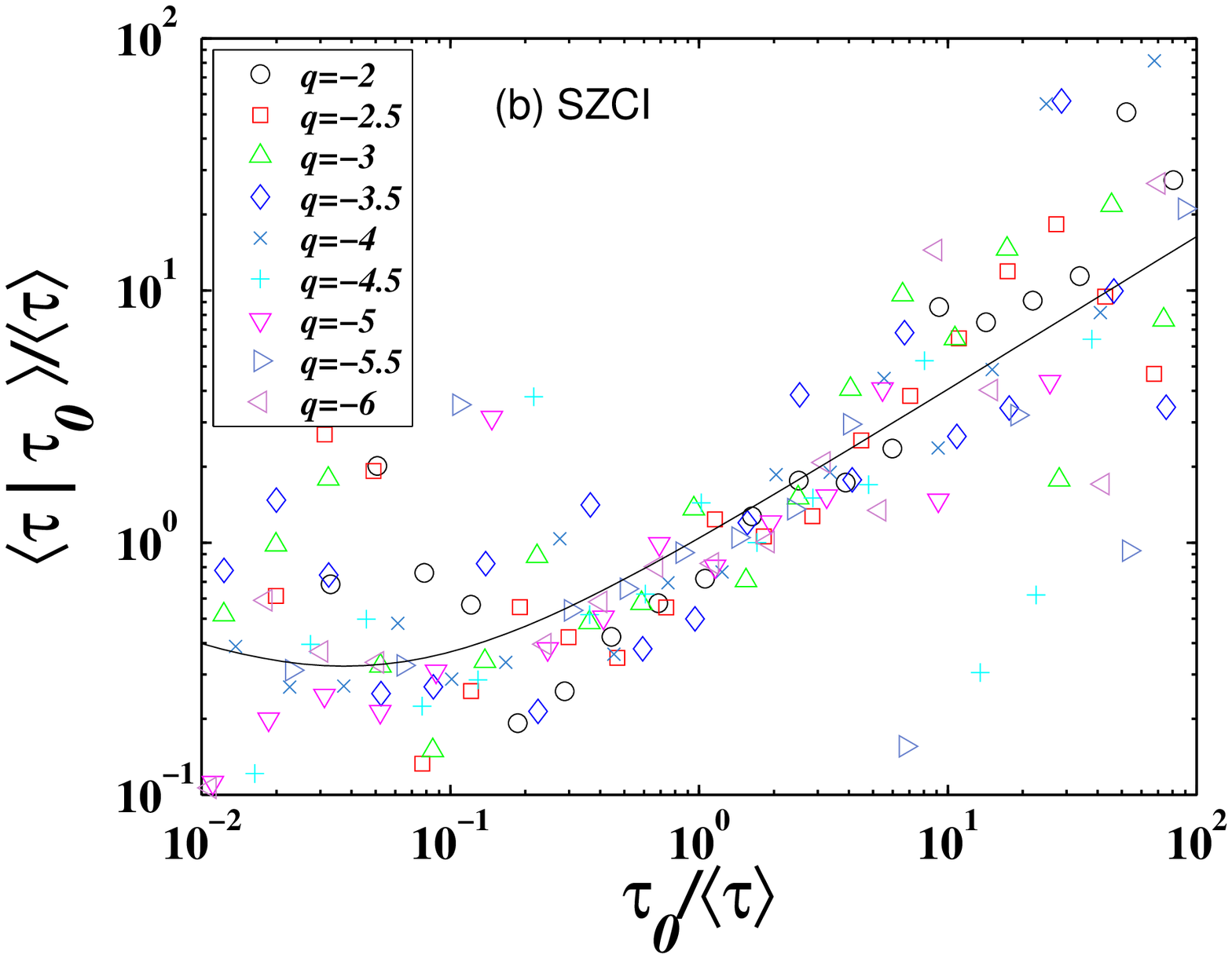}
\includegraphics[width=5cm]{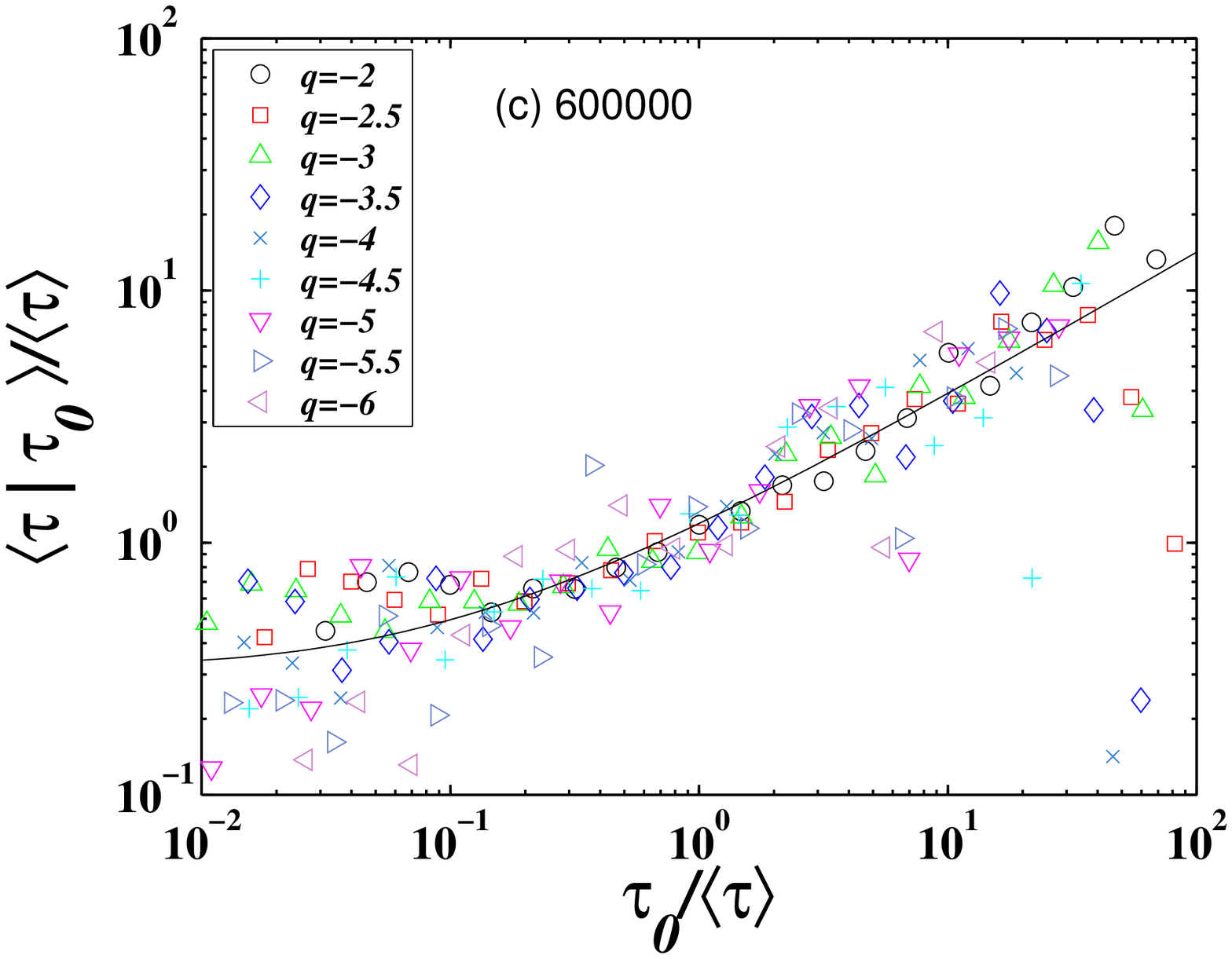}
\includegraphics[width=5cm]{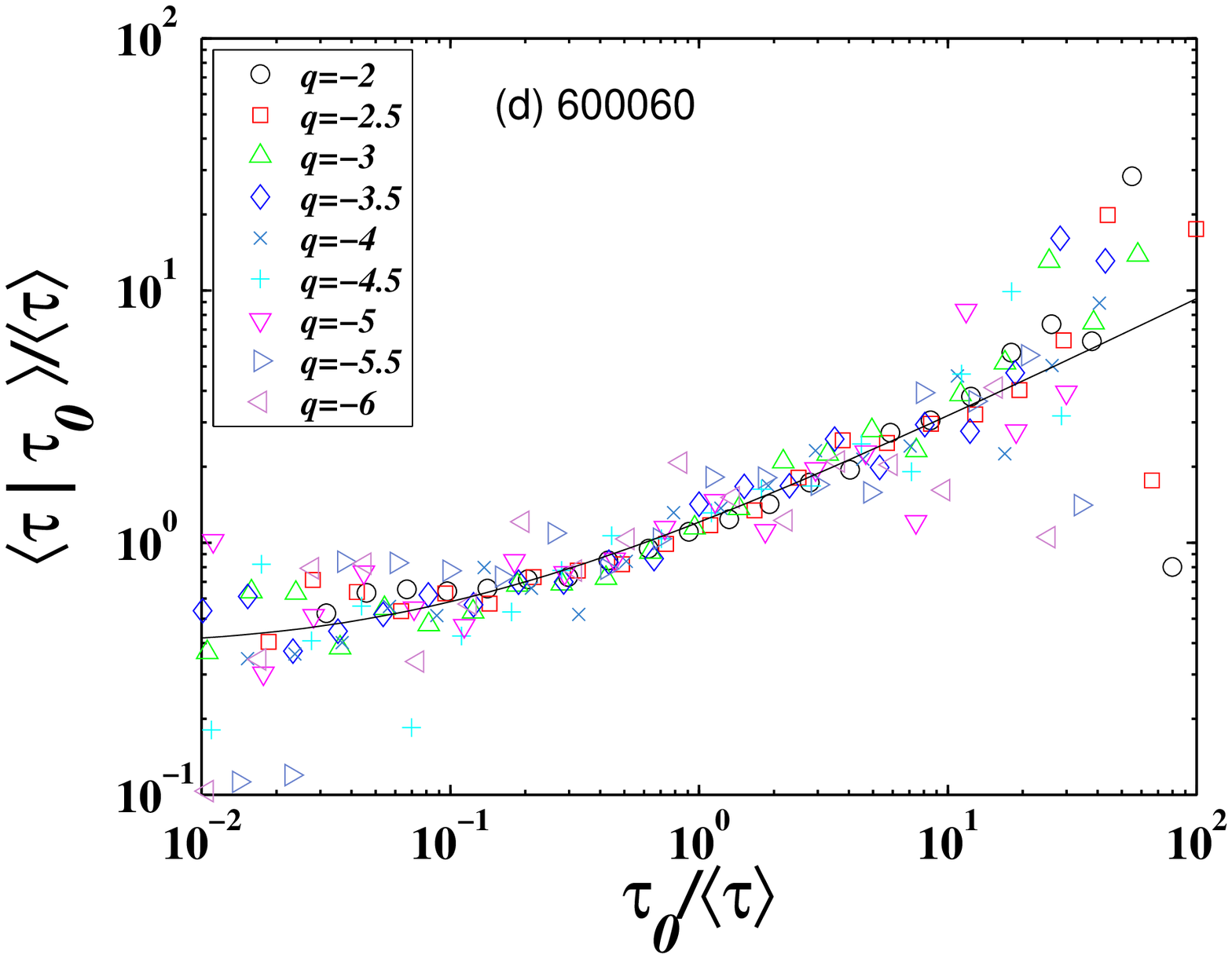}
\includegraphics[width=5cm]{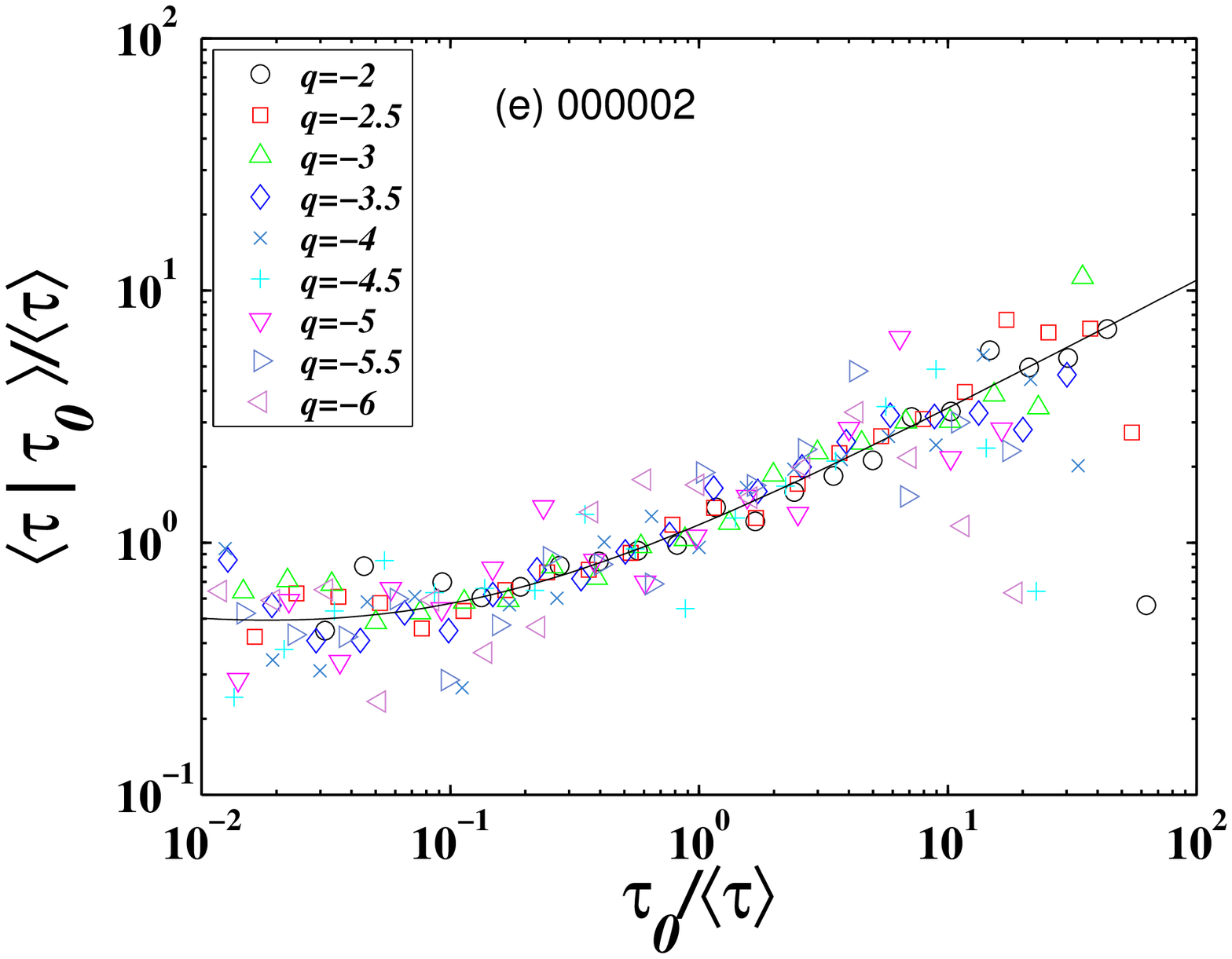}
\includegraphics[width=5cm]{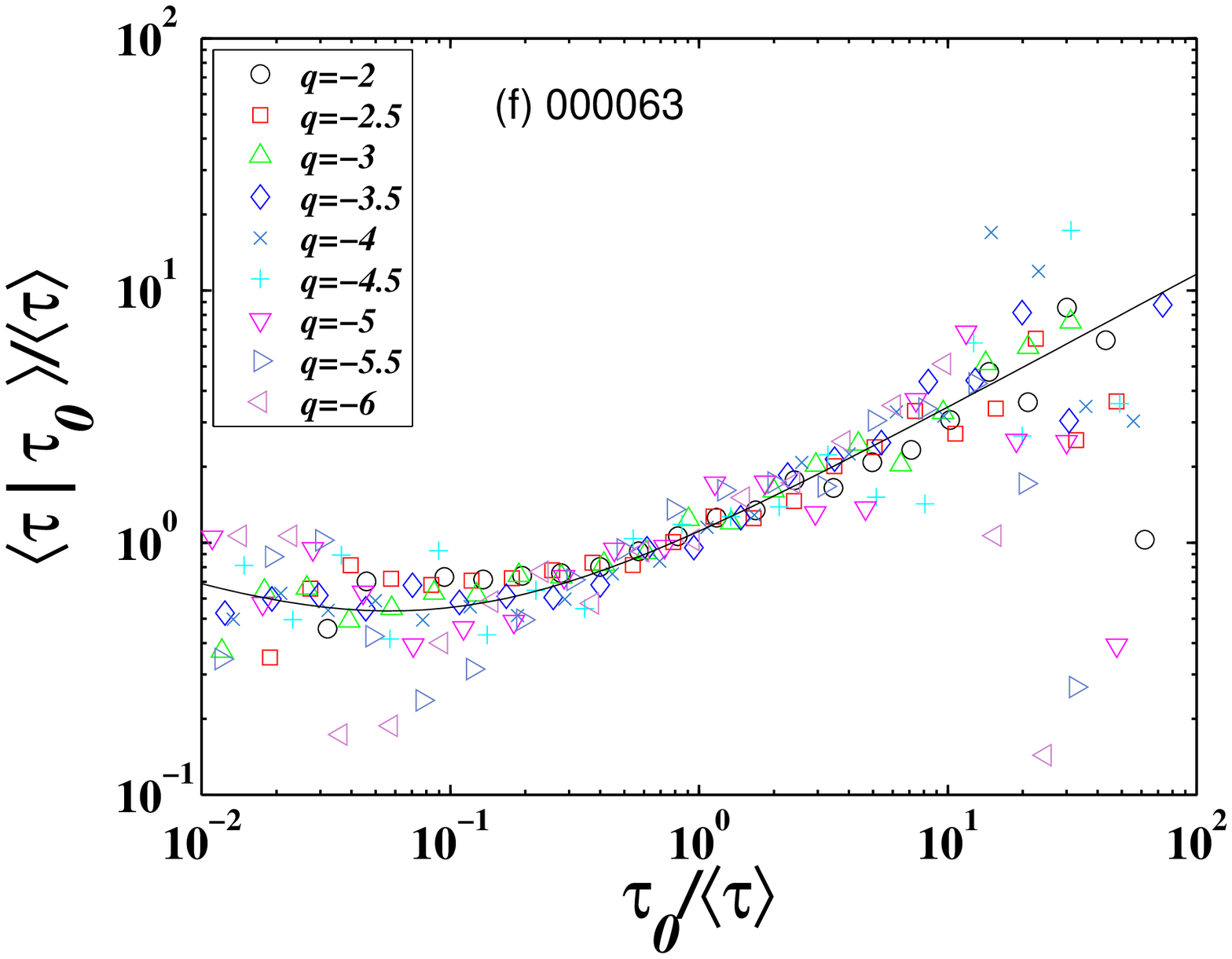}
\caption{\label{Fig:MeanRI1:tau0}  Scaled mean conditional
recurrence interval $\langle \tau| \tau_0 \rangle/ \langle \tau
\rangle$ as a function of scaled recurrence intervals $\tau_0/
\langle \tau \rangle$ for (a) SSEC, (b) SZCI, and (c)-(e) four
representative stocks 600009, 600060, 000002 and 000063.}
\end{figure}

The conditional mean recurrence interval $\langle \tau|\tau_0
\rangle$ is defined as mean recurrence interval conditioned on the
preceding interval $\tau_0$. In Figure~\ref{Fig:MeanRI1:tau0}, we
plot the scaled conditional mean recurrence interval $\langle
\tau|\tau_0 \rangle/\langle \tau \rangle$ as a function of the
scaled preceding interval $\tau_0/\langle \tau \rangle$ for the two
Chinese indices and four representative stocks. We assume that
$\langle \tau|\tau_0 \rangle/\langle \tau \rangle$ follows
\begin{equation}
 \frac{\langle \tau|\tau_0\rangle}{\langle \tau \rangle}= \left[1+\gamma \left(\frac{\tau_0}{\langle \tau \rangle}\right)^{-\mu} \right]\left(\frac{\tau_0}{\langle \tau
 \rangle}\right)^\nu.
 \label{Eq:con:mean:tau}
\end{equation}
As shown in figure~\ref{Fig:MeanRI1:tau0}, the relationship between
$\langle \tau|\tau_0 \rangle/\langle \tau \rangle$ and
$\tau_0/\langle \tau \rangle$ could be nicely described by
Eq.~(\ref{Eq:con:mean:tau}) in the medium region of $\tau_0/\langle
\tau \rangle \in (0.1,10]$. For large and small $\tau_0/\langle \tau
\rangle$, $\langle \tau|\tau_0 \rangle/\langle \tau \rangle$ for
different $q$ values evidently diverges, and could not be well
fitted by Eq.~(\ref{Eq:con:mean:tau}).

\begin{figure}[htb]
\centering
\includegraphics[width=5cm]{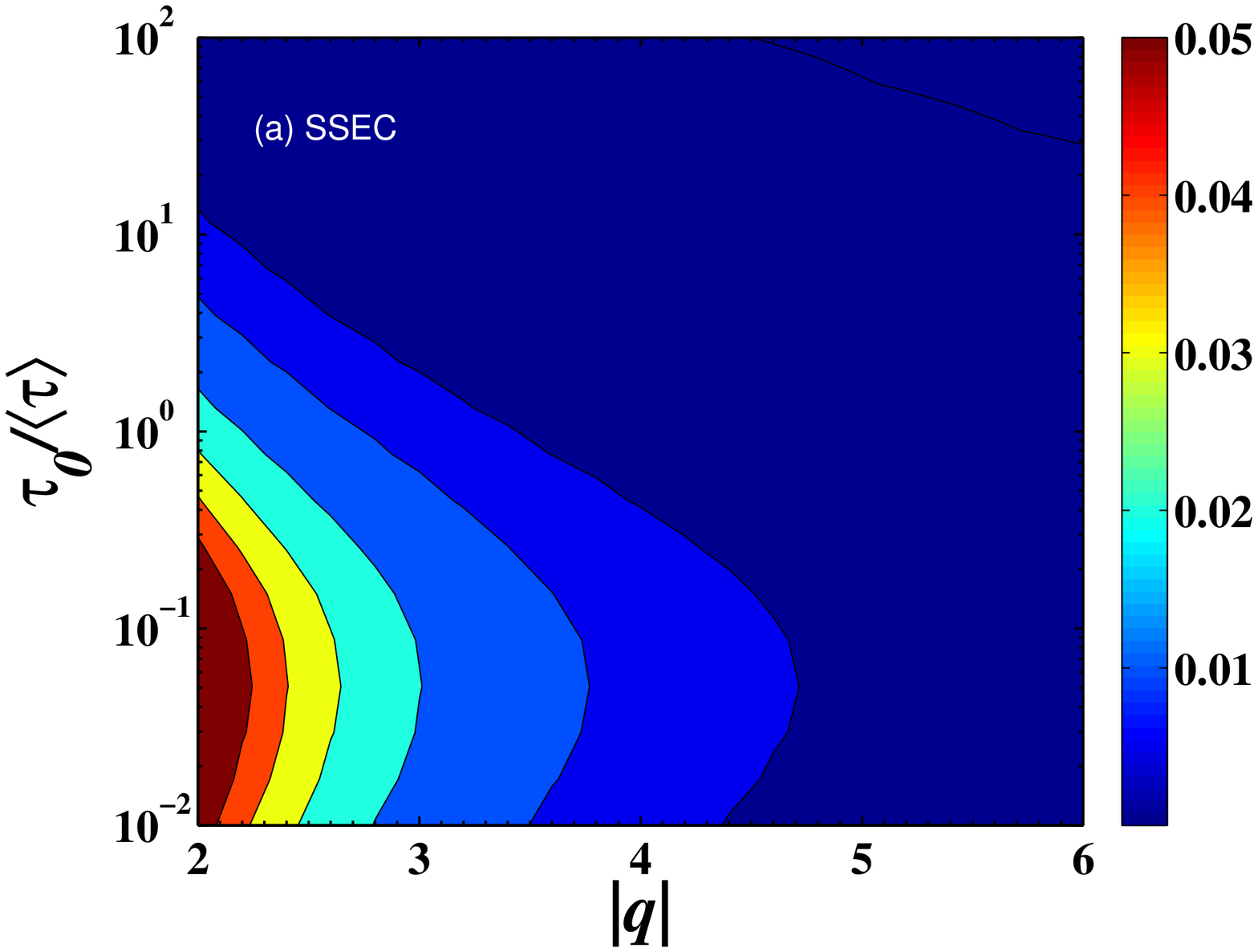}
\includegraphics[width=5cm]{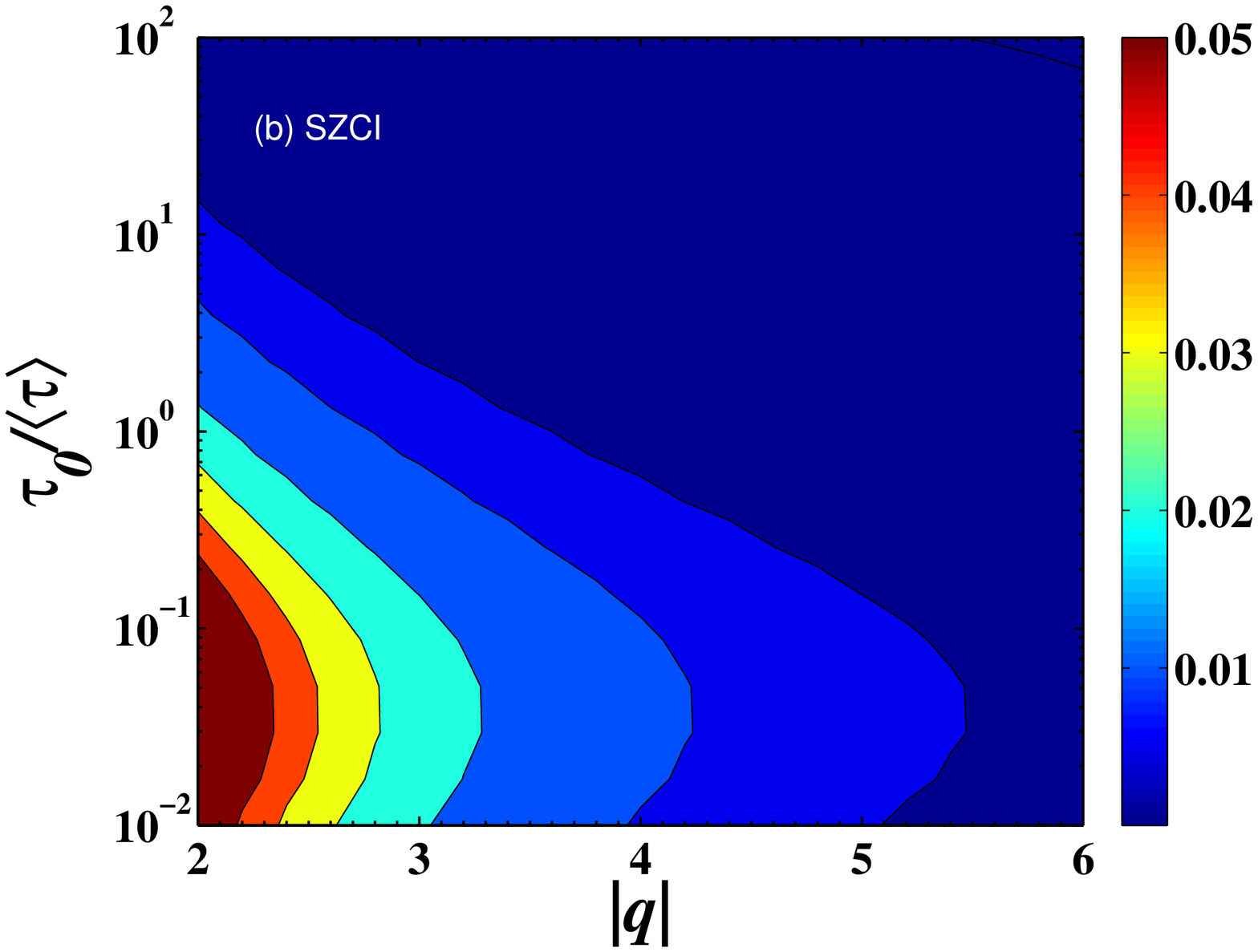}
\includegraphics[width=5cm]{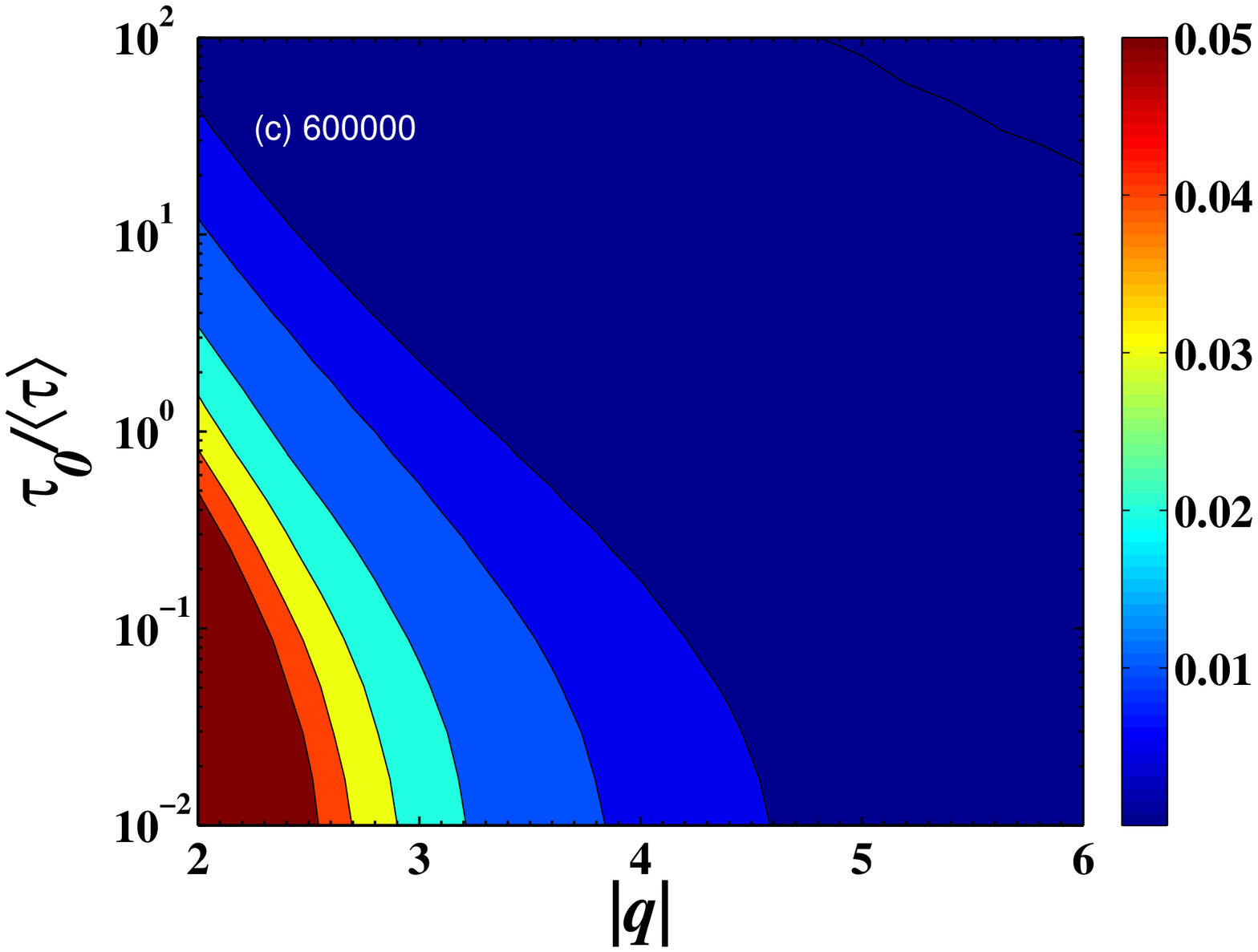}
\includegraphics[width=5cm]{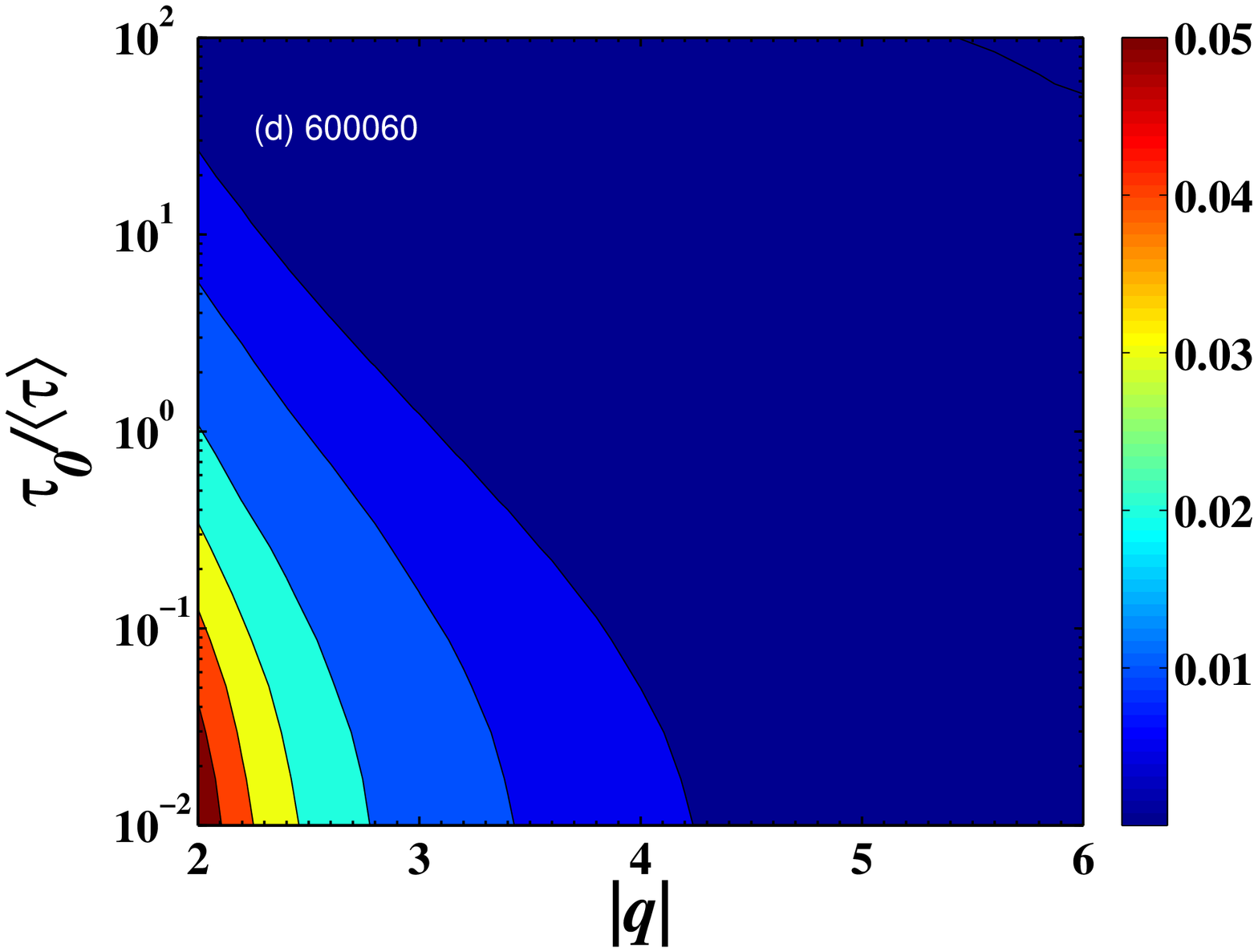}
\includegraphics[width=5cm]{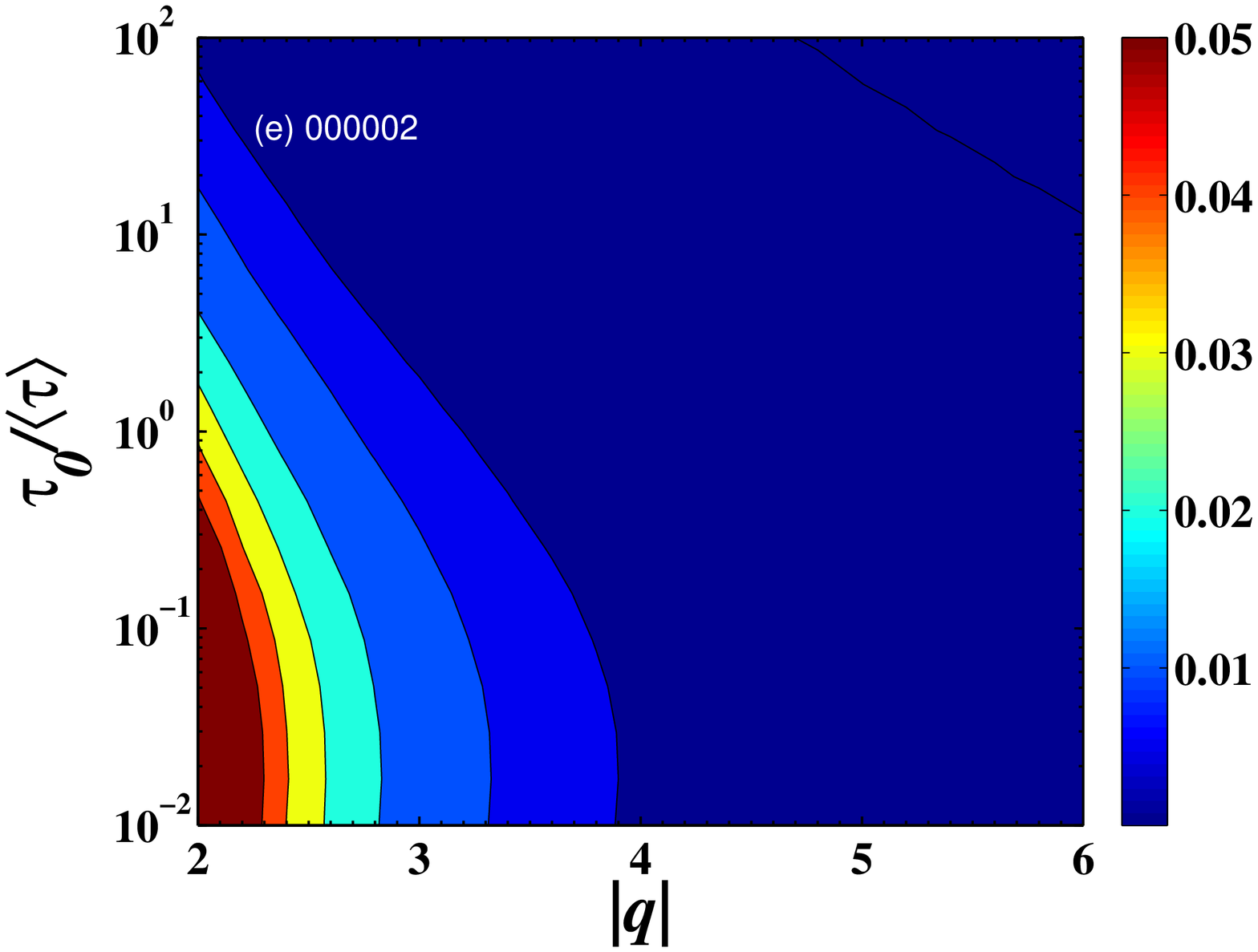}
\includegraphics[width=5cm]{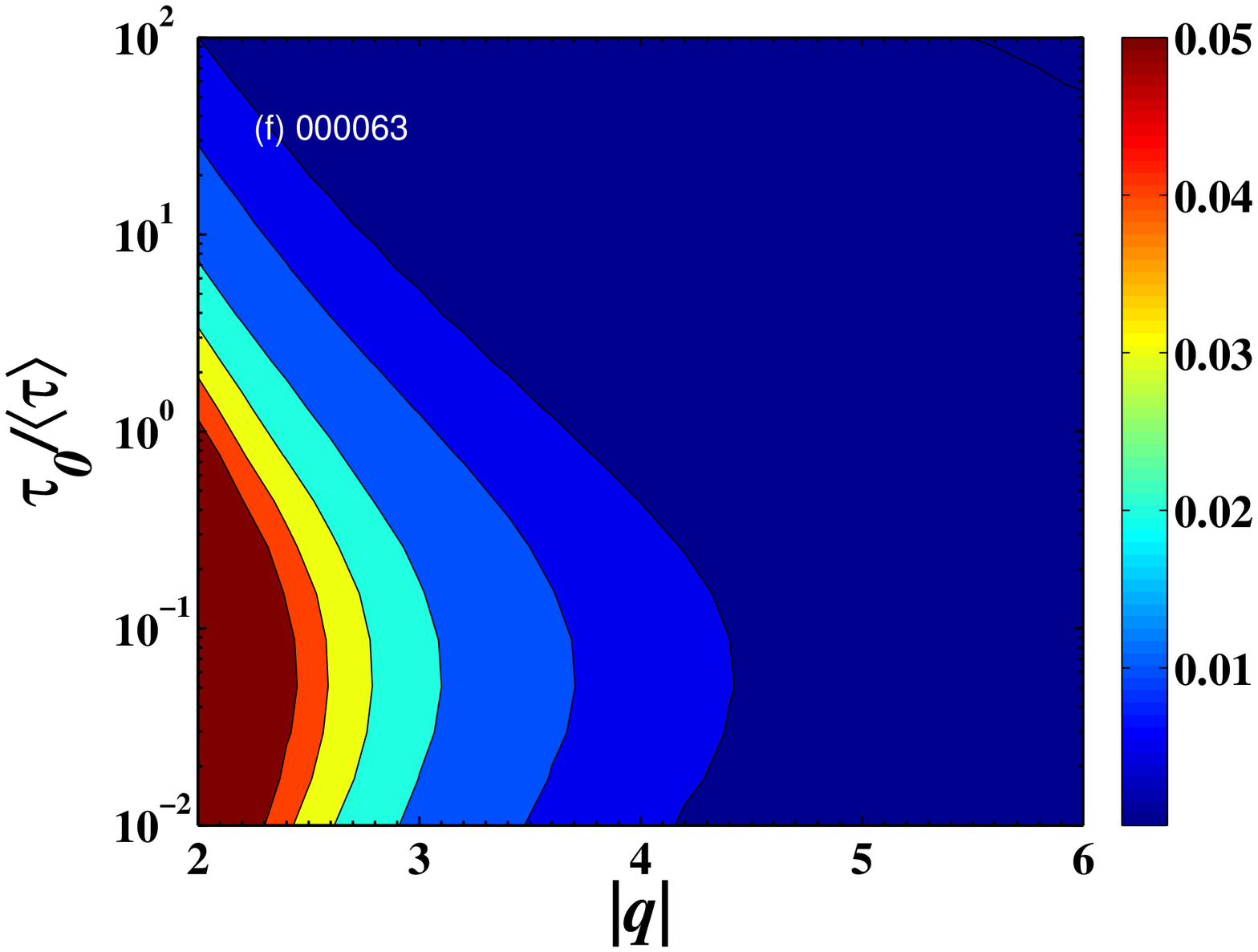}
\caption{\label{Fig:MeanRI4:tau0}  Contour maps of theoretical loss
probability $p^*$ for (a) SSEC, (b) SZCI, and (c)-(e) four
representative stocks 600009, 600060, 000002 and 000063.}
\end{figure}

\begin{figure}[htb]
\centering
\includegraphics[width=5cm]{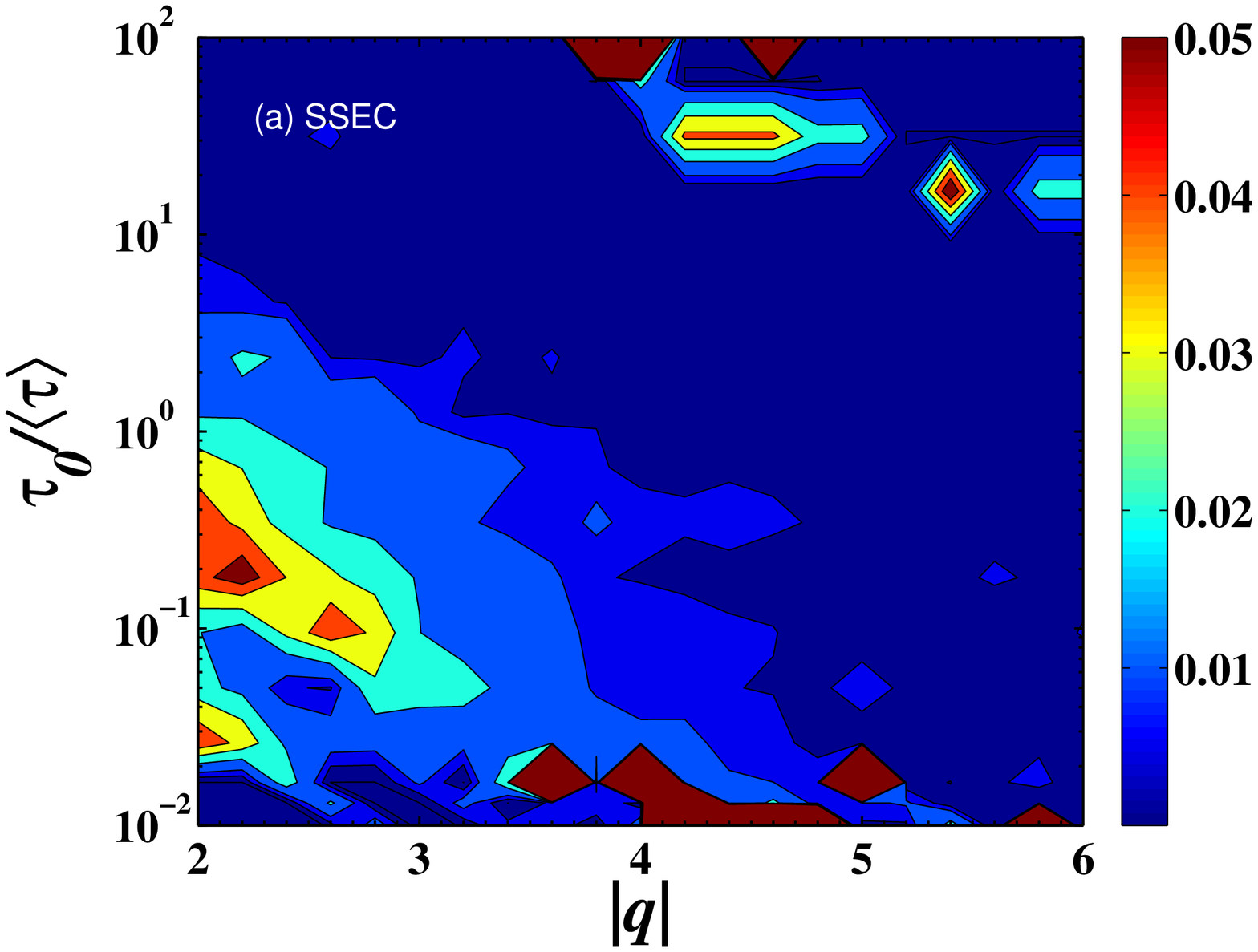}
\includegraphics[width=5cm]{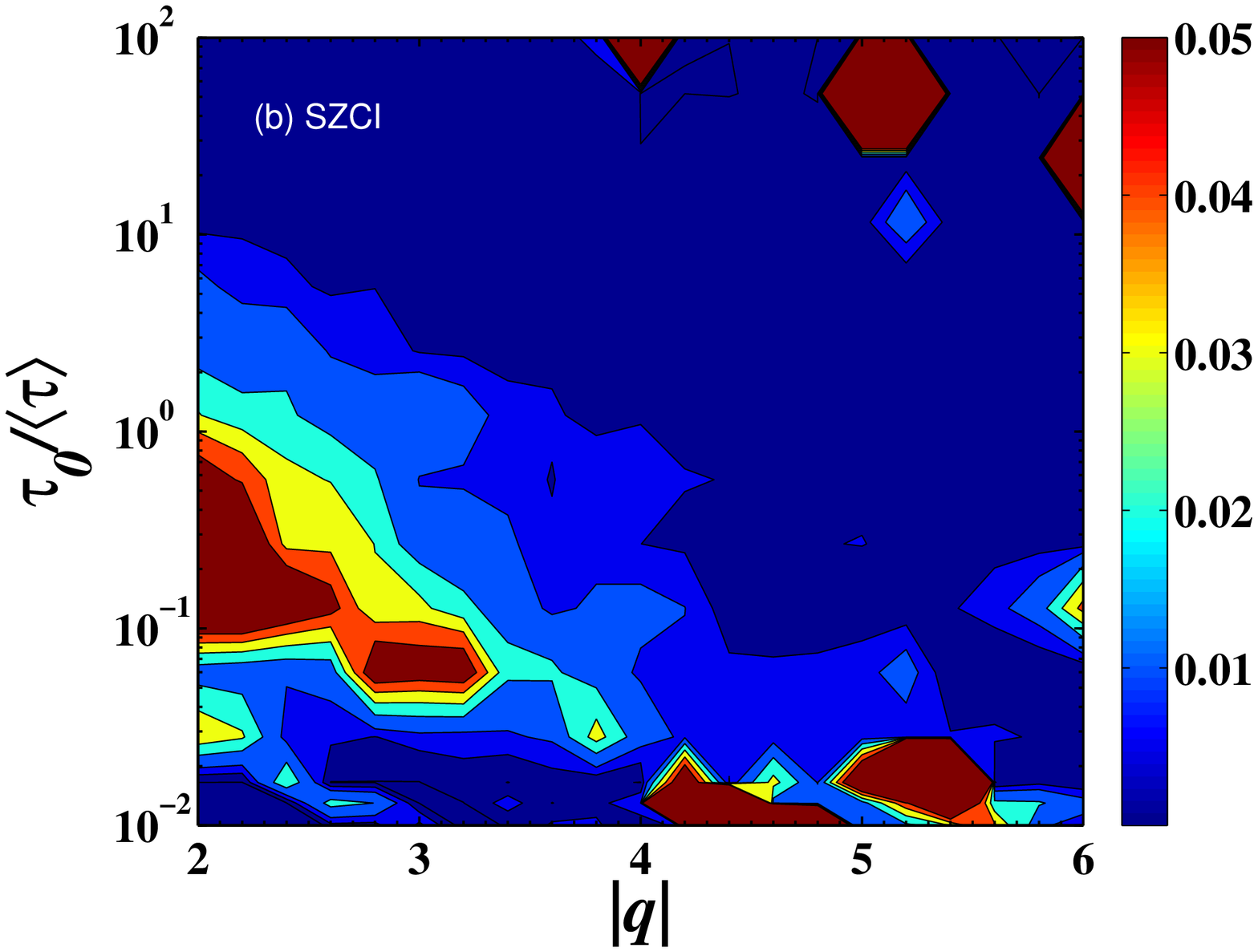}
\includegraphics[width=5cm]{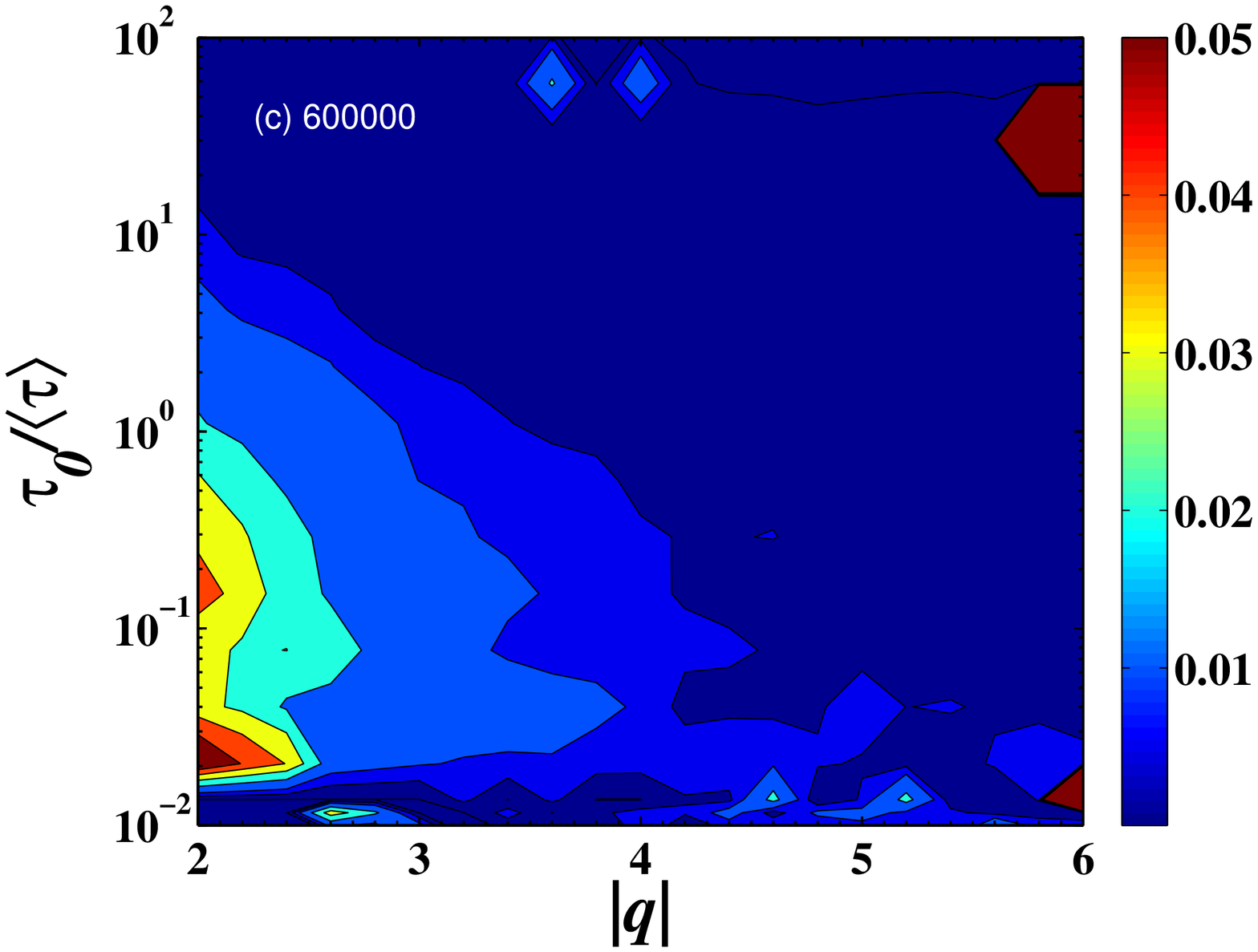}
\includegraphics[width=5cm]{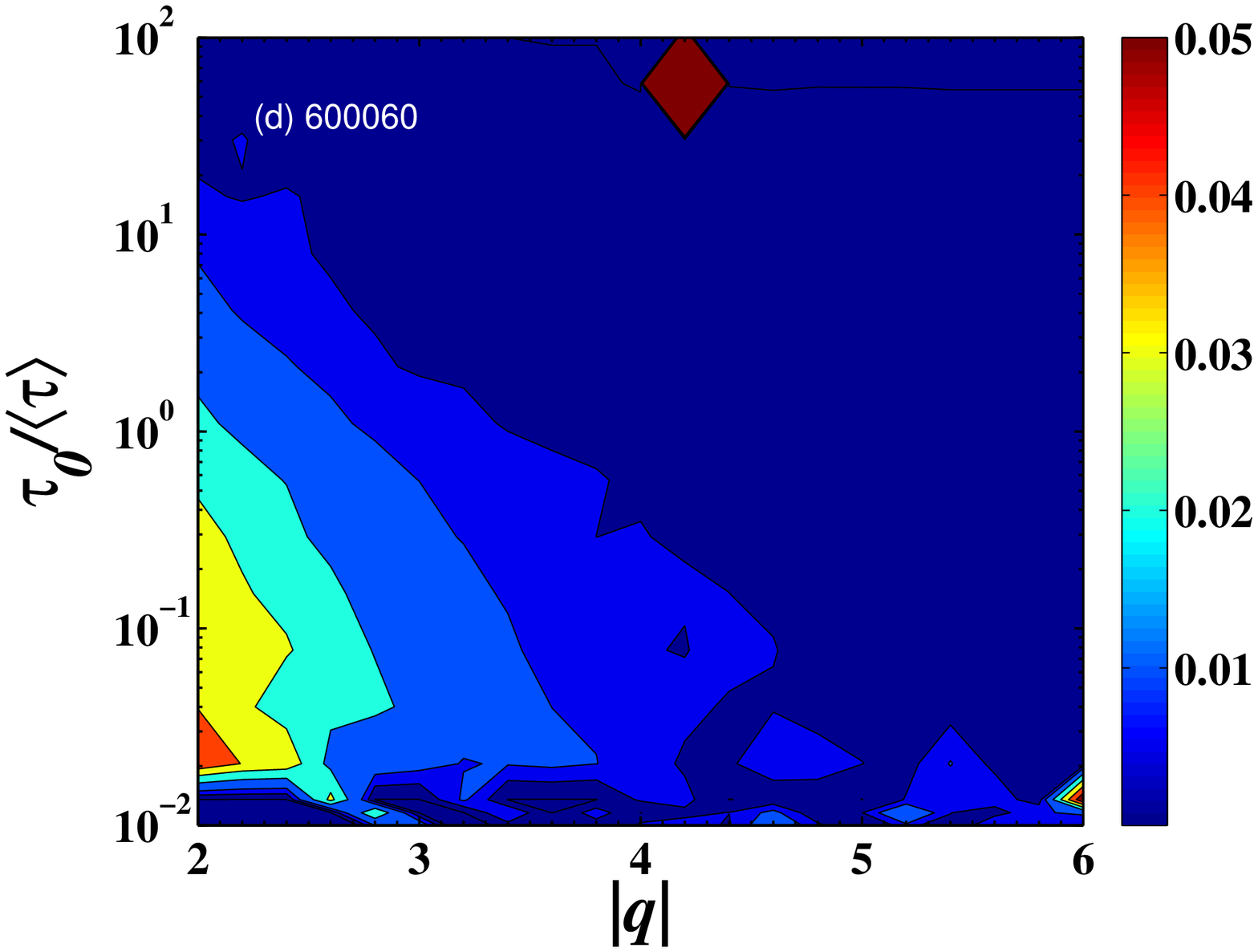}
\includegraphics[width=5cm]{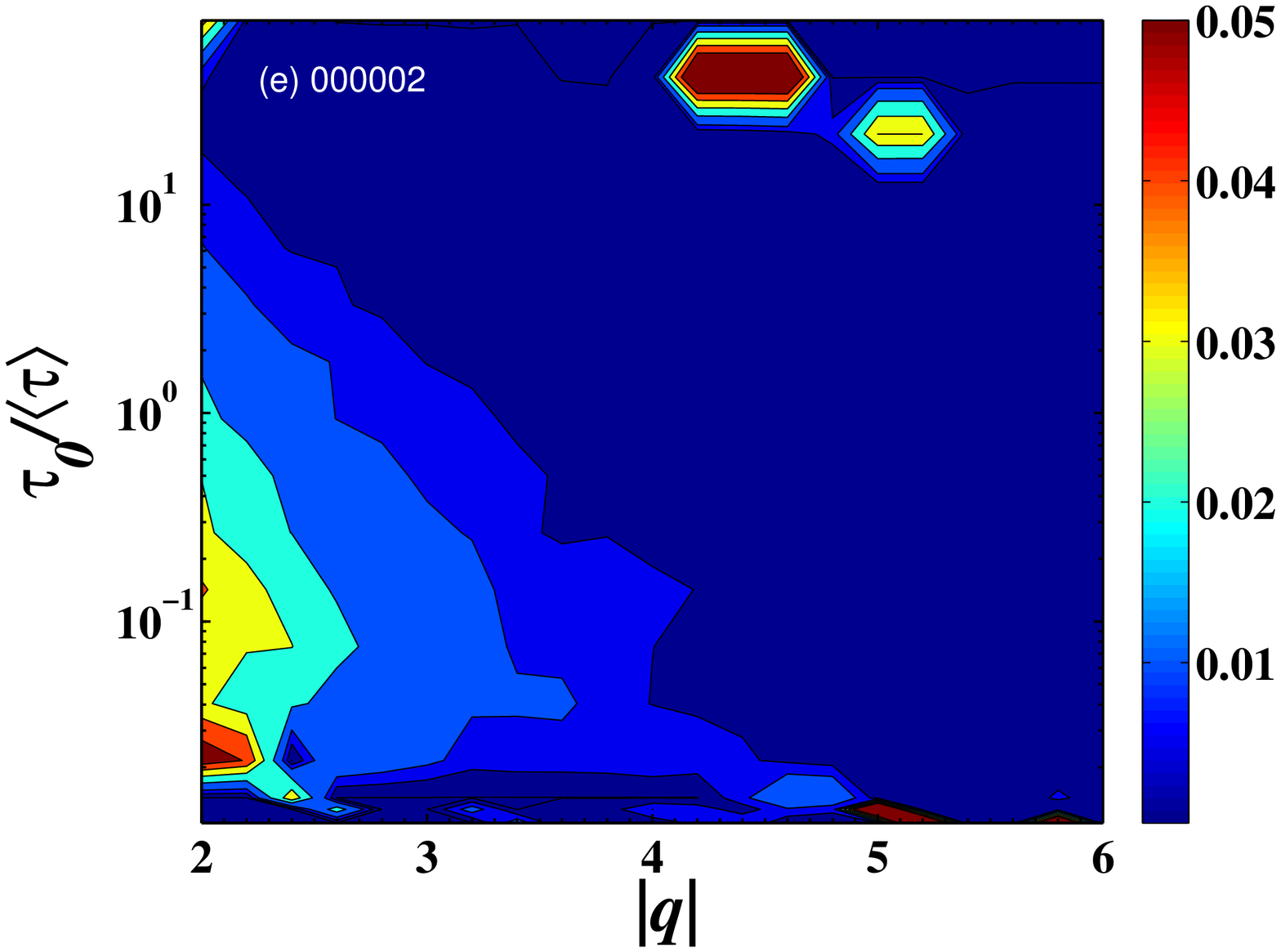}
\includegraphics[width=5cm]{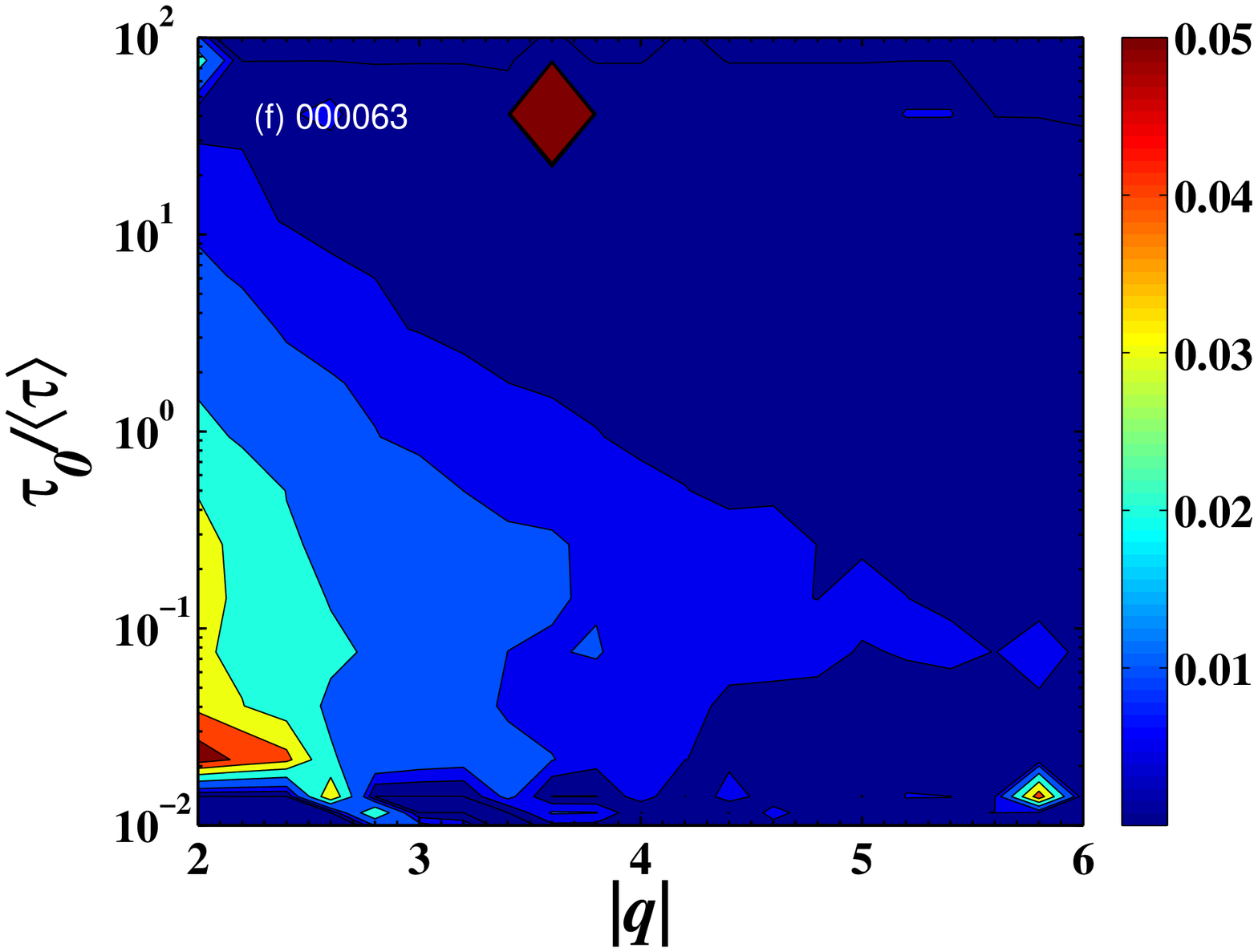}
\caption{\label{Fig:VaR:Emp}  Contour maps of empirical loss
probability $p^*$ for (a) SSEC, (b) SZCI, and (c)-(e) four
representative stocks 600009, 600060, 000002 and 000063.}
\end{figure}

Based on the conditional mean recurrence interval, we further
calculate the loss probability conditioned on the preceding interval
$\tau_0$. Similar to Eqs.~(\ref{Eq:VaR}) and ~(\ref{Eq:mean:tau:q}),
we expect that
\begin{equation}
 p^*=\frac{1}{\langle \tau |
 \tau_0\rangle}=\int_{-\infty}^{q}p(r|\tau_0)dr,
 \label{Eq:Con:VaR}
\end{equation}
where $p(r|\tau_0)$ is the probability that a return $r$ immediately
follows interval $\tau_0$, and $p^*$ is the loss probability for a
risk level of $VaR=q<0$ conditioned on the preceding interval
$\tau_0$ of losses below $q$. If we know the preceding interval
$\tau_0$, we can estimate the risk level corresponding to a certain
loss probability $p^*$. Substituting Eq.~(\ref{Eq:mean:tau:q}) into
Eq.~(\ref{Eq:con:mean:tau}), we obtain the expression of $1/\langle
\tau | \tau_0\rangle$ depends on $\tau_0/\langle \tau \rangle$ and
$q$. In figure~\ref{Fig:MeanRI4:tau0}, we plot the contours of the
theoretical conditional loss probability $p^*=1/\langle \tau |
\tau_0\rangle$ using the parameters estimated from
figures~\ref{Fig:MeanRI:q} and ~\ref{Fig:MeanRI1:tau0}. For
comparison purposes, we also plot the the contours of empirical
conditional loss probability in figure~\ref{Fig:VaR:Emp}. Patterns
of the two contour maps are similar except for large and small
$\tau_0/\langle \tau \rangle$. The large values of empirical loss
probability for large or small $\tau_0/\langle \tau \rangle$
indicate the high risk of extreme loses event conditioned on a long
or short elapsed time since the previous extreme event. To better
theoretically estimate the conditional loss probability, we need to
further refine the relationship between $\langle \tau|\tau_0
\rangle/\langle \tau \rangle$ and $\tau_0/\langle \tau \rangle$ in
Eq.~(\ref{Eq:con:mean:tau}) for large and small $\tau_0/\langle \tau
\rangle$.

\section{Conclusion}
\label{S1:Conclusion}

We have studied the probability distribution and the memory effect
of recurrence intervals between two consecutive returns above a
positive threshold $q>0$ or below a negative threshold $q<0$ for
SSEC, SZCI and 20 liquid stocks of the Chinese stock markets.  Since
our data sets are 1-min high-frequency returns, the statistics of
our findings are much better than previous results using daily data
\cite{Yamasaki-Muchnik-Havlin-Bunde-Stanley-2006-inPFE,Bogachev-Eichner-Bunde-2007-PRL,Bogachev-Bunde-2008-PRE}.

We found that the PDFs of recurrence interval with different
thresholds collapse onto a single curve for each index and
individual stocks and are symmetric with respect to positive and
negative returns. The tails of the recurrence interval distributions
for different values of threshold $q$ follow scaling behavior, and
the goodness-of-fit test shows that the scaling function of 16
stocks and SZCI could be approximated by power law under the
significance level of 1\% using the $KS$, $KSW$ and $CvM$
statistics.

The investigation of the conditional PDF $P_q(\tau|\tau_0)$ and the
detrended fluctuation function $F(l)$ demonstrates the existence of
both short-term and long-term correlation in the recurrence
intervals. To the best of our knowledge, the memory effects of
recurrence intervals of financial returns have not been studied
previously, although the memory effects of volatility recurrence
intervals are well documented. The long-term correlation in the
recurrence intervals is usually attributed to the long-term
correlations in the original time series
\cite{Yamasaki-Muchnik-Havlin-Bunde-Stanley-2005-PNAS}. In the
current case, although the returns are uncorrelated, the positive
returns are strongly correlated, and so are the negative returns.
Therefore, the long memory in the recurrence intervals can also be
attributed to the long memory in the positive and negative return
series.

We also apply the recurrence interval analysis to the risk
estimation for Chinese stock markets, and provide relatively
accurate estimations of the probability $W_q(\Delta{t}|t)$ and the
conditional loss probability $p^*$. These results are useful for the
assessment and management of risks in financial engineering. We
submit that recurrence interval analysis of financial returns
pioneered in
Ref.~\cite{Yamasaki-Muchnik-Havlin-Bunde-Stanley-2006-inPFE,Bogachev-Eichner-Bunde-2007-PRL,Bogachev-Bunde-2008-PRE,Bogachev-Bunde-2009-PRE}
has the potential power to forging a link between econophysics and
financial engineering and more work should be done in this
direction.

\section*{Acknowledgments:}

We are grateful to Kun Guo (Research Center on Fictitious Economics
\& Data Science, Chinese Academy of Sciences) for retrieving the
data analyzed in this work and Gao-Feng Gu (School of Business, East
China University of Science and Technology) for preprocessing the
data. This work was partially supported by the Shanghai Educational
Development Foundation (2008CG37 and 2008SG29) and the Program for
New Century Excellent Talents in University (NCET-07-0288).

\bibliographystyle{iopart-num} 
\bibliography{E:/Papers/Auxiliary/Bibliography} 

\end{document}